\documentclass[11pt,a4paper]{article}
\usepackage{jheppub}
\usepackage[utf8]{inputenc}
\usepackage{graphicx,fancybox,float,comment}
\usepackage{pdflscape}
\usepackage{units}
\usepackage{multirow,array,arydshln}
\usepackage[dvipsnames]{xcolor}
\usepackage{braket,tensor}
\usepackage{amsmath,amssymb,amsfonts,mathrsfs}
\usepackage{tikz-cd}
\usepackage{mdframed}

\usepackage{ulem}

\global\interfootnotelinepenalty=1000

\newcommand{\exd}{\mathrm{d}}
\DeclareMathOperator{\arctanh}{arctanh}

\preprint{\texttt{IFT-UAM/CSIC-24-121}}

\title{Perturbing a quantum black hole}

\author[a]{Casey Cartwright,}
\emailAdd{c.c.cartwright@uu.nl}
\author[a]{Umut G\"{u}rsoy,}
\emailAdd{u.gursoy@uu.nl}
\author[b]{Juan F. Pedraza}
\emailAdd{j.pedraza@csic.es}
\author[a]{and Guim Planella Planas}
\emailAdd{g.planellaiplanas1@uu.nl}
\affiliation[a]{Institute for Theoretical Physics, Utrecht University, Princetonplein 5, 3584 CC Utrecht,\\The Netherlands}
\affiliation[b]{Instituto de F\'{i}sica Te\'{o}rica UAM/CSIC, Calle Nicol\'{a}s Cabrera 13-15, Madrid 28049, Spain}

\abstract{We analyze the analytic structure of correlators in the field theory dual to the quantum Ba\~{n}ados-Teitelboim-Zanelli (qBTZ) black hole, a braneworld model incorporating exact backreaction from quantum conformal matter. We first compute the quasi-normal mode (QNM) spectrum of operators with dimension $\Delta$ and spin $s=0,\pm 1/2$. The leading QNMs and their overtones display qualitatively different behavior depending on the branch of qBTZ solution, which corresponds to distinct CFT states: branch 1 is a conical singularity dressed with a horizon while branch 2 is a quantum-corrected BTZ black hole. Consequently, the relaxation of probe matter effectively differentiates the CFT states and identifies the corresponding bulk descriptions. We then turn to pole-skipping locations where Green's functions are not unique. At these points, frequency is proportional to temperature, but momentum exhibits complex temperature dependence due to quantum effects. Under the assumption that the pole-skipping point closest to the origin reflects quantum chaos, we infer the likely behavior of the quantum Lyapunov exponent and butterfly velocity in the dual theory. Finally, we examine pole collisions in complex momentum space, showing that quantum corrections imprint a unique signature on the analytic structure of the poles in retarded Green's functions, resulting in level-crossing phenomena that differ notably from the level-touching phenomena in the uncorrected BTZ geometry.
}

\begin{document}
\newcommand{\rb}{\bar{r}}
\newcommand{\phib}{\bar{\phi}}
\newcommand{\w}{\mathfrak{w}}
\newcommand{\q}{\mathfrak{q}}
\newcommand{\vev}[1]{\braket{#1}}
\maketitle

\section{Introduction and summary}
\label{sec:intro}

\subsection*{Motivation and antecedents}
The AdS/CFT correspondence has become an invaluable tool in the study of quantum field theory and gravitational physics. In its early development, the analysis of CFT correlation functions was pivotal in validating the correspondence, laying the groundwork for the holographic dictionary~\cite{Witten:1998qj,Gubser:1998bc}. These concepts were swiftly extended to include geometries with black holes in the bulk, which correspond to thermal states of the dual CFT~\cite{Horowitz:1999jd}. This led to the conjecture that the return of perturbations to thermal equilibrium in the CFT is captured by the quasi-normal modes (QNMs) of the corresponding black hole. Owing to its analytical tractability, the QNMs of the BTZ black hole were derived~\cite{Birmingham:2001hc,Govindarajan:2000vq,Cardoso:2001hn} and later identified as the poles of retarded Green's functions~\cite{Birmingham:2001pj}, thereby confirming their role in describing the CFT's relaxation toward thermal equilibrium.

Since then, a vast body of literature has emerged, focusing on the interpretation of QNMs within CFTs and their application to studying strongly coupled phenomena. This research encompasses a range of topics, including real-time formulations~\cite{Son:2002sd}, lower bounds on transport coefficients~\cite{Kovtun:2004de}, hydrodynamic approximations of Green's functions~\cite{Policastro:2002se,Policastro:2002tn}, and extensions to higher dimensions~\cite{Kovtun:2004de}. Additionally, numerous studies, too many to cite individually, have explored the properties of strongly coupled CFTs at finite temperature across different dimensions and field contents.

In the past decade, there has been a renewed surge of interest in the interpretation of QNMs and the structure of Green's functions in holographic CFTs. This research has focused on intricate aspects of holographic Green's functions, including pole-skipping~\cite{Natsuume:2019sfp,Natsuume:2019xcy,Blake:2019otz} and its connection to quantum chaos~\cite{Shenker:2013pqa,Shenker:2015keq,Schalm:2018lep}, the analytic structure of dispersion relations, especially regarding pole collisions and bounds on the convergence radius of hydrodynamic expansions~\cite{Withers:2018srf,Grozdanov:2019kge,Grozdanov:2019uhi,Heller:2020hnq,Grozdanov:2020koi,Carballo:2024kbk}, interactions between QNMs and higher-point functions of CFT operators beyond linear response~\cite{Pantelidou:2022ftm}, spectral reconstruction~\cite{Grozdanov:2022npo,Grozdanov:2023tag}, and new insights into the stability of the QNM spectrum~\cite{Arean:2023ejh,Cownden:2023dam}.

The AdS/CFT correspondence, however, proves useful in more than just the analysis of correlation functions and their geometric counterparts. As a duality between supersymmetric CFTs and string theory, it provides a powerful framework for exploring quantum gravity beyond classical limits. A notable area of interest is the study of `quantum' black holes, first introduced in~\cite{Emparan:1999wa,Emparan:1999fd,Emparan:2002px}. Following early ideas by Randall-Sundrum and Karch-Randall~\cite{Randall:1999vf, Randall:1999ee,Karch:2000ct,Karch:2000gx}, this research employs so-called braneworld models to study \textit{semi-classical} black holes localized on a brane within an asymptotically AdS bulk. Recently, renewed interest has arisen in these models \cite{Emparan:2020znc,Emparan:2022ijy,Panella:2023lsi,Feng:2024uia,Climent:2024nuj,Panella:2024sor}, particularly with the advent of `double holography' and its implications for the information paradox \cite{Almheiri:2019hni,Geng:2020qvw,Geng:2020fxl,Chen:2020uac,Chen:2020hmv}. Further advancements and applications of quantum black holes have been investigated in~\cite{Emparan:2021hyr,Frassino:2022zaz,Chen:2023tpi,Kolanowski:2023hvh,Frassino:2023wpc,Johnson:2023dtf, HosseiniMansoori:2024bfi,Wu:2024txe,Frassino:2024fin,Frassino:2024bjg,Xu:2024iji}.

Braneworld models have been instrumental in exploring quantum effects in gravity. This has been accomplished by establishing a precise holographic correspondence between: (i) classical solutions of the $(d+1)$-dimensional Einstein bulk theory coupled to a brane, (ii) a $d$-dimensional semi-classical gravity theory on the brane coupled to a quantum matter sector, and (iii) a boundary description without gravity, in terms of a $d$-dimensional conformal field theory with boundary (BCFT$_d$) coupled to a $(d-1)$-dimensional defect conformal field theory (DCFT$_{d-1}$). Of particular interest to us is that, in the intermediate description, the following field equations are satisfied:
\begin{equation}\label{eq:scEE}
    G_{\mu\nu}+\Lambda_d\, g_{\mu\nu}+\cdots=8\pi G_{d}\braket{T_{\mu\nu}}_{\text{CFT}}\, .
\end{equation}
Introducing a brane into an AdS bulk imposes a UV cutoff on the dual CFT, which in turn introduces a normalizable $d$-dimensional graviton on the brane. This cutoff splits the CFT degrees of freedom, revealing two distinct types of corrections in the $d$-dimensional brane theory. High-energy modes above the cutoff contribute to brane gravity, producing higher curvature corrections to the effective theory, as indicated by the ellipses in (\ref{eq:scEE}). Meanwhile, low-energy modes below the cutoff generate a large-$N$ CFT sector whose stress tensor expectation value sources the corrected field equations.

A couple of points are worth emphasizing. First, backreaction effects on classical geometries can be much larger than the Planck length, $L_P$. This is because the expectation value of the brane CFT stress-tensor is proportional to its central charge (number of species), $c\gg1$, and even though $G_d$ is small in units of the AdS radius $L$ (so that $L^{d-1}/G_d\sim L/L_P\sim N^2$ is large), quantum backreaction effects are enhanced by $c$. Second, braneworld black holes are in fact \textit{exact} in the parameter $\nu\sim c\cdot(G_d/L^{d-1})$, and thus account for all orders in backreaction (all order matter loops). This is in stark contrast to standard perturbative analyses of quantum backreaction, where one is limited to perturbatively small quantum effects. This implies that braneworld black holes are robust against quantum gravitational effects due to graviton loops, which are suppressed by $L_P$.

A prominent example of an exact quantum black hole is the quantum BTZ (qBTZ) solution \cite{Emparan:2020znc}, which appears as an induced black hole within a Karch-Randall AdS$_3$ brane embedded in a slice of the AdS$_4$ C-metric. Due to its relative simplicity, the qBTZ geometry provides an excellent framework for investigating quantum backreaction effects in a controlled setting. While previous studies have predominantly focused on equilibrium properties (see, e.g., \cite{Emparan:2021hyr,Frassino:2022zaz,Chen:2023tpi,Kolanowski:2023hvh,Frassino:2023wpc,Johnson:2023dtf,HosseiniMansoori:2024bfi,Wu:2024txe,Frassino:2024fin,Frassino:2024bjg,Xu:2024iji}), our work aims to extend these investigations by exploring how quantum corrections influence the correlation functions of the CFT dual to the qBTZ geometry, particularly slightly away from thermal equilibrium. Specifically, we will focus on the QNM spectrum of spin\footnote{In $(2+1)$-dimensions, a spin-$1$ field with a 2-form field strength $F_{\mu\nu}\exd x^\mu\wedge \exd x^\nu$ is Poincar\'{e} dual to a spin-$0$ scalar field $\Phi$, so we will consider only the non-trivial cases of $s=0$ and $s=1/2$.} $s=0$ and $s=1/2$ probe matter confined to the brane,\footnote{Previous studies~\cite{Chung:2015mna} explored whether a brane-bound observer could detect the additional dimension of the ambient bulk spacetime through local measurements.} which naturally map to the poles of the retarded Green's functions in the dual CFT.

%%%%%%%%%%%%%%%%%%%%%%%%%%%%%%%%%%%%%%%%%%%%%%%%%%%%%%%%%%%5
\subsection*{Summary and main results}

Our study focuses on the impact of coupling gravity to quantum conformal matter on the analytic structure of the poles of correlators for operators $\mathcal{O}$ with conformal weights $(h_L, h_R)$, satisfying $h_L + h_R = \Delta$ and $h_R - h_L = 0, \pm \frac{1}{2}$. Already here we find something new; a novel feature of the AdS braneworld construction in AdS$_{3+1}$ is an explicit realization of the \textit{quantum censorship of conical singularities}~\cite{Emparan:2002px}, discussed in more detail~\footnote{See~\cite{Bhattacharjee:2020nul} for a related discussion detailing scalar quasinormal modes in JT gravity to probe the rescuing of strong cosmic censorship by quantum effects.}  in~\cite{Casals:2016ioo,Casals:2016odj,Emparan:2020znc}. Specifically, for a certain range of parameters, the low-energy degrees of freedom of the brane CFT act to `dress' conical singularities with a horizon, thereby shielding local observers from these singularities. Our findings reveal that the quasinormal mode (QNM) spectrum exhibits qualitatively distinct behavior depending on whether the qBTZ geometry corresponds to the quantum-dressed conical singularity (qCone) branch or the quantum-corrected BTZ branch. As expected, the state of the CFT is sensitive to the nature of the geometry which recieved the quantum corrections, and the correlation functions of single-trace operators can differentiate between these geometrical scenarios. For earlier studies on using CFT correlators to probe black hole singularities, see~\cite{Maldacena:2001kr, Festuccia:2005pi, Festuccia:2006sa, Fidkowski:2003nf,Hubeny:2006yu,Horowitz:2023ury,Caceres:2023zft}.

The sensitivity of the CFT to the geometry is further demonstrated by its influence on the detailed properties of Green's functions, particularly at the pole-skipping points ---locations in the complex momentum plane where the Green's function becomes multivalued and ceases to be uniquely determined. These points correspond to specific values of frequency and momentum where no unique ingoing solution exists at the horizon. This phenomenon is noteworthy because it suggests that local horizon dynamics carry information that imposes constraints on boundary correlators. Consequently, it is reasonable to expect that pole-skipping points would be affected by quantum backreaction on the horizon. In the qBTZ branch, for masses below the maximum, each pole-skipping point in the Matsubara frequency tower splits, resulting in two distinct momenta at which pole-skipping occurs. In contrast, the qCone regime shows no such splitting. This behavior is observed for operators with spin $s=0$ and $s=\pm 1/2$

As stated above, a connection between quantum chaos and hydrodynamics has emerged through comparisons of sound modes ---specifically, energy-energy correlation functions--- which shift into the upper half of the complex plane at imaginary momentum, indicating an instability. This link is further demonstrated by calculations of out-of-time-order correlators (OTOCs) involving the scattering of high-energy particles near the horizon~\cite{Schalm:2018lep}. The operator responsible for the Eikonal phase shift in this process is identical to the one found in the fluctuation equations governing stress-energy tensor correlators. At the pole-skipping locations, it is possible to extract both the maximal Lyapunov exponent and butterfly velocity. Subsequent research has revealed that these pole-skipping points are not exclusive to energy-energy correlators but occur across a broader spectrum of cases~\cite{Natsuume:2019sfp, Natsuume:2019xcy, Blake:2019otz, Grozdanov:2019uhi}. Interestingly, the pole-skipping point closest to the origin in all energy-momentum channels can define the Lyapunov exponent and butterfly velocity, despite the direct association with shockwave computations being primarily linked to the sound channel.

Moreover, pole-skipping in the Green's function of single-trace scalar operators with $\Delta = 2$ has been found to accurately predict the maximal Lyapunov exponent and butterfly velocity in $1+1$ holographic CFTs. Although it remains unclear whether this is a definitive indicator of quantum chaos in holographic theories, we apply this reasoning here and obtain intriguing results. While the maximal Lyapunov exponent continues to saturate the MSS bound~\cite{Maldacena:2015waa} even as quantum corrections are introduced, the butterfly velocity exhibits non-trivial behavior. In the qBTZ branch, below the maximal mass, the butterfly velocity splits into two forward values, $v_+^i$ ($v_{\text{BTZ}} > v_+^1 > v_+^2$). In contrast, in the qCone branch, the butterfly velocity remains single-valued and falls below the conformal value.

As a final indication of the distinction between the different CFT states, we examine the analytic structure of the Green's functions through pole collisions. These collisions, occurring at specific locations in the complex frequency and momentum plane, involve merging of Green's function poles. When a gapped mode collides with a hydrodynamic mode, these events can bound the radius of convergence for the linearized hydrodynamic expansion~\cite{Grozdanov:2019kge,Grozdanov:2019uhi,Heller:2020hnq}. The critical momentum at which two poles collide marks the branch point singularity of the hydrodynamic dispersion relation. While the dispersion relations for poles of the retarded Green's function of single-trace scalar operators in the CFT dual to the BTZ black hole are analytically known, it has been established~\cite{Grozdanov:2019kge,Grozdanov:2019uhi,Heller:2020hnq,Grozdanov:2020koi,Abbasi:2020xli} and clarified in~\cite{Cartwright:2024rus} that such pole collisions do not indicate a singularity in the dispersion relation $\omega(k)$.
Instead, this scenario, known as level-touching, suggests a locally analytic branch of the dispersion relation $\omega\sim (k^2-k_*^2)^v$ where $v$ is a positive integer.
When parameterizing the two interacting modes as $\omega^n(|k_*|e^{i\phi})$ and $\omega^m(|k_*|e^{i\phi})$, the modes touch at $(\omega_*, k_*)$ during a full circle traced in the complex plane $\phi \in [0, 2\pi]$ (the monodromy), momentarily becoming a second-order pole before continuing along their respective trajectories. In the BTZ case, these modes trace out perfect circles in the complex plane.

However, our study shows that quantum corrections alter this picture. With non-zero quantum backreaction, the poles of the retarded Green's functions of single-trace operators exhibit level-crossing rather than level-touching. This means that at certain points in the complex frequency and momentum plane, quasinormal modes (QNMs) from different overtones collide and exchange their overtone numbers. This results in a non-trivial local dependence where $v$ is no longer a positive integer. Moreover, we demonstrate that, for a given parameter $x_1$, the locations of these collision points differ between the qCone and qBTZ branches, providing a distinct fingerprint that distinguishes between the CFT states dual to these two different quantum corrected geometries.

\subsection*{Organization of the paper}

The structure of this paper is as follows. In section~\ref{sec:background_geometry} we review the AdS braneworld model in $(3+1)$-dimensions leading to the qBTZ solution. This section covers some basics of the AdS C-metric, the bulk parameters, the inclusion of the brane, the induced gravitational theory on the $(2+1)$-dimensional brane, and the various branches of the solution. In section~\ref{sec:probe_matter_and_QNMs} we begin our discussion of brane localized matter probes in the quantum-corrected geometries. We first analyze the QNM spectrum for scalar probes in section~\ref{sec:scalar_QNM}, followed by the QNM spectrum for spin-$1/2$ probes in section~\ref{sec:fermion_QNM}. Next, we take a more detailed look at the dispersion relations obtained in the previous section. section~\ref{sec:pole_skipping} examines pole-skipping points for scalar (section~\ref{sec:scalar_pole_skipping}) and fermionic (section~\ref{sec:fermion_pole_skipping}) probes. Finally, in section~\ref{sec:critical_points} we investigate the singular points of the curves defining the dispersion relations, again separating our analysis in terms of scalar (section~\ref{sec:scalar_critical}) and fermionic (section~\ref{sec:fermion_critical}) probes. We conclude with a discussion of our key results in section~\ref{sec:discussion}.

\section{Quantum black holes on the brane}
\label{sec:background_geometry}
The information about this geometry is available in other places, hence we will not take the space here to describe it in full detail.  However in an effort to be self--contained, we will in appendix~\ref{app:black_holes_on_brane} collect the necessary information about the ambient background geometry, the inclusion of the brane and the gravitational theory induced on that brane. In addition, we will also briefly review the BTZ solution since our efforts are focused on the difference between the CFT dual of the BTZ black hole and the quantum BTZ black hole.

The line element of the qBTZ geometry, in canonically normalized coordinates $(\rb,\bar{t},\bar{\phi})$, is given as
\begin{equation}\label{eq:quantum_blackhole_main}
   \exd s^2= \frac{\exd \rb^2}{ H\left(\rb\right)}-H\left(\rb\right)\exd \bar{t}^2 +\rb^2\exd\phib^2\, , \quad H(\rb)= \frac{\rb^2}{\ell_3^2}+\Delta ^2 \kappa -\frac{\Delta^3  \ell \mu }{\rb}\, .
\end{equation}
Here $H$ is the blackening factor, $\mu\ge 0$ is a mass parameter, $\kappa$ is a discrete constant $\kappa=\pm1,0$, $\ell_3$ is naive three dimensional AdS radius, $\ell$ is proportional to the brane tension and measures the strength of the quantum backreaction, and $\Delta=\frac{2 x_1}{3-\kappa  x_1^2}$ is a function which is related to the regularity of the geometry in the $\bar{\phi}$ direction. As discussed further in the appendix~\ref{app:black_holes_on_brane}, the solutions can be characterized into 3 branches based on the value of the parameter $x_1$
\begin{subequations}
\begin{align}
    \text{Branch 1a: } \quad & \kappa=+1,\quad 0< x_1 \leq 1\, , \\
    \text{Branch 1b: } \quad & \kappa=-1,\quad 0< x_1 < \sqrt{3}\, , \\
    \text{Branch 2: } \quad & \kappa=-1,\quad \sqrt{3}< x_1 < \infty \, .
\end{align}
\end{subequations}
These branches cover the range $-\frac{1}{8\mathcal{G}_3}\leq M \leq \frac{1}{24\mathcal{G}_3}$. Beginning with the negative mass range, for which $\kappa=+1$ and $x_1\in (0,1]$ and non-vanishing $\ell$ the solution in this range can be thought of as quantum-corrected conical singularities and we will refer to it as the qCone solution.  The remaining branches of the solution, branch 1b and 2, exist within the range $0<M\leq1/(24\mathcal{G}_3)$. We will refer to them as the qBTZ solution in an effort to distinguish them from the backreacted conical singularities. They are separated into two distinct branches due to the primary source of corrections to the BTZ black hole, further details about this range of solutions is also contained within appendix~\ref{app:black_holes_on_brane}.

\begin{figure}
    \centering
    \includegraphics[width=0.9\linewidth]{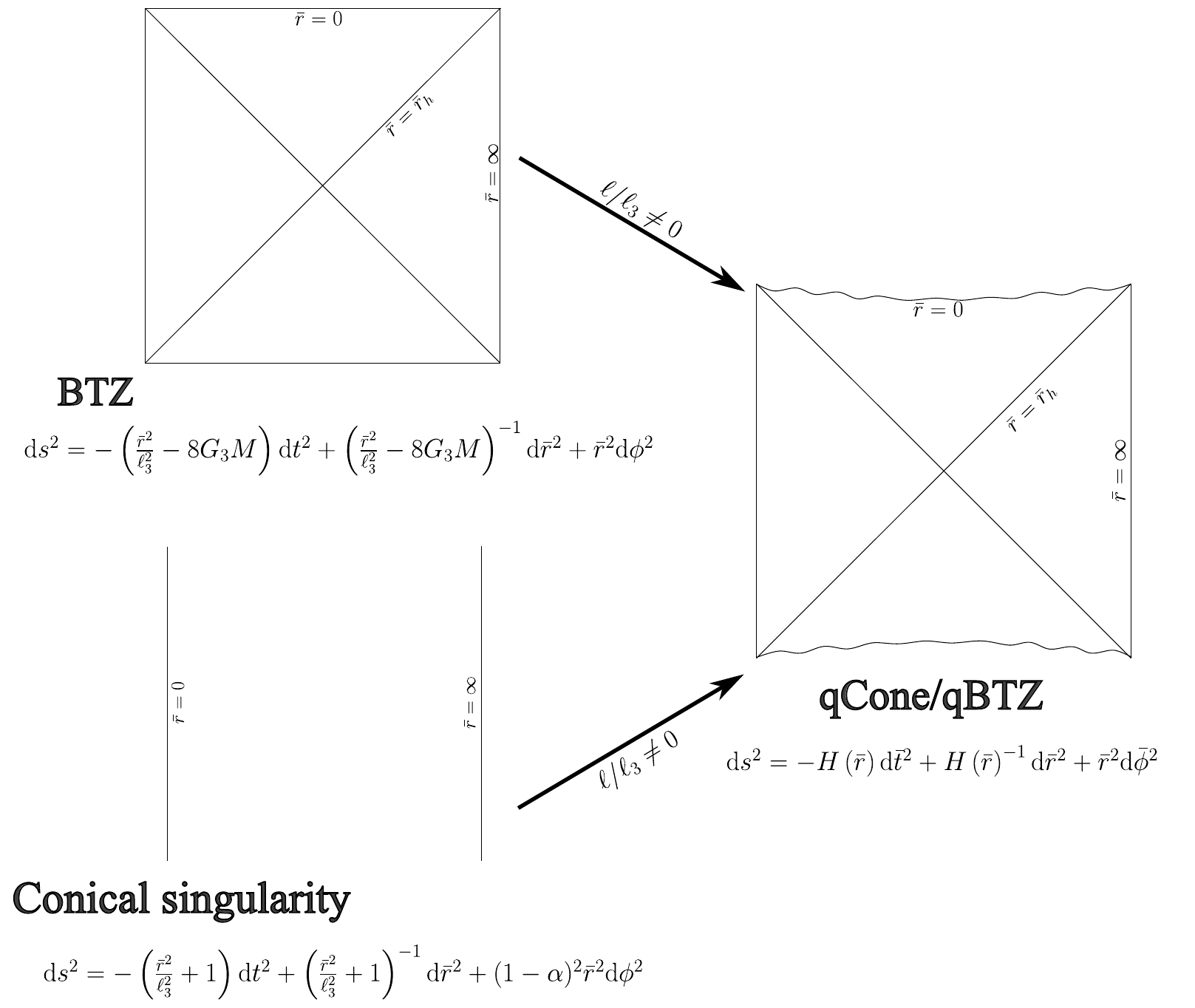}
    \caption{Penrose diagrams for the geometries we consider in this work. Of note, after backreaction both the qCone and qBTZ solutions have the same conformal structure. The conical singularity is displayed as a line while the genuine curvature singularity of the qCone/qBTZ solution is displayed as a wavy line. Their CFT interpretation, however, differs: before backreaction, the conical singularity is dual to a one-particle state $\mathcal{O}(z)|0\rangle$ in the DCFT$_2$, which is pure, while the BTZ solution corresponds to a thermal state. When backreaction is turned on, the DCFT$_2$ couples with the BCFT$_3$ and the state of the DCFT$_2$ is thus described by a density matrix. In this case, thermality arises due to entanglement between the DCFT$_2$ and the BCFT$_3$.
    \label{fig:penrose}}
\end{figure}
To aid in the interpretation of our results, it is worth noting some features of the causal structure of the geometries we consider in this work. In figure~\ref{fig:penrose} we display Penrose diagrams for the geometries in this work. Notably, the AdS conical singularity has a diagram which is identical to global AdS. This can be seen by taking the metric of the AdS conical singularity, displayed in the appendix in eq.\ (\ref{eq:conical_defect}), and performing the transformation $r=\ell_3 \sinh{\rho}, t=\ell_3 \tau$ leading to
\begin{equation}
    \exd s^2=\ell _3^2 \left(-\exd \tau^2 \cosh ^2\rho +\exd \rho^2+(\alpha -1)^2 \sinh ^2\rho \exd \phi^2  \right)\, ,
\end{equation}
with $\tau \in \mathbb{R}$, $\rho\in \mathbb{R}_+$ and $\phi\in[0,2\pi)$. For $\alpha=0$ this is the universal covering space of AdS$_{2+1}$.

Notice that the conical defect does not feature in the Penrose diagram of the conical singularity, as it eliminates the angular component, and the tortoise coordinate, which is obtained from the blackening factor in eq.\ (\ref{eq:conical_defect}) as
\begin{equation}
    r_*=\int_a^r \frac{\exd r}{f(r)}\,,
\end{equation}
is finite at both boundaries ($r=0$ and $r=\infty$)~\cite{Schindler:2018wbx}.

The Penrose diagram for the quantum corrected solution is almost identical to the BTZ diagram, except for the ``bowing'' of the singularity, as expected~\cite{Fidkowski:2003nf,Kolanowski:2023hvh}. Consider a null infalling geodesic beginning at the boundary, we can compute the time taken for the geodesic to reach the singularity of the BTZ black hole (whose blackening factor we will take to be $g(r)=r^2-r_h^2$ to remain consistent with~\cite{Kolanowski:2023hvh})
\begin{align}
    \lim_{r\rightarrow 0} t(r) &=   \lim_{r\rightarrow 0} \left(t_0- \int_r^{\infty}\frac{\exd y}{g(y)}\right) \\
    &=  \lim_{r\rightarrow 0}\left( t_0+\frac{1}{r_h}\arctanh\left(\frac{r}{r_h}\right)-\frac{i \pi}{2r_h}\right) \\
    &=- \frac{i\pi}{2r_h}
\end{align}
when starting at $t_0=0$. The vanishing real part of the expression implies that the null geodesic sent from the other boundary will meet this geodesic at precisely the conical singularity, in the center of the Penrose diagram, hence the diagram is a square. For the quantum corrected geometry, making use of the parameterization
\begin{equation}\label{eq:simple_parametization_H}
    H(\bar{r})=\frac{\bar{r}^2-\bar{r}_h^2}{\ell_3^2}+\frac{\Delta ^3 \ell \mu  (\bar{r}-\bar{r}_h)}{\bar{r }\bar{r}_h}
\end{equation}
a similar calculation can be done giving (again beginning at $t_0=0$)
\begin{equation}
   \lim_{\bar{r}\rightarrow 0}\bar{t}(\bar{r})= -i \pi  \gamma_2 \ell_3^2+\frac{1}{2} \gamma_2 \ell_3^2 (\log (m)-3 \log (\bar{r}_h))+\frac{1}{2} \gamma_1 \ell_3^2 \left(\pi  \alpha -2 \alpha  \arctan(\alpha  \bar{r}_h)\right)
\end{equation}
with $m=\Delta ^3 \ell  \ell_3^2 \mu $ and the following parameters
\begin{equation}
    \gamma_1=\frac{2 m+\bar{r}_h^3}{m+2 \bar{r}_h^3},\quad \gamma_2=\frac{\bar{r}_h^2}{2 m+\bar{r}_h^3} \gamma_1,
    \quad \alpha=\frac{\sqrt{\bar{r}_h}}{\sqrt{4m-\bar{r}_h^3}}\, .
\end{equation}
There are subtleties to this equation, more detail is presented in~\cite{Kolanowski:2023hvh}. The non-trivial real part of this expression implies null geodesics fired from the asymptotic boundary hit the singularity ``off center'' in the Penrose diagram and hence the diagram is bowed~\cite{Fidkowski:2003nf,Kolanowski:2023hvh} as displayed in figure~\ref{fig:penrose}. Notice that it is independent of the choice of $\kappa$.

Turning to the nature of the singularity, for small $r$ one may write the metric for  conical singularity (see eq.\ (\ref{eq:conical_defect})) as
\begin{equation}
    \exd s^2\approx-\exd t^2+\exd x^2+\exd y^2 = -\exd t^2+\exd r^2+r^2\exd \phi_{12}
\end{equation}
with $r=\sqrt{x^2+y^2}$ and $\phi=(1-\alpha)^{-1}\phi_{12}=(1-\alpha)^{-1}\arctan(y/x)$. It is easy to see that $x^{\mu}=(r_0-\lambda,t_0,\phi_{0})$ is a geodesic which cannot be extended past $\lambda=r_0$ and the vector tangent to the curve $k^{\mu}=\exd x^{\mu}/\exd\lambda$ is spacelike $k^\mu k_\mu=1$, hence the singularity may be described as timelike. Likewise, consider the line element for BTZ ($J=0$) for $r\ll \ell_3$
\begin{equation}
    \exd s^2 \approx -\frac{\ell_3^2 }{r_h^2}\exd r^2+\frac{r_h^2 }{\ell_3^2}\exd t^2+r^2 \exd\phi^2
\end{equation}
It is easy to see that $x^{\mu}=(r_0-\lambda,t_0,\phi_{0})$ is a geodesic which cannot be extended past $\lambda=r_0$ and the vector tangent to the curve $k^{\mu}=\exd x^{\mu}/\exd\lambda$ is timelike $k^\mu k_\mu=-\ell_3^2/r_h^2$, hence the singularity may be described as spacelike. Furthermore in both cases, the conical singularity at the origin of AdS and the BTZ black hole, making use of a distributional identity~\cite{Jackiw:1991nb} or by the holonomies of a vector parallel transported in a closed loop around the singularity~\cite{Briceno:2024ddc}  shows that central singularity of the geometry is due to a delta point source; for $M>0$ (BTZ)
\begin{align}
    R^{01}+\frac{e^0e^1}{\ell_3^2}&=-2\pi \sqrt{8G_3M}\delta(\Sigma_{M>0})\, , \\
    T^0&=\pi \frac{|J|}{\sqrt{M}}\delta(\Sigma_{M>0})\, ,\\
    T^2&=-\pi \frac{J}{\sqrt{M}}\delta(\Sigma_{M>0})\, ,
\end{align}
while for $M<0$ (conical singularity)
\begin{align}
      R^{12}+\frac{e^1e^2}{\ell_3^2}&=-2\pi \sqrt{-8G_3M}\delta(\Sigma_{M<0}) \, ,\\
    T^0&=\pi \frac{J}{\sqrt{-M}}\delta(\Sigma_{M<0})\, ,\\
    T^2&=\pi \frac{|J|}{\sqrt{-M}}\delta(\Sigma_{M<0})\, ,
\end{align}
where we have left $J\neq0$ to display the relations in all generality. Here $R^{ab}$ and $T^a$ are the curvature and torsion two-forms and $\Sigma_x$ is a delta two-form on a hypersurface which intersects the singularity at the point $r=0$.

This is notably different for the quantum corrected solution, where curvature invariants reveal a true curvature singularity for all backreacted solutions; for instance the Kretschmann scalar is
\begin{equation}
K=R_{\alpha\beta\lambda\rho}R^{\alpha\beta\lambda\rho}=\frac{6 \Delta ^6 \ell^2 \mu ^2}{\rb^6}+\frac{12}{\ell_3^4}    \, .
\end{equation}
Interestingly, the central singularity for the quantum corrected solution is spacelike for both $\kappa=\pm 1$. This is most easily seen from the simple parameterization of the blackening factor in eq.\ (\ref{eq:simple_parametization_H}), for $\bar{r}/\ell_3\ll 1$
\begin{equation}
\exd s^2\approx-\frac{\exd \bar{r}^2 \rb}{\Delta ^3 \mu  \ell }+\exd \bar{t}^2 \left(-\frac{\Delta ^3 \mu  \ell }{\rb_h}+\frac{\rb_h^2}{\ell _3^2}+\frac{\Delta ^3 \mu  \ell }{\rb}\right)+\exd\bar{\phi}^2 \rb^2
\end{equation}
It is easy to check that $x^{\mu}=(2^{-1}\sqrt{\Delta ^3 \ell \mu}\left(r_0^{3/2}-3 \lambda \right)^{2/3},t_0,\phi_{0})$ is a geodesic which cannot be extended past $\lambda=r_0^{3/2}/3$ and the vector tangent to the curve $k^{\mu}=\exd x^{\mu}/\exd\lambda$ is timelike $k^\mu k_\mu=-1$, hence the singularity may be described as spacelike.

Despite the above, the remainder of this paper is dedicated to showing that black hole perturbations can correctly diagnose whether the geometry contained an initially timelike delta singularity or an initially spacelike delta singularity. From a CFT perspective, this distinction is clear: before backreaction, the conical singularity is dual to a one-particle state $\mathcal{O}(z)|0\rangle$ in the DCFT$_2$. This is a pure state, which explains the absence of a horizons in the bulk geometry. In contrast, the BTZ solution corresponds to a thermal state in the DCFT$_2$. When backreaction is turned on, the DCFT$_2$ couples with the BCFT$_3$; this is mirrored in the bulk by the fact that the brane is now placed at some finite distance into the bulk, implying that it has a non-negligible effect on the geometry. In this case, the state of the DCFT$_2$ is thus described in terms of a density matrix, and thermality arises due to entanglement between the DCFT$_2$ and the BCFT$_3$.

\begin{figure}[t!]
    \centering
\includegraphics[width=0.75\textwidth,trim={0 1mm 0 0},clip]{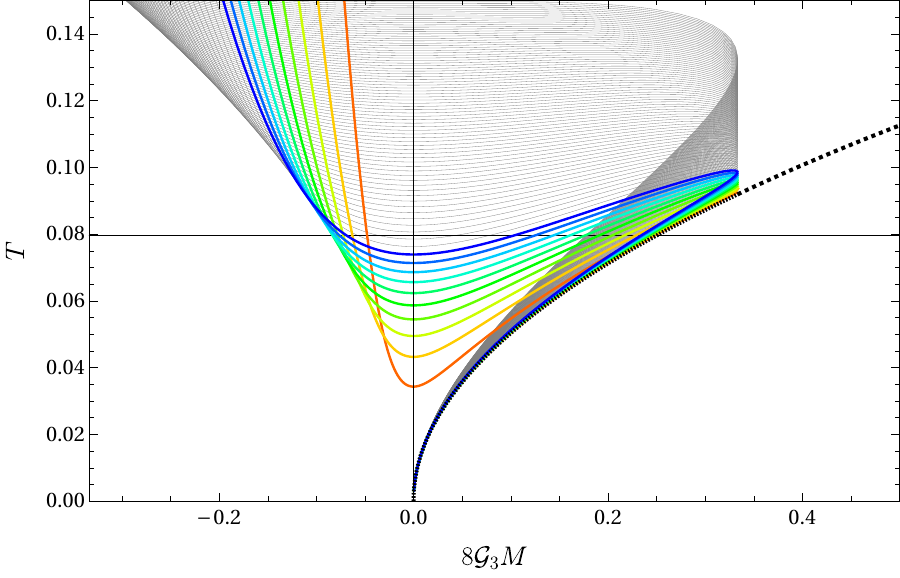}
\vspace{-3mm}
    \caption{\textbf{Temperature: }The temperature of the quantum corrected black hole is displayed as a function of the $8\mathcal{G}_3 M$ with $\ell_3=1$. The colors display different values of the quantum backreaction, from $\ell=1/100$ in red, to $\ell=1/10$ in blue in steps of $\delta \ell=1/100$. The thin dashed black line displays the temperature of the BTZ black hole. Notice that the BTZ black hole on the brane is limited to a value of $8\mathcal{G}_3M=1/3$, even at vanishing quantum backreaction. Hence we continue this line outward with a different dashing to emphasize that the temperature of the BTZ black hole continues for larger black hole mass, but the uncorrected BTZ black hole on the brane is limited to $8\mathcal{G}_3M=1/3$. The thin gray lines continue to display the temperature in steps of $\delta \ell=1/100$ but are grayed out to emphasize that we will not consider solutions with that particular value of $\ell$ in this work. The horizontal line in the figure corresponds to the choice of temperature used throughout this work and corresponds to $4\pi T=1$.
    \label{fig:temperature}}
\end{figure}

Naturally, given that corrected geometry smoothly connects with the uncorrected conical singularity and the BTZ geometry, it will be highly interesting to understand how the quantum backreaction affects the QNMs of the geometry (see eq.\ (\ref{eq:Exact_BTZ_omega})). We have several parameters we can vary, $(\ell_3,\ell,x_1,m)$. The equations of motion of the scalar field expanded near the AdS boundary show that the operator dimension of the associated field theory operator depends on the choice of $m$ and $\ell_3$. For the remainder of this work we will pick $\ell_3=1$ in our numerical calculations and $m=0$. Additionally, all dimensionful quantities will be measured in units of $\ell_3$. That leaves us with two remaining parameters, $\ell$ and $x_1$ which control the strength of the quantum backreaction and the mass of the black hole respectively. Additionally, the choice of these remaining parameters will also fix the value of the temperature. An image of the temperature of the quantum-corrected black hole is displayed in figure~\ref{fig:temperature}.

The image displays the temperature as a function of $8\mathcal{G}_3M$ for a variety of different values of $\ell$ holding $\ell_3=1$. The colored lines display the temperature with $\ell=1/100$ in red, to $\ell=1/10$ in blue in steps of $\delta \ell=1/100$. Recall that the mass in the mass is restricted to the range $-1\leq 8\mathcal{G}_3M \leq \frac{1}{3}$ and that $-1\leq 8\mathcal{G}_3M<0$ is associated with the qCone and $0< 8\mathcal{G}_3M \leq \frac{1}{3}$ is associated with the qBTZ. Notably, we see that there are two temperatures for any given positive value of $8\mathcal{G}_3M$ associated with what we refer to as branch 1b and branch 2 qBTZ solutions. Branch 1b is always the hotter of the two branches for positive masses. The thin dashed line connected to $T=0$ and continuing outward represents the uncorrected BTZ geometry. When the quantum correction is set to zero ($\ell=0$) the BTZ solution is limited by $8G_3M=1/3$ and hence we continue this line outward past this point with a different dashing to indicate the behavior of the BTZ black hole which is not obtained from this brane construction, to help guide the eye. The remaining gray curves are further temperature/mass relations in increasing steps of $\delta\ell=1/100$. They are represented by gray lines since we will not consider these values of $\ell$ in this work. Notice the following feature of the plot: for a given temperature, such as the one indicated by the horizontal line, there are values of $\ell$ where no corrected black hole from branch 1a or 1b exists at that temperature. Additionally, if the temperature is too high, another issue arises: only corrected black holes from branches 1a and 1b exist, while no black hole from branch 2 can be found at the chosen temperature and $\ell$.

In what follows, the value of the quantum correction will have a strong influence on the state of the putative dual CFT, in order to have some control when comparing the QNMs between branches we will be, at the very least, interested in comparing them in a thermal state with the same temperature. Additionally, we will be interested primarily in the regime of small quantum backreaction where the effective action given in eq.\ (\ref{eq:effective_action}) is a good description of the theory. To this end we will impose the temperature slice depicted in figure~\ref{fig:temperature}, which occurs at $4\pi T=1$, for the rest of this work since connects smoothly to $\ell=0$ and occurs at $x_1=1$ for which the uncorrected mass is given by $8G_3M=1/4$. The remaining values of the parameter $x_1$ used in this work are displayed in table~\ref{tab:parameter_values}

\begin{table}[t!]
    \centering
    \begin{tabular}{c|ccc}
    $\ell$ & Branch 1a & Branch 1b & Branch 2 \\
    \hline
 0    & $\times$    & 1 & 3 \\
 0.01 & 0.316841 & 0.959867 & 3.02188 \\
 0.02 & 0.359959 & 0.919327 & 3.04311 \\
 0.03 & 0.3833 & 0.878142 & 3.06374 \\
 0.04 & 0.397283 & 0.836022 & 3.08381 \\
 0.05 & 0.405284 & 0.792600 & 3.10337 \\
 0.06 & 0.408675 & 0.747389 & 3.12243 \\
 0.07 & 0.40793 & 0.699702 & 3.14104 \\
 0.08 & 0.402899 & 0.648507 & 3.15923 \\
 0.09 & 0.39273 & 0.592107 & 3.17701 \\
 0.1 & 0.375286 & 0.527327 & 3.1944
    \end{tabular}
    \caption{The parameter values used for the calculation of QNM in this work. Each value is selected such that the temperature of the black hole is fixed so that $4\pi T=1$ and corresponds to the colored lines of figure~\ref{fig:temperature}. Note that branch 1a for $\ell=0$ does not exist so we label it with an $\times$ in the table.
    \label{tab:parameter_values}}
\end{table}

Before closing the section, we mention in passing that the rest of the analysis of this work will be concerned with the behavior of probe matter confined to the brane, and in particular solutions to the probe matter equations of motion with infalling boundary conditions. Hence, we find it much simpler to work with coordinates adapted to the choice of boundary conditions we wish to impose. For this reason, we will make great use of the infalling Eddington-Finkelstein coordinates, in which the line element is given by,
\begin{equation}
   \exd s^2 =2\exd v  \exd \bar{r}-\exd v^2 H(\bar{r})+\rb^2\exd \phib^2 \, .
\end{equation}
where $H$ is given by the expression in eq.\ (\ref{eq:quantum_blackhole}).

%%%%%%%%%%%%%%%%%%%%%%%%%%%%%%%%%%%%%%%%%%%%%%%%%%%%%%%%%%%%%%%%%%%%%%%%%%%%%%%%%%%%%%
\section{Probe matter on the brane and QNM}\label{sec:probe_matter_and_QNMs}
In this section we will begin our study of the response of matter probes constrained to exist on the brane. We will be interested in matter whose dual CFT interpretation is of operators with dimension $\Delta$ and spin $s=0,1/2$. In particular, in this section we will focus on the poles of the retarded Green's functions of such operators, the bulk interpretation of which is given by QNM. As is already well known, QNMs are solutions to the field equations with Dirichlet boundary conditions at the conformal boundary and infalling boundary conditions at the horizon. The QNM of a scalar perturbation as seen by an observer localized to the brane has been described in the work~\cite{Chung:2015mna} but we repeat this exercise to establish notation, especially for the analysis of scalar pole-skipping and level-crossing and to point out an interesting feature of the modes not discussed in~\cite{Chung:2015mna}. To save on space, we will directly begin displaying the results of our analysis. The analytic setup of the field equations etc. is contained in appendix~\ref{app:fieldEQ} with appendix~\ref{app:fieldEQ_Scalar} dedicated to the scalar field equations and with appendix~\ref{app:fieldEQ_Fermion} dedicated to the spinor field equations.

%%%%%%%%%%%%%%%%%%%%%%%%%%%%%%%%%%%%%%%%%%%%%%%%%%%%%%%%%%%%%%%%%%%%%%%%%%%%%%%%%%%%%%%%%
\subsection{Scalar field QNM}\label{sec:scalar_QNM}
\begin{figure}[t!]
    \centering
\includegraphics[width=0.75\textwidth,trim={0 2mm 0 0},clip]{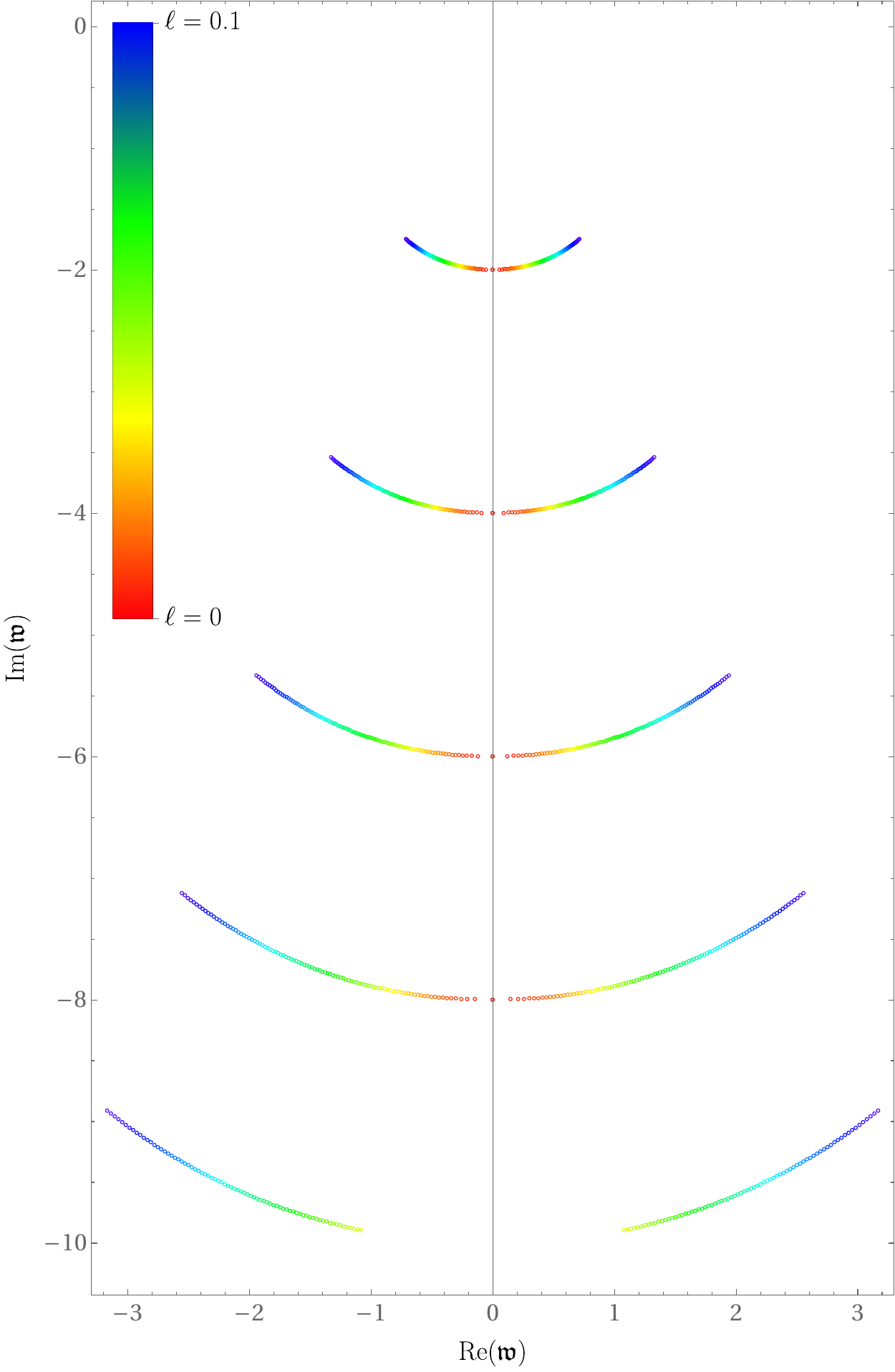}
\vspace{-2mm}
    \caption{\textbf{Mode transition from BTZ to qBTZ: }The QNM frequencies of a BTZ black hole are displayed as one slowly turns on quantum backreaction with each different curve representing a different overtone. The modes are displayed for fixed $n=0$ and $4\pi T=1$ while $\ell$ varies from $\ell=0$ to $\ell=0.1$. The figure appears to display that there are no modes for $\ell=0$ of the fourth overtone, however, this is just a reflection of the numerics breaking down. The curves are shown at a fixed value of temperature.
    \label{fig:BTZ_TO_QUBTZ}}
\end{figure}

As discussed previously, we will work at fixed temperature for values of the quantum backreaction $\ell$ that smoothly connect to the uncorrected geometry (see table~\ref{tab:parameter_values}). We begin by setting $\kappa=-1$, and show in figure~\ref{fig:BTZ_TO_QUBTZ} the behavior the QNMs as one switches on the quantum backreaction at zero momentum ($n=0$).
The color coding represents the quantum correction parameter $\ell$, ranging from red for $\ell=0$ to blue for $\ell=0.1$. As backreaction increases, the QNMs transition from purely dissipative modes (on the imaginary axis with no real part) to propagating modes (with a finite real part). We give the values~\footnote{The modes computed for a fixed parameter ($x_1=1/2$) for $\kappa=-1$, without holding the temperature fixed, as a function of a larger range of $\ell$ is displayed in appendix~\ref{app:extra_modes}. The results for large values of $\ell$, at fixed mass, do not display any qualitative difference with those at smaller values of $\ell$ and fixed temperature and hence we chose to work with the range of $\ell$ displayed in the main text for simplicity.} of the QNM for the leading and first three overtones at zero momentum in table~\ref{tab:qubtzModes}

\begin{figure}[t!]
    \centering
\includegraphics[width=0.9\textwidth,trim={0 2mm 0 0},clip]{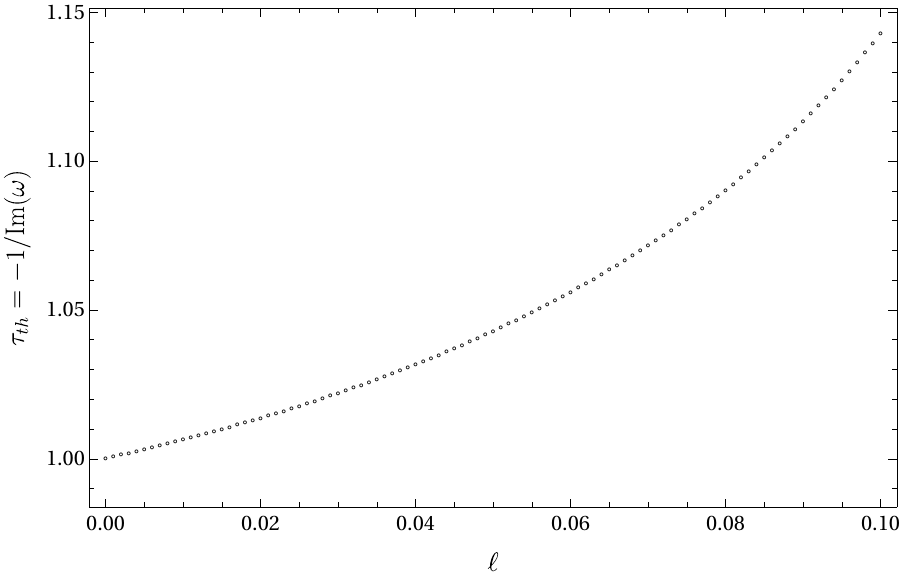}
\vspace{-2mm}
    \caption{\textbf{Mode transition from BTZ to qBTZ: Thermalization time } The imaginary part of the lowest QNM frequency of a BTZ black hole is displayed as one slowly turns on quantum backreaction. The modes are displayed for fixed $n=0$ and $4\pi T=1$ as $\ell$ varies from $\ell=0$ to $\ell=0.1$. The curve is displayed at fixed temperature.
\label{fig:BTZ_TO_QUBTZ_Thermalization_time}}
\end{figure}

Looking both at figure~\ref{fig:BTZ_TO_QUBTZ} and table~\ref{tab:qubtzModes} one notices that not only does the mode attain a finite real part as a result of the quantum back reaction, but that as one increases the quantum backreaction the imaginary part decreases in magnitude i.e. the QNMs move upwards towards the upper half-plane. This movement of the QNMs in the complex plane has interesting consequences for the retarded Green's function in dual CFT. In figure~\ref{fig:BTZ_TO_QUBTZ_Thermalization_time} we display the motion of the imaginary part of the lowest QNM ($n_z=0$), at zero momentum, as one switches on the quantum backreaction.
This quantity $\tau_{th}=-1/Im(\omega)$ provides a thermalization timescale for CFT dual to the quantum-corrected black hole. Recall that we have $e^{-i \omega t}=e^{-i (\omega_R+i \omega_I)t}=e^{-i\omega_Rt + \omega_It}$ since $\omega_I<0$ put $\omega_I=-1/\tau_{th}$ then we have $F(t)\equiv e^{-i\omega_Rt}e^{ -t/\tau_{th}}$. Interestingly, we find that the timescale increases as the quantum backreaction increases. That is, notice as $\tau_{th}$ gets smaller the time $t$ needed for the amplitude of $F(t)$ to reach $e^{-1}F(0)$ increases, hence the system thermalizes slower.

So far we have focused on $\kappa=-1$, displaying how the modes are shifted in the complex plane as we allow quantum backreaction on the BTZ geometry. As we have discussed in the introduction and the previous section, the qBTZ construction provides us with the opportunity to understand how the modes behave in geometries with horizons that cloak a different type of singularity. For this we turn to the qCone part of the spectrum with $\kappa=+1$. In table~\ref{tab:conSingModes} we display the values~\footnote{The modes computed for a fixed parameter ($x_1=1/2$) for $\kappa=-1$, without holding the temperature fixed, as a function of a larger range of $\ell$ is displayed in appendix~\ref{app:extra_modes}.} of the QNMs for the leading and first three overtones at zero momentum of the qCone. It is important to notice that unlike table~\ref{tab:qubtzModes}, table~\ref{tab:conSingModes} begins at $\ell=0.01$. The reason we begin our table at $\ell=0.01$ is that our numerical construction relies on a horizon being present in the geometry. This is only true for the conical singularity at non-zero $\ell$. Looking at the table, we can notice that, unlike the quantum-corrected BTZ black hole, the imaginary part of the frequency moves deeper into the complex plane for the quantum-dressed conical singularity as $\ell $ increases. Here we see the first instance of the knowledge the CFT dual has about the original type of singularity.

\newpage
\begin{landscape}
\begin{table}[t!]
    \centering
    \begin{tabular}{c |c c c c}
    $\ell$ & $n_z=0$ &  $n_z=1$&  $n_z=2$&  $n_z=3$\\
    \hline
 $0   $&$  -2 i $&$ -4 i $&$-6 i$ &$ -8 i$ \\
 $0.01 $&$ \pm 0.177385-1.9875 i $&$ \pm 0.299424\, -3.98337 i $&$ \pm  0.404523\, -5.97955 i $&$ \pm  0.502899\, -7.97569 i $\\
 $0.02 $&$ \pm 0.256934-1.9733 i $&$ \pm 0.440672-3.96203 i $&$ \pm 0.606715-5.95032 i $&$ \pm 0.767963-7.93802 i $\\
 $0.03 $&$ \pm 0.322348-1.95716 i $&$ \pm 0.560796-3.93568 i $&$ \pm 0.783453-5.9125 i $&$ \pm 1.00346-7.88831 i $\\
 $0.04 $&$ \pm 0.381387\, -1.93884 i $&$ \pm  0.672\, -3.90397 i $&$ \pm  0.9495\, -5.86609 i $&$ \pm 1.22587-7.82709 i$ \\
 $0.05 $&$ \pm 0.437091-1.91798 i $&$ \pm 0.778969-3.8664 i $&$ \pm  1.11047\, -5.8107 i $&$ \pm  1.44176\, -7.75399 i $\\
 $0.06 $&$ \pm 0.491116\, -1.89413 i $&$ \pm  0.88427\, -3.82224 i $&$ \pm  1.26957-5.74544 i $&$ \pm 1.65511-7.66783 i$ \\
 $0.07 $&$ \pm 0.5446-1.86667 i $&$ \pm  0.98975\, -3.77039 i $&$ \pm  1.42925\, -5.66876 i $&$ \pm  1.86914\, -7.56654 i$ \\
 $0.08 $&$ \pm 0.598538-1.83466 i $&$ \pm  1.09714\, -3.70914 i $&$ \pm  1.59195\, -5.57814 i $&$ \pm 2.08714-7.44671 i $\\
 $0.09 $&$ \pm 0.654038\, -1.79663 i $&$ \pm 1.20853-3.63566 i $&$ \pm  1.76076\, -5.46934 i $&$ \pm  2.31327\, -7.30272 i$ \\
 $0.1 $&$ \pm  0.71272\, -1.74989 i $&$ \pm  1.32715\, -3.54473 i $&$ \pm 1.94055-5.3346 i $&$ \pm 2.55411-7.12421 i $\\
    \end{tabular}
    \caption{\textbf{QNM of the qBTZ black hole ($s=0$):} The QNM frequencies, $\mathfrak{w}=\omega/(2\pi T)$, displayed here were computed with $4\pi T=1$ and for $\kappa=-1$ at zero momentum using the values of $x_1$ tabulated in table~\ref{tab:parameter_values}.
    \label{tab:qubtzModes}}
\end{table}
\begin{table}[b!]
    \centering
    \begin{tabular}{c |c c c c}
    $\ell$ & $n_z=0$ &  $n_z=1$&  $n_z=2$&  $n_z=3$\\
    \hline
 $0.01 $&$ \pm 0.755078-0.275791 i $&$ \pm 1.48074\, -0.593145 i $&$ \pm  2.20996\, -0.91007 i $&$ \pm 2.93972-1.2269 i $\\
 $0.02 $&$ \pm 0.86332\, -0.398271 i $&$ \pm 1.69176-0.845315 i $&$ \pm  2.5232\, -1.29122 i $&$ \pm 3.35512-1.73702 i $\\
 $0.03 $&$ \pm 0.927769\, -0.496676 i $&$ \pm  1.81633\, -1.04616 i $&$ \pm 2.70744-1.59422 i $&$ \pm  3.59901-2.14214 i $\\
 $0.04 $&$ \pm 0.972041\, -0.583716 i $&$ \pm 1.901-1.22289 i $&$ \pm 2.83218-1.8605 i $&$ \pm 3.76377-2.49793 i $\\
 $0.05 $&$ \pm 1.00407\, -0.664525 i $&$ \pm 1.96142\, -1.38635 i $&$ \pm 2.9207-2.10654 i $&$ \pm  3.88037\, -2.82652 i$ \\
 $0.06 $&$ \pm 1.02744\, -0.741996 i $&$ \pm 2.00461\, -1.54259 i $&$ \pm 2.98351-2.3415 i $&$ \pm  3.96275\, -3.14018 i $\\
 $0.07 $&$ \pm 1.04395\, -0.818205 i $&$ \pm 2.0341-1.69589 i $&$ \pm 3.02581-2.57186 i $&$ \pm  4.01781\, -3.44758 i $\\
 $0.08 $&$ \pm 1.05444\, -0.89502 i $&$ \pm 2.05146-1.85005 i $&$ \pm 3.0499-2.80334 i $&$ \pm 4.0486\, -3.75636 i$ \\
 $0.09 $&$ \pm 1.059-0.974576 i $&$ \pm  2.05679\, -2.00936 i $&$ \pm 3.05589-3.04239 i $&$ \pm 4.05521-4.07511 i $\\
 $0.1 $&$ \pm 1.05686-1.06002 i $&$ \pm  2.04839-2.18011 i $&$ \pm 3.04115\, -3.29837 i $&$ \pm 4.03408-4.41631 i $\\
    \end{tabular}
    \caption{\textbf{QNM of the qCone ($s=0$):} The QNM frequencies, $\mathfrak{w}=\omega/(2\pi T)$, displayed here were computed with $n=0$ and $4\pi T=1$ and for $\kappa=+1$ at zero momentum using the values of $x_1$ tabulated in table~\ref{tab:parameter_values}. Notice, there is no horizon for $\ell=0$, hence this row is omitted.
    \label{tab:conSingModes}}
\end{table}
\end{landscape}
\newpage

\begin{mdframed}
    %The poles of the retarded Green's function of single trace scalar operators in the CFT dual to semi-classical Einstein gravity are sensitive to the type of singularity which was cloaked behind the horizon. The poles move towards the real axis for increasing backreaction of the quantum-corrected BTZ black hole, and move away from the real axis for quantum-corrected conical singularity.
    The poles of the retarded Green's function of single trace scalar operators in the CFT dual to semi-classical Einstein gravity are naturally sensitive to the type of singularity originally present in the geometry, which is  encoded in the CFT state. The poles move towards the real axis for increasing backreaction of the quantum-corrected BTZ black hole, and move away from the real axis for quantum-corrected conical singularity and hence may be used to distinguish between these two bulk geometries.
\end{mdframed}

\begin{figure}[t!]
    \centering
\includegraphics[scale=0.85,trim={0 1mm 0 0},clip]{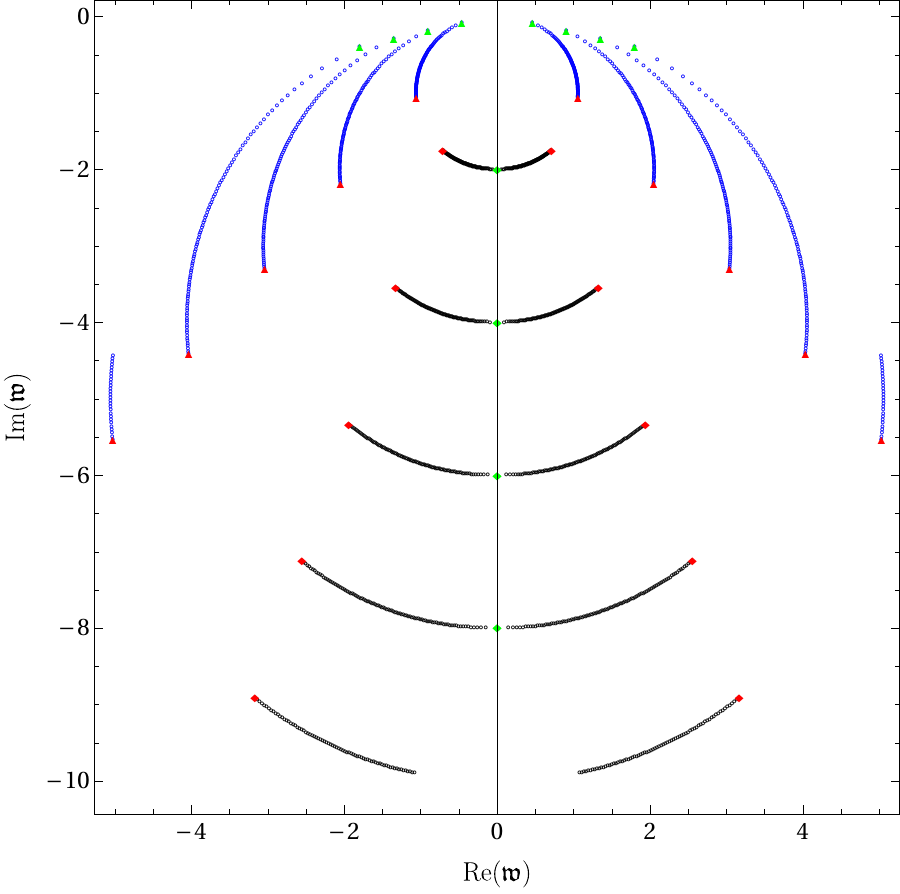}
\vspace{-3mm}
    \caption{\textbf{Mode transition from quantum dressed conical singularity to quantum corrected BTZ: }The lowest QNM frequencies are displayed as one slowly turns on quantum backreaction. The modes are displayed for fixed $n=0$ and $4\pi T=1$ as $\ell$ varies from $\ell=1/100$ displayed in green (triangles for qCone, diamonds for qBTZ) to $\ell=1/10$ displayed in red (triangles for qCone, diamonds for qBTZ).
    Blue dots represent the quantum dressed conical singularity ($\kappa=1$) while the black dots represent the quantum corrected BTZ geometry ($\kappa=-1$).
\label{fig:quBTZ_TO_QDConical}}
\end{figure}

To make this more concrete we display in figure~\ref{fig:quBTZ_TO_QDConical} the motion of the lowest QNMs, at zero momentum, in quantum BTZ black hole and the quantum dressed conical singularity as we increase the quantum backreaction.
The blue dots correspond to the quantum-dressed conical singularity and the black dots to the quantum-corrected BTZ black hole. Beginning with here $\ell=1/100$, the green (triangles for qCone, diamonds for qBTZ), the QNMs are well separated. However, we see that as we increase the quantum back reaction, or decrease the acceleration, swinging the brane into the bulk from near the conformal boundary, the QNMs move towards one another ending at $\ell=1/10$ (triangles for qCone, diamonds for qBTZ). At any finite $\ell$ a gap remains between the QNMs of the quantum dressed conical singularity and the quantum correct BTZ black hole, which increases in size with increasing overtone.

Finally, before closing the section, although we have looked at how the thermalization time changes as we turn on the quantum correction it is also interesting to ask how the thermalization time of the dual CFT responds to the different physical origins of the quantum correction as well as the type of singularity originally present in the geometry. As stated in section~\ref{sec:background_geometry} not only is there a distinction between the quantum-dressed cone and quantum-corrected BTZ black hole, but the primary origin of the quantum correction differs depending on the branch of the solution. The quantum-dressed cone (branch 1a) and the quantum-corrected BTZ black hole (in branch 1b) receive their quantum corrections due to the backreaction of the Casimir stress tensor, while the quantum black hole (branch 2) receives its quantum corrections primarily from the backreaction of Hawking radiation sitting in thermal equilibrium with the black hole.
 \begin{figure}[t!]
    \centering
\includegraphics[scale=0.85,trim={0 2mm 0 0},clip]{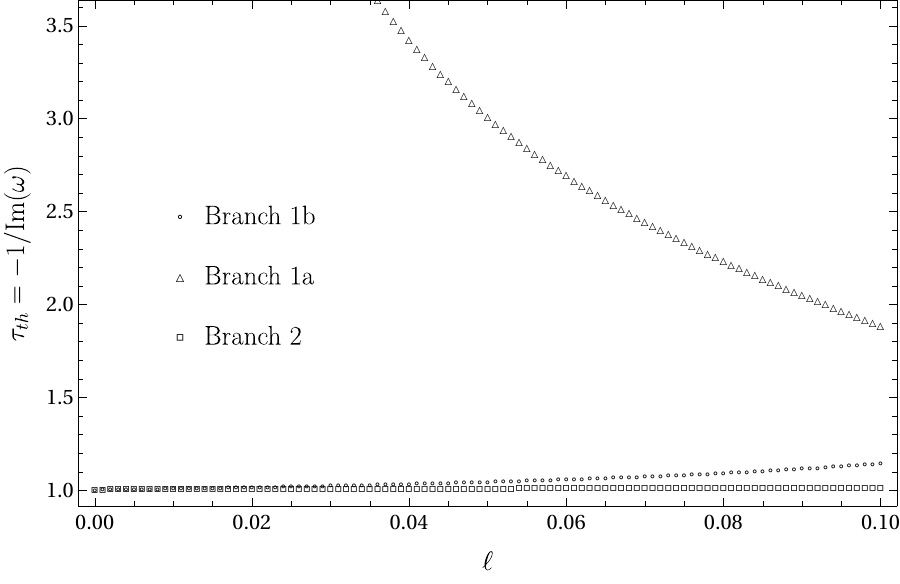}
\vspace{-2mm}
    \caption{\textbf{Origin of the quantum correction: Thermalization time} The imaginary part of the lowest QNM frequency of a quantum corrected geometry is displayed as one slowly increases the quantum backreaction. The modes are displayed for fixed $n=0$ and $4\pi T=1$ as $\ell$ varies from $\ell=0$ to $\ell=0.1$.
\label{fig:Thermal_branches}
}
\end{figure}
 In figure~\ref{fig:Thermal_branches} we display thermalization time in each of the three distinct parameter regimes of the quantum corrected geometry. Interestingly, as we may well have surmised from our findings of the mode behavior in the different branches, we find that the thermalization time of the Green's functions of single trace scalar operators in the CFT dual to the semi-classical geometry are strongly influenced by the type of singularity which was present before backreaction.

\subsection{Spinor field QNM}\label{sec:fermion_QNM}
We will begin by looking at how the quantum backreaction affects the QNM's of the unperturbed geometry by taking $\kappa=-1$. Before we follow the same strategy of study as the scalar case, it is interesting to make the following observation. In table~\ref{tab:qubtzModes_s_1/2} we give the values of the QNM for the leading and first three overtones at zero momentum at a fixed value of the parameter $x_1$ (allowing the temperature to vary). One immediately notices that, unlike the $s=0$ case, where turning on the backreaction of the quantum fields on the geometry generates a non-trivial real part of the mode, the $s=1/2$ case does not have this occur. That is, at zero momentum, the fermionic QNMs are purely dissipative.
\begin{table}[t!]
    \centering
    \begin{tabular}{c |c c c c}
    $\ell$ & $n_z=0$ &  $n_z=1$&  $n_z=2$&  $n_z=3$\\
    \hline
 $0$ & $-0.25 i $ & $  -0.75 i $ & $ -1.25 i $ & $ -1.75 i$ \\
 $0.1$ & $-0.293127 i $ & $ -0.87938 i $ & $ -1.46563 i $ & $ -2.05189i $ \\
 $0.2$ & $-0.327103 i $ & $ -0.98131 i $ & $ -1.63552 i $ & $ -2.28972 i $\\
 $0.3$ & $-0.35559 i $ & $ -1.06677 i $ & $ -1.77795 i $ & $ -2.48913 i $\\
 $0.4 $& $-0.380357 i $ & $ -1.14107 i $ & $ -1.90178 i $ & $ -2.6625i $ \\
 $0.5$ & $-0.40241 i $ & $ -1.20723 i $ & $ -2.01205 i $ & $ -2.81687i $ \\
 $0.6$ & $-0.42238 i $ & $ -1.26714 i $ & $ -2.1119 i $ & $ -2.95666 i $\\
 $0.7$ & $-0.440695 i $ & $ -1.32208 i $ & $ -2.20347 i $ & $ -3.08486 i $\\
 $0.8$ & $-0.457656 i $ & $ -1.37297 i $ & $ -2.28828 i $ & $ -3.20359i $ \\
 $0.9$ & $-0.473486 i $ & $ -1.42046 i $ & $ -2.36743 i $ & $ -3.3144 i $\\
 $1$ & $-0.488355 i $ & $ -1.46507 i $ & $ -2.44178 i $ & $ -3.41849 i $\\
    \end{tabular}
    \caption{\textbf{QNM of the qBTZ black hole:} The QNM frequencies, $\omega$, displayed here were computed with $x_1=1$ and for $\kappa=-1$ at zero momentum for $s=1/2$.
    \label{tab:qubtzModes_s_1/2}}
\end{table}
Furthermore, one can notice that the normalization given in table~\ref{tab:qubtzModes_s_1/2} is not the same as table~\ref{tab:qubtzModes}. This is no mistake. The interesting pattern of the modes is actually given precisely by eq.\ (\ref{eq:btz_fermion_poles}) with $n=0$
\begin{equation}\label{eq:qbtz_zero_modes}
    \omega= -4\pi i T(n_z+h_R)\, , \quad \omega= -4\pi i T(n_z+h_L)\,, \quad n_z\in\mathbb{Z}_+\, .
\end{equation}
As discussed in the previous section, the leading mode will set a thermalization time scale of CFT operators of conformal dimension $(h_L,h_R)$, and in this case, it is given precisely by
\begin{equation}\label{eq:qbtz_zero_modes_therm}
    \tau_{th}=-1/\omega=1/(4\pi T h_L)
\end{equation}
As we discussed in section~\ref{sec:background_geometry} mass relation $8 \mathcal{G}_3 M =\kappa \Delta^2$ admits two solutions for $x_1$ for a given value of $M$. One can repeat the exercise of completing table~\ref{tab:qubtzModes_s_1/2} for the second value of $x_1$ (which would correspond to branch 2) and again find the QNMs are identical to those displayed in table~\ref{tab:qubtzModes_s_1/2}. And, perhaps even more surprising, if we instead choose $\kappa=+1$ we find that the zero momentum QNMs of qCone are identically equal in form to the qBTZ black hole, i.e., they are given by eq.\ (\ref{eq:qbtz_zero_modes}). We therefore summarize this as follows:
\begin{mdframed}
    The functional form of the thermalization time of correlations of single trace operators of spin $s=\pm 1/2$ in the field theory dual to the braneworld construction (at zero momentum) are completely insensitive to the original type of singularity present in the geometry. Furthermore, they are completely insensitive to the physical origin of the quantum correction in the bulk. Despite their functional form, their dependence on the temperature may be used to distinguish the state of the CFT.
\end{mdframed}
It is very interesting that at zero momentum the pole structure of the retarded Green's functions is unchanged. It is very likely that these modes of zero momentum can be regarded as fermion zero modes, and are protected from quantum correction. Further, it is important to note that while the form eq.\ (\ref{eq:qbtz_zero_modes_therm}) has not changed, the behavior of the temperature as a function of the parameters of the system has. Hence although the form of the thermalization time has not changed, it can display very different behavior then that of its uncorrected counterpart, as a function of the quantum backreaction, for instance.

However, this story does not remain the same once we turn on a non-zero momentum of the fermion. With this in mind, we now follow the same procedure as the previous section, working with the fixed temperature data displayed in table~\ref{tab:parameter_values}. Beginning with $\kappa=-1$ the results for non-zero momentum are displayed in table~\ref{tab:qubtzModes_s_1/2_non-zero-momentum}. One again finds non-trivial modifications to the real part of the mode i.e. to the propagation of the mode.
One can see that the real part of the mode increases in magnitude, moving away from the imaginary axis as we increase the quantum backreaction. The imaginary part decreases in magnitude as we increase the quantum backreaction, moving towards the real axis.

With some understanding of the behavior of the modes in the quantum corrected black hole we now turn to the behavior of the modes in the quantum dressed conical singularity with $\kappa=+1$.
In table~\ref{tab:quconeModes_s_1/2_non-zero-momentum} we display the values of the QNMs for the leading and first three overtones at non-zero momentum of the qCone. Here we again note table~\ref{tab:quconeModes_s_1/2_non-zero-momentum} begins at $\ell=1/100$ since our numerics require a horizon. Looking at the table, we again notice that unlike the black hole corrected by the backreaction of the Casimir stress-energy tensor, the real parts of the modes decrease in magnitude, moving towards the imaginary axis, as the quantum backreaction is increased. However, the imaginary parts now show the opposite trend, increasing in magnitude, moving further away from the real line, deeper into the complex plane.

\begin{figure}[t!]
    \centering
\includegraphics[scale=0.85,trim={0 1mm 0 0},clip]{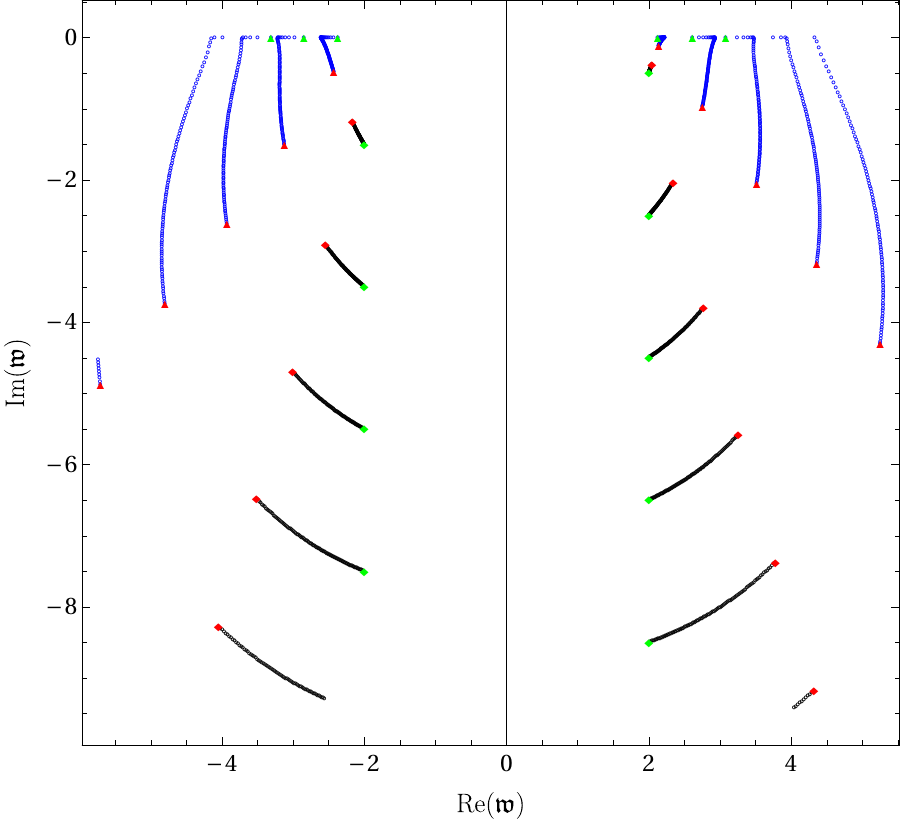}
\vspace{-3mm}
    \caption{\textbf{Mode transition from quantum dressed conical singularity to quantum corrected BTZ: }The lowest QNM frequencies are displayed as one slowly turns on quantum backreaction for operators obeying $h_R-h_L=\pm 1/2$. The modes are displayed for fixed $n=1$ and $4\pi T=1$ as $\ell$ varies from $\ell=1/100$ (green triangles) to $\ell=1/10$ (red triangles). Blue dots represent the quantum dressed conical singularity ($\kappa=1$) while the black dots represent the quantum corrected BTZ geometry ($\kappa=-1$). Note that the Green dots on the black curves are at $\text{Re}(\w)=2$ due to our choice of units.
\label{fig:quBTZ_TO_QDConical_s_1/2}}
\end{figure}

This contrasting behavior is made more apparent visually in figure~\ref{fig:quBTZ_TO_QDConical_s_1/2}. Here we see the motion of the QNMs, at non-zero momentum, in the background of both the quantum-dressed conical singularity and the quantum-corrected BTZ geometry. As in the scalar case (figure~\ref{fig:quBTZ_TO_QDConical}), the blue dots correspond to the quantum-dressed conical singularity and black dots to the quantum-corrected BTZ black hole. The general trend of the modes is similar, beginning with here $\ell=1/100$, the green triangles, the QNMs are well separated. And, as we increase the quantum back reaction, the QNMs move towards one another ending at $\ell=1/10$ (red triangles). However, the shape the QNM trace through the complex plane is quite different than the scalar case. Notice that at any finite $\ell$ a gap remains between the QNMs of the quantum dressed conical singularity and the quantum correct BTZ black hole, which increases in size as one moves deeper in the complex frequency plane.

\newpage
\begin{landscape}
    \begin{table}[t!]
    \centering
    \begin{tabular}{c |c c c c}
    $\ell$ & $n_z=0$ &  $n_z=1$&  $n_z=2$&  $n_z=3$\\
    \hline
 $0 $&$ 2.\, -0.5 i $&$ -2.-1.5 i $&$ 2.\, -2.5 i $&$ -2.-3.5 i $\\
 $0.01 $&$ 2.00274\, -0.492698 i $&$ -2.01148-1.4804 i $&$ 2.02505\, -2.47153 i $&$ -2.04176-3.4645 i$ \\
 $0.02 $&$ 2.0057\, -0.484765 i $&$ -2.02375-1.45914 i $&$ 2.05158\, -2.44064 i $&$ -2.0856-3.42588 i $\\
 $0.03 $&$ 2.0089\, -0.476128 i $&$ -2.03688-1.43603 i $&$ 2.07968\, -2.40707 i $&$ -2.13162-3.38378 i $\\
 $0.04 $&$ 2.01238\, -0.466693 i $&$ -2.05096-1.41082 i $&$ 2.10948\, -2.37042 i $&$ -2.17998-3.33768 i $\\
 $0.05 $&$ 2.01616\, -0.456336 i $&$ -2.06611-1.38319 i $&$ 2.14115\, -2.33024 i $&$ -2.23092-3.28694 i $\\
 $0.06 $&$ 2.02032\, -0.44489 i $&$ -2.08247-1.3527 i $&$ 2.17494\, -2.28585 i $&$ -2.28477-3.23068 i $\\
 $0.07 $&$ 2.02492\, -0.432116 i $&$ -2.10028-1.31872 i $&$ 2.21121\, -2.23632 i $&$ -2.34202-3.16764 i $\\
 $0.08 $&$ 2.03008\, -0.417662 i $&$ -2.11988-1.28033 i $&$ 2.25052\, -2.18025 i $&$ -2.40343-3.09599 i $\\
 $0.09 $&$ 2.03599\, -0.40096 i $&$ -2.14182-1.23602 i $&$ 2.29376\, -2.11541 i $&$ -2.47028-3.01271 i $\\
 $0.1 $&$ 2.043\, -0.380986 i $&$ -2.1671-1.18308 i $&$ 2.34257\, -2.03774 i $&$ -2.54483-2.91244 i $\\
    \end{tabular}
    \caption{\textbf{QNM of the qBTZ black hole ($s=1/2$):} The QNM frequencies, $\mathfrak{w}=\omega/(2\pi T)$, displayed here were computed with $n=1$ and $4\pi T=1$ for $\kappa=-1$ at zero momentum using the values of $x_1$ tabulated in table~\ref{tab:parameter_values}.
    \label{tab:qubtzModes_s_1/2_non-zero-momentum}}
\end{table}
\begin{table}[b!]
    \centering
    \begin{tabular}{c |c c c c}
    $\ell$ & $n_z=0$ &  $n_z=1$&  $n_z=2$&  $n_z=3$\\
    \hline
 $0.01 $&$ 2.19975\, -8.018758\times 10^{-7} i $&$ -2.57824-0.000041219 i $&$ 2.91853\, -0.000926899 i $&$ -3.21791-0.0103489 i $ \\
 $0.02 $&$ 2.21778\, -0.000146734 i $&$ -2.60838-0.00521028 i $&$ 2.91845\, -0.0461654 i $&$ -3.1929-0.160675 i $\\
 $0.03 $&$ 2.2199\, -0.00166429 i $&$ -2.59215-0.032306 i $&$ 2.88529\, -0.149339 i $&$ -3.18464-0.352944 i $\\
 $0.04 $&$ 2.2134\, -0.00677473 i $&$ -2.56672-0.0785592 i $&$ 2.86104\, -0.266672 i $&$ -3.18633-0.531388 i $\\
 $0.05 $&$ 2.20253\, -0.0165768 i $&$ -2.54184-0.134412 i $&$ 2.84356\, -0.384207 i $&$ -3.18735-0.697639 i $\\
 $0.06 $&$ 2.19001\, -0.0307109 i $&$ -2.51859-0.195352 i $&$ 2.82881\, -0.499599 i $&$ -3.18494-0.856729 i $\\
 $0.07 $&$ 2.17713\, -0.048431 i $&$ -2.49642-0.259907 i $&$ 2.81414\, -0.613886 i $&$ -3.17815-1.01279 i $\\
 $0.08 $&$ 2.16431\, -0.0692502 i $&$ -2.47452-0.328121 i $&$ 2.7978\, -0.729241 i $&$ -3.16621-1.16963 i $\\
 $0.09 $&$ 2.15154\, -0.0931449 i $&$ -2.45193-0.401121 i $&$ 2.77829\, -0.848703 i $&$ -3.14797-1.33159 i $\\
 $0.1 $&$ 2.13848\, -0.120747 i $&$ -2.42733-0.481475 i $&$ 2.75366\, -0.976992 i $&$ -3.12127-1.50505 i $\\
  \end{tabular}
    \caption{\textbf{QNM of the qCone ($s=1/2$):} The QNM frequencies, $\mathfrak{w}=\omega/(2\pi T)$, displayed here were computed with $n=1$ and $4\pi T=1$ for $\kappa=+1$ at zero momentum using the values of $x_1$ tabulated in table~\ref{tab:parameter_values}. Notice, there is no horizon for $\ell=0$, hence this row is omitted.
    \label{tab:quconeModes_s_1/2_non-zero-momentum}}
\end{table}
\end{landscape}
\newpage

%%%%%%%%%%%%%%%%%%%%%%%%%%%%%%%%%%%%%%%%%%%%%%%%%%%%%%%%%%%%%%%%%%%%%%%%%%%%%%%%%%%%%%%%%%%%%%%%
\section{Pole-skipping}\label{sec:pole_skipping}
There is a curious situation that can occur in the Green's functions where the pole can be ``skipped'' , that is, a general Green's function can be decomposed as
\begin{equation}
G_R(\omega,q)=\frac{A(\omega,q)}{B(\omega,q)}
\end{equation}
the poles of this function correspond to the points where $B(\omega_*,q_*)=0$. Pole-skipping refers to those points that would be poles of the Green's function if it was not the case that $A(\omega_*,q_*)=0$ as well. It has been demonstrated that these points are related to a situation where there exists an additional infalling solution (a trivial solution) to the near horizon solution of the perturbative equations of motion. In this section we will obtain the pole-skipping relations for the various CFT operators discussed in the previous section. Before we begin we make note of the fact that we will follow in the footsteps of previous studies (see for instance ~\cite{Birmingham:2001pj,Son:2002sd,Castro:2014tta,Blake:2019otz,Liu:2020yaf,Natsuume:2020snz}) and extend the range of $\phi\in (-\pi,\pi]$ to $\phi\in (-\infty,\infty)$ replacing the mode expansion~\footnote{This is a good assumption in the large temperature limit for instance, see for example~\cite{Amano:2023bhg} for an in-depth example of this. } given in eq.\ (\ref{eq:mode_expansion}) by \begin{equation}
      \Phi(t,r,\phi)=\int \exd^2 q e^{-i\omega t+iq\phi}\Phi(r;\omega,q)\, ,\quad (\omega,q)\in\mathbb{C}\, ,
\end{equation} where notably the sum over $n$ has been replaced, with some foresight, to an integral over the complex-valued momentum $q$.

As with previous section, further details about pole-skipping can be found in appendix~\ref{app:pole} with appendix~\ref{app:pole_scalar} dedicated to information about pole-skipping for scalars and appendix~\ref{app:pole_spinor} dedicated to information about pole-skipping for spinors.

%%%%%%%%%%%%%%%%%%%%%%%%%%%%%%%%%%%%%%%%%%%%%%%%%%%%%%%%%%%%%%%%%%%%%%%%%%%%%%%%%%%%%%%%%%%%%%%
\subsection{Operators with dimension $\Delta$ and spin $s=0$.}\label{sec:scalar_pole_skipping}
In principle the equation of motion for $\Phi$, expanded order by order near the horizon relates the coefficients $\Phi^{(0)}$ and $\Phi^{(n)}$ together. However, there is a separate solution which leaves $\Phi^{(1)}$ undetermined, this is given by,
\begin{equation}
    \omega_* = -\frac{1}{2} i H'(\bar{r}_h)\, , \qquad q_*=\pm \frac{\sqrt{-\bar{r}_h \left(H'(\bar{r}_h)+2 m^2 \bar{r}_h\right)}}{\sqrt{2}}
\end{equation}
Using the relation between the blackening factor $H$ and the temperature as well as the relation between the mass of the bulk scalar field and the operator dimension we can rewrite this more succinctly as,
\begin{equation}
  \omega_*= -2\pi i T\, , \qquad q_*^2= \rb_h \left(-\frac{(\Delta -2) \Delta  \rb_h}{\ell_3^2}-2 \pi  T\right)
\end{equation}
This story is directly related to the singularity structure of the ODE and as explained in appendix~\ref{app:pole_scalar}, from a brane observer's perspective, the backreaction of quantum fields leading to a semi-classical metric preserves the singularity structure of massive scalar probes, consistent with previous studies on higher curvature corrections to pole-skipping points~\cite{Natsuume:2019vcv}. Pole-skipping thus occurs for correlators of the scalar operators dual to $\Phi$.

In~\cite{Schalm:2018lep} a crucial observation is made relating the hydrodynamic behavior of sound modes to the scrambling of information in holographic theories. The energy-energy correlation function's would-be pole, which defines scalar hydrodynamic modes, has vanishing residue, and is skipped at complexified momentum given by (in the case of $\mathcal{N}=4$ SYM)~\cite{Grozdanov:2019uhi}
\begin{equation}
    \mathfrak{w}_*=i\frac{\lambda_L}{2\pi T}=i\mathfrak{D}_*\, ,\quad \mathfrak{q}_*=i \mathfrak{l}_*
\end{equation}
where
\begin{equation}
  \mathfrak{D}_*=1\, , \quad \mathfrak{l}_*=\sqrt{3/2}\, ,\quad v_b=\frac{\mathfrak{D}_*}{\mathfrak{l}_*}=\sqrt{\frac{2}{3}}\, .
\end{equation}
These correspond to the same values obtained from the out-of-time-order four-point correlators that diagnose quantum chaos, between a pair of local operators $V$ and $W$. Specifically, for quantum chaotic theories with many degrees of freedom one expects
\begin{equation}
\frac{\braket{V(0,0)W(\vec{x},t)V(0,0)W(\vec{x},t)}_\beta}{\braket{V(0,0)V(0,0)}_\beta\braket{W(\vec{x},t)W(\vec{x},t)}_\beta}= 1- e^{\lambda_L(t-t_*-|\vec{x}|/v_b)}\,,
\end{equation}
where $t_*\sim\beta \log S$ is the scrambling time, $\lambda_L$ is the so-called quantum Lyapunov exponent and $v_b$ is the butterfly velocity. The quantum Lyapunov exponent is universal for holographic theories with classical gravity duals, saturating the MSS bound on chaos \cite{Maldacena:2015waa}, $\lambda_L= 2\pi T$. We expect $\lambda_L$ to remain robust against the quantum corrections in our braneworld setup for several reasons. First, the saturation of the bound can be linked to graviton scattering in the near-horizon region \cite{Shenker:2014cwa}, which should be unaffected by matter loop corrections. Second, the bound has been proven to hold in the open string sector, where it arises from the scattering of an infinite tower of string excitations, effectively mimicking a single graviton exchange \cite{deBoer:2017xdk}.
Although the braneworld theory includes a massive graviton with an infinite series of higher derivative corrections, we expect the same rationale to apply, partly because one should be able to recast the same process in terms of graviton scattering in the higher-dimensional Einstein gravity bulk. Conversely, we anticipate that $v_b$ will receive corrections due to quantum backreaction, making it an intriguing quantity to investigate in our context. Initially, a bound on $v_b$ was conjectured in~\cite{Mezei:2016zxg}, similar to the MSS bound, given by
\begin{equation}
v_b \leq \sqrt{\frac{d}{2(d-1)}}\,,
\end{equation}
which is saturated by the Schwarzschild-AdS$_{d+1}$ black hole. This bound was established for holographic theories with Einstein gravity duals, under the assumptions of relativistic invariance and the null energy condition (NEC). However, subsequent studies~\cite{Giataganas:2017koz,Fischler:2018kwt,Gursoy:2020kjd,Eccles:2021zum} have demonstrated that this bound can be violated in systems undergoing non-trivial RG flows that break relativistic symmetries, which can in principle be induced by backreaction effects in our braneworld models. Additionally, the NEC may be violated at the semi-classical level due to Casimir effects, which are prominent at least for the qCone branch.

\begin{figure}[t!]
    \centering
\includegraphics[width=0.8\textwidth,trim={0 1mm 0 0},clip]{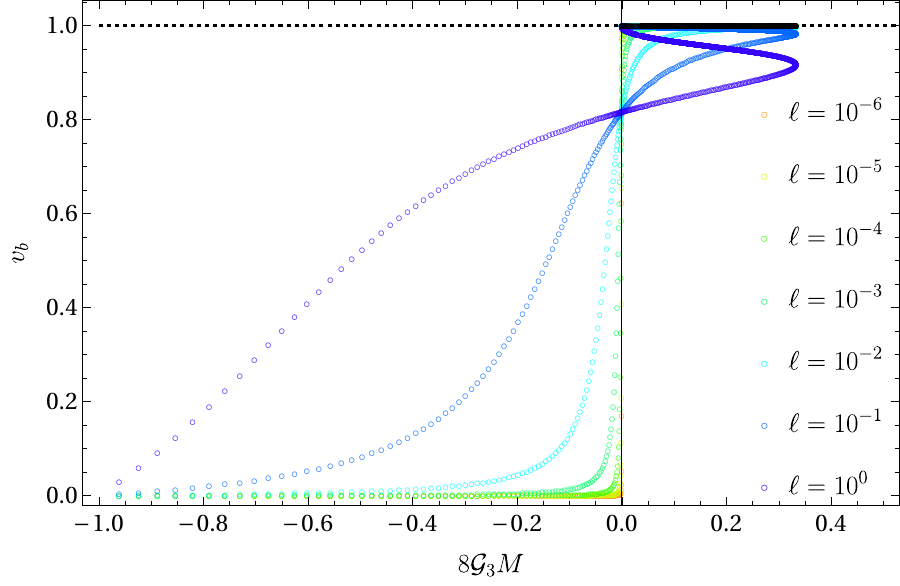}
\vspace{-3mm}
    \caption{Ratio of the frequency to momentum of the nearest pole-skipping point to the origin of the complex plane $v_b=|\mathfrak{w}_*/\mathfrak{q}_*|$. The black line refers to the value of the ratio for $\ell=0$. \label{fig:qubtz_pole_skipping_ratio}}
\end{figure}

In addition to energy-energy correlators, a quantum Lyapunov exponent and a butterfly velocity can be defined for each channel (spin-2, spin-1, and spin-0) in a holographic theory. These quantities are determined by the pole-skipping location closest to the origin of the complex frequency plane, by
\begin{equation}
    \lambda_L=2\pi T |\mathfrak{w}_*|, \qquad v_b=\left| \frac{\mathfrak{w}_*}{\mathfrak{q}_*}\right|
\end{equation}
Curiously, for scalar fluctuations around the BTZ black hole, pole-skipping, of operator dimension $\Delta=2$ (corresponding to a marginal operator) occurs at
\begin{equation}
    \mathfrak{w}_*=-i(n+1)\, ,\quad \mathfrak{q}_*=\pm i(n+1) \, , \quad n\in\mathbb{Z}
\end{equation}
for which the closest to the origin is $(\mathfrak{w}_*,\mathfrak{k}_*)=(-i,\pm i)$. This yields $(\lambda_L,v_b)=(2\pi T, 1)$, in agreement with the OTOC calculation.
Naively extending this relation to the qBTZ black hole
is displayed in figure~\ref{fig:qubtz_pole_skipping_ratio}.
Recall that setting $\ell=0$ and $\kappa=-1$, the line element on the brane reduces to BTZ with blackening factor $H(r)=r^2/\ell_3^2-M$ for which the mass function $M=4 x_1^2/\left(x_1^2+3\right)^2$ is bounded above by $1/3$. The black points of figure~\ref{fig:qubtz_pole_skipping_ratio} correspond to this branch where we expect a butterfly velocity of $1$ in natural units. Notice that this does not extend to negative values of $8\mathcal{G}_3M$ since we have computed this from a near horizon expansion and there is no horizon for $\ell=0$ and $8\mathcal{G}_3M<0$. Furthermore, there is no mass/temperature dependence of the butterfly velocity, it is a constant. With the colored dots we display the same ratio $|\w_*/\q_*|$ for various values of $\ell$. We can see that turning on the quantum backreaction we begin to see a non-trivial dependence of the ratio on the temperature. For small backreaction $\ell=10^{-6}$ the ratio is roughly the same as before, however, the presence of a horizon allows us to continue this ratio to the quantum-dressed conical singularity. Here we see that there is a sharp transition at $x_1=0$ from $|\w_*/\q_*|\approx 1$ to $|\w_*/\q_*|\approx 0$. If we can identify this ratio as the butterfly velocity then this shows that for small quantum backreaction in dressed conical singularities, the butterfly velocity is approximately zero indicating a lack of local dependence on the growth of operators in the CFT dual.
As we increase the quantum backreaction we see that this ratio behaves in a different way for each branch of the solution and we see that for all branches the ratio is less than the conformal value. For branch 1b and 2, with $8\mathcal{G}_3M>0$ we find that the ratio splits such that for each value of $8\mathcal{G}_3 M$ there are two values of the ratio. The source of the quantum backreaction distinguishes the two, with a larger magnitude of the ratio for the situation when the primary source of the quantum backreaction is Hawking radiation in thermal equilibrium with the black hole. The ratio for branch 1a, for the range of $\ell$ displayed, is always smaller than the other two branches and monotonically increases as the magnitude of $8\mathcal{G}_3 M$ decreases.

\textit{Higher Matsubara Frequencies:} It should be noted that these are not the only pole-skipping points that occur. The pattern in the equations will continue at higher orders in the expansion. An important observation, as made in a list in~\cite{Natsuume:2019vcv}, is that one requirement for this particular form of the equations of motion near the horizon is a static geometry. That this, there exists a one-parameter family of isometries whose orbits are timelike curves (a time translation) and a spacelike hypersurface orthogonal to the orbits of the isometry~\cite{Wald:1984rg}. Since we can extend this isometry to the conformal boundary it implies those holographic CFTs, in thermal states, with time-reversal invariance have pole-skipping occurring in the retarded Green's functions at Matsubara frequencies~\footnote{Interestingly, pole-skipping generically does not occur at zero temperature for $1+1$ CFTs, like the ones we are interested in, only for very specific conditions (massless scalar perturbations) of extremal BTZ black holes does pole-skipping occur~\cite{Natsuume:2020snz}. In this case, it is intersector pole-skipping, left-moving zeros cancel right-moving poles. However, it has been shown to occur in the ``confining'' phase of SYM theory at frequencies similar to the Matsubara frequencies~\cite{Natsuume:2023lzy} }. A stationary geometry, lacking the spacelike hypersurface orthogonal to the time translations, will naturally not have pole-skipping frequencies coinciding with Matsubara frequencies. Key examples of this include the BTZ geometry~\cite{Jeong:2023rck} with non-vanishing angular momentum~\footnote{We therefore expect this to occur in the rotating qBTZ solution.} $J\neq 0$, Kerr-AdS$_4$ black holes~\cite{Blake:2021hjj} and Myers-Perry AdS$_5$ black holes~\cite{Amano:2022mlu}. However, in each of these examples, one can go to ``comoving'' coordinates, which restore the Matsubara form of the pole-skipping frequencies.

\begin{figure}[t!]
    \centering
\includegraphics[width=0.85\textwidth,trim={0 1mm 0 0},clip]{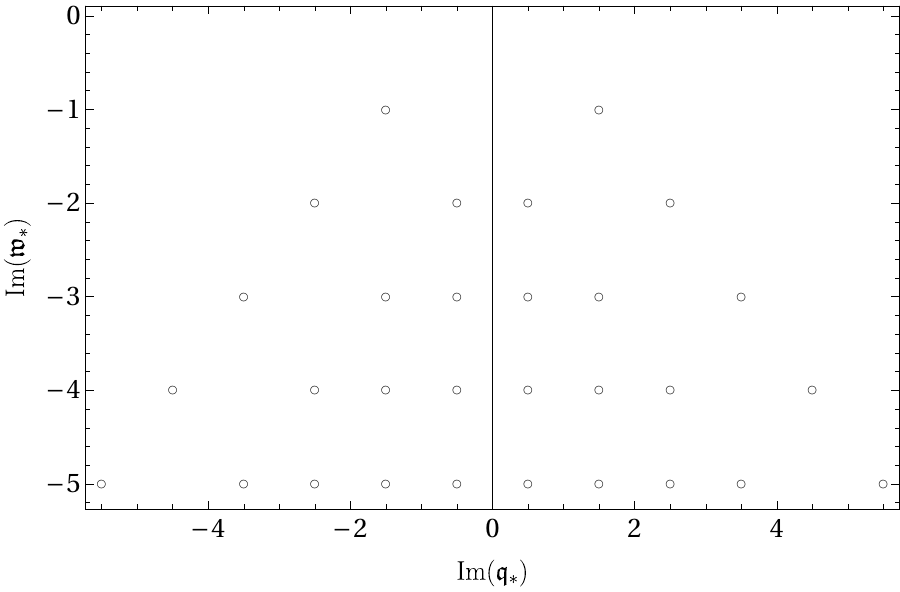}
\vspace{-3mm}
    \caption{Pole-skipping points of Green's functions of single trace scalar operators dual to massive scalar fields in BTZ geometry
    For this image $\Delta=2.5$, $\ell=1$ and $\rb_h=1$.
\label{fig:btz_Matsubara}}
\end{figure}

Before investigating what this procedure leads to in the case of qBTZ it is useful to first remind oneself of the results in a simpler case. Shown in figure~\ref{fig:btz_Matsubara} are the first four sets of higher pole-skipping frequencies. These have been investigated in detail in many works including~\cite{Blake:2019otz}. In particular one can notice that they occur at imaginary frequency and imaginary momentum. And, for a given Matsubara frequency a pair of two additional momenta appear at which a pole is skipped.
\begin{figure}[t!]
    \centering
    \includegraphics[width=0.49\textwidth]{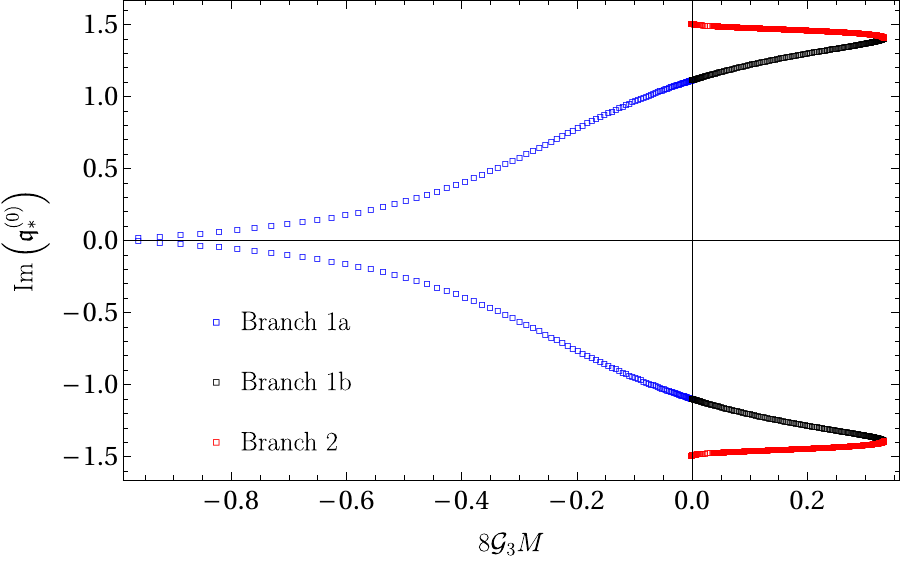} \hfill
    \includegraphics[width=0.49\textwidth]{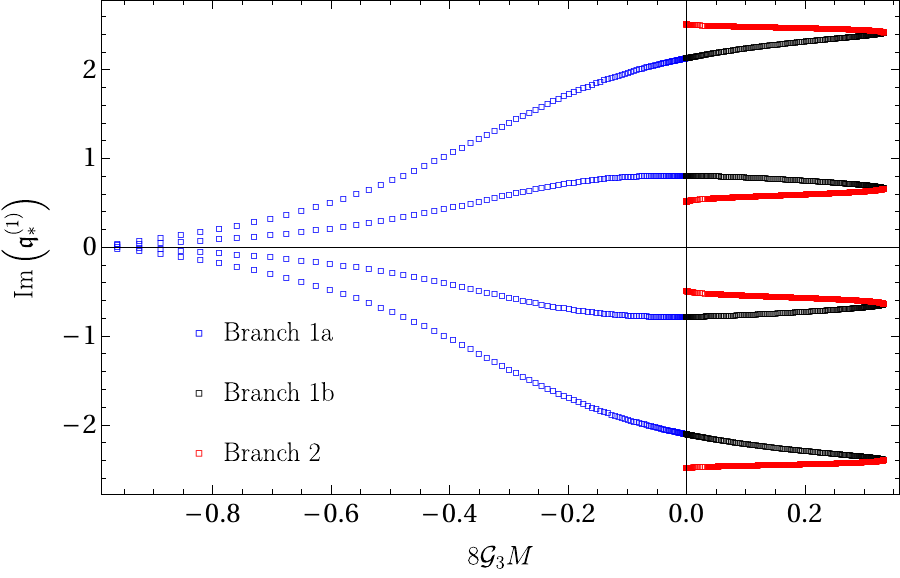} \\
    \includegraphics[width=0.49\textwidth]{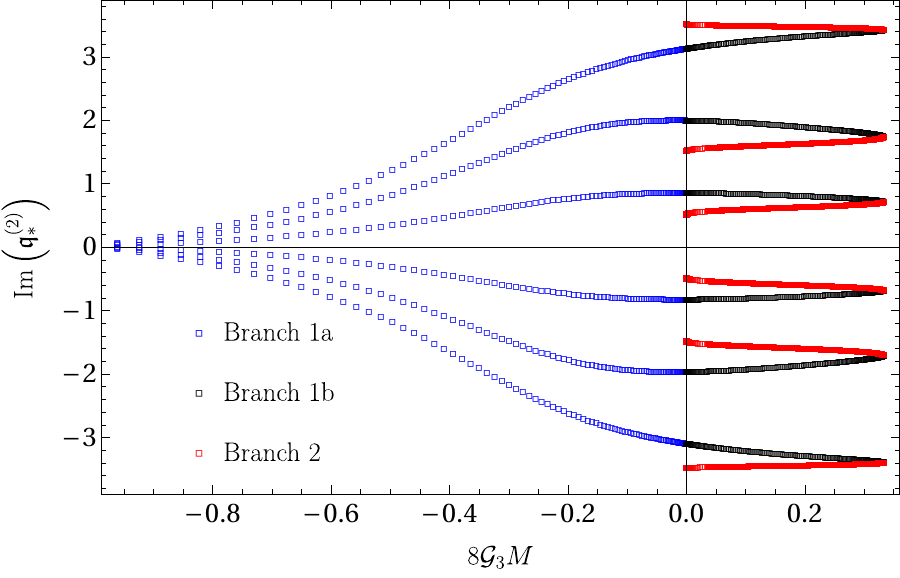} \hfill
    \includegraphics[width=0.49\textwidth]{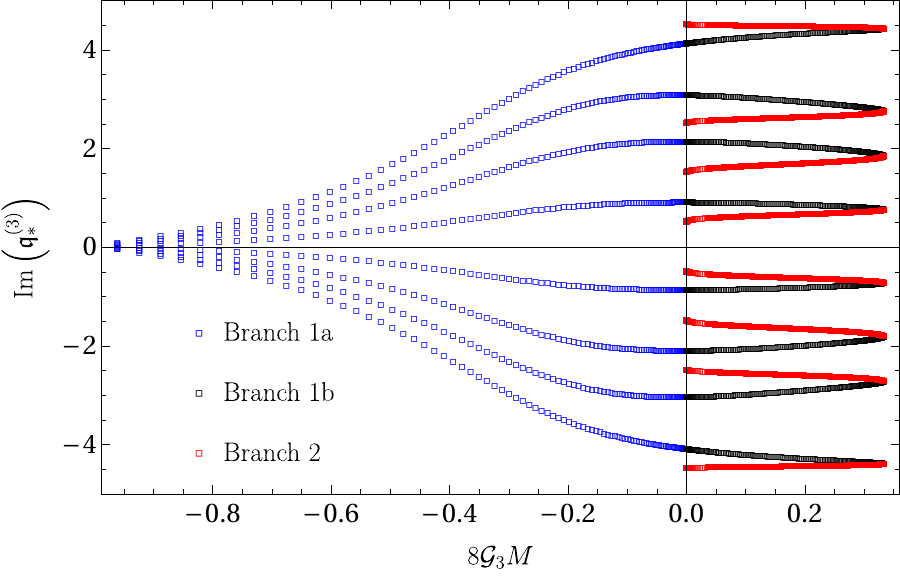}
\vspace{-4mm}
    \caption{\textbf{Scalar pole-skipping: }Lowest pole-skipping momentum of Green's functions of single trace scalar operators dual to massive scalar fields in qBTZ geometry. Here we have taken $\Delta=2.5$ and $\ell=1/3$ and hence the strength of the quantum backreaction $\nu=1/3$.
    \label{fig:quBTZ_Lowest_pole_skipping}}
\end{figure}

In figure~\ref{fig:quBTZ_Lowest_pole_skipping} we display the momentum associated with the first four Matsubara frequencies as a function of the mass of the quantum corrected black hole, labeled as $\q^{(n-1)}_*$ for $\w_n=- i n$. In each image, we use the same color scheme with blue representing branch 1a, black representing branch 1b and red representing branch 2. There are a few features that are worth pointing out. First, for all pole-skipping frequencies, as $8\mathcal{G}_3M$ approaches its lower bound, the associated momentum all approach zero and the momentum increases in magnitude as $8\mathcal{G}_3M$ approaches zero from below.
As we move from the quantum-dressed conical singularity to the quantum-corrected black hole the number of pole-skipping points doubles, this is due to the fact that there are two branches for $
\kappa=-1$.
This doubling of the pole-skipping points exists for all the positive values of $8\mathcal{G}_3M$ except for the location where branch 1b and branch 2 meet at the maximum value of $8\mathcal{G}_3M=1/3$ where there are again $2n$ pole-skipping momenta.
\begin{mdframed}
Hence one can distinguish between the qCone phase and the qBTZ phase by the number of pole-skipping points of the retarded Green's function at a given mass $M$.
\end{mdframed}
In addition, one can notice the following subtle feature of the plots. The value of the momentum can be organized by magnitude~\footnote{Here we are considering only what is happening in the images provided for figure~\ref{fig:quBTZ_Lowest_pole_skipping}. It can happen the pole-skipping momentum are degenerate, e.g.for $n=1$ and $\{q^{(0)}_{i\,\, *}\}$ we have
\begin{equation}
    q^{(1)}_{1\,\, *}=q^{(1)}_{2\,\, *}<q^{(1)}_{3\,\, *}=q^{(1)}_{4\,\, *} \, ,
\end{equation} which are referred to as anomalous pole-skipping points. This situation typically indicates that although these points appear in a near horizon analysis they do not actually correspond to points of the retarded Green's function where pole-skipping occurs.} e.g. for $n=1$ and $\{q^{(0)}_{i\,\, *}\}$ we have
\begin{equation*}
    q^{(0)}_{1\,\, *}<q^{(0)}_{2\,\, *}<q^{(0)}_{3\,\, *}<q^{(0)}_{4\,\, *}
\end{equation*}
for $0\leq 8\mathcal{G}_3M<1/3$ and $q^{(0)}_{1\,\, *}<q^{(0)}_{2\,\, *}$ for $-1/8\leq \mathcal{G}_3M<0$.
\begin{figure}[t!]
    \centering
    \includegraphics[width=0.85\textwidth,trim={0 1mm 0 0},clip]{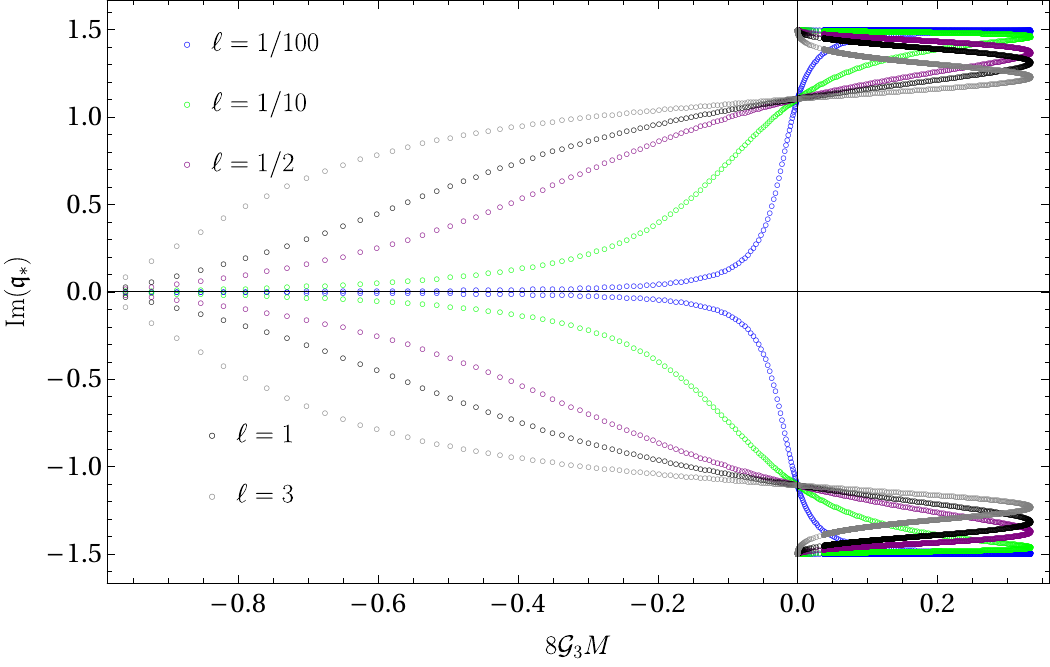}
    \vspace{-3mm}
    \caption{Lowest pole-skipping momentum of Green's functions of single trace scalar operators dual to massive scalar fields in qBTZ geometry. Here we have taken $\Delta=2.5$ while allowing the strength of the quantum backreaction $\ell$ to vary.
    \label{fig:quBTZ_nudep}}
\end{figure}
One can notice that the pole-skipping momentum of branch 2 is larger in magnitude than branch 1b. However, when we move to the next Matsubara frequency, where we have 8 possible pole-skipping momentum, something changes. In the quantum dressed conical singularity regime the pole-skipping momentum can be ordered $q^{(1)}_{1\,\, *} < \cdots <q^{(1)}_{4\,\, *}$. Following the trajectory of $q^{(1)}_{1\,\, *}\,\, \&\,\, q^{(1)}_{4\,\, *}$ we see that the same phenomenon occurs, the magnitude of $q^{(1)}_{1\,\, *}\,\, \&\,\, q^{(1)}_{4\,\, *}$ is larger for branch 2. However, for $q^{(1)}_{2,3\,\, *}$ the inverse is true, the magnitude of $q^{(1)}_{2,3\,\, *}$ in branch 1b is larger. In fact this continues to occur for all higher Matusubara frequencies, at $n$ there are $4n$ possible momentum organized into $2n$ branches, $q^{(n-1)}_{j\,\, *}$ for $j=1,2n$ will be largest in magnitude in branch 2 while $q^{(n-1)}_{j\,\, *}$ for $j=2\cdots 2n-1$ will be largest in magnitude in branch 1b.

\begin{figure}[t!]
    \centering
    \includegraphics[width=0.5\linewidth]{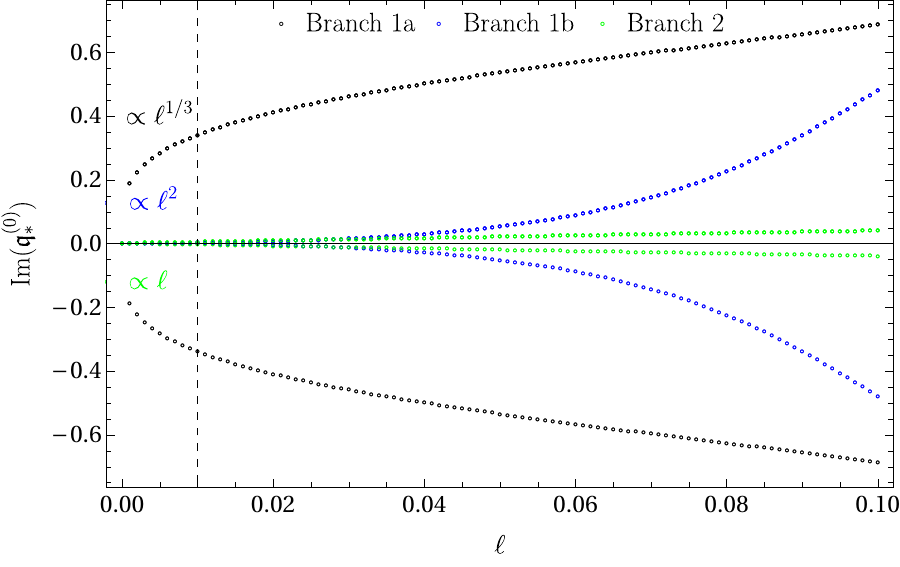}\hfill
    \includegraphics[width=0.49\linewidth]{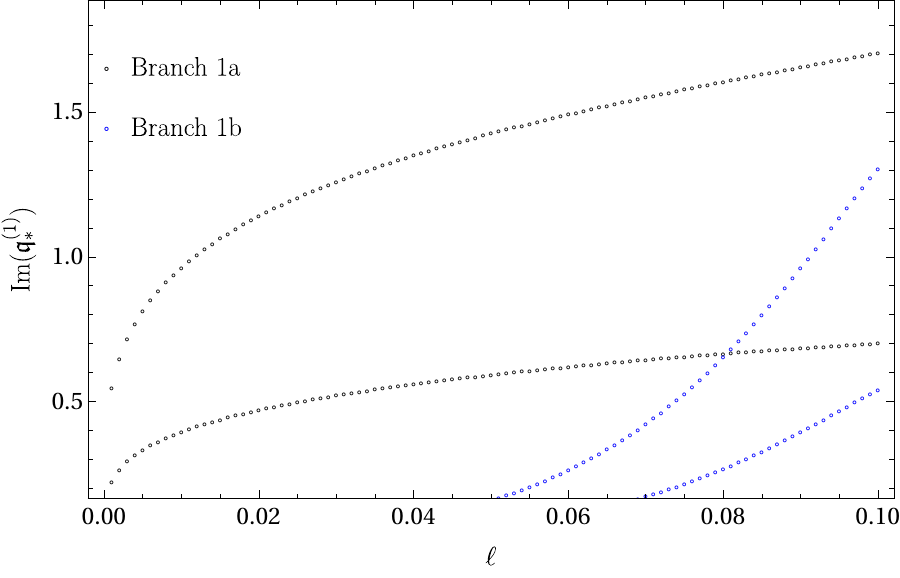}
    \vspace{-5mm}
    \caption{Pole-skipping momentum of Green's functions of single trace scalar operators dual to massive scalar fields in qBTZ geometry displayed as a function of $\ell$. Here we have taken $\Delta=2$ and $4\pi T=1$ while allowing the strength of the quantum backreaction $\ell$ to vary. \textit{Left:} The lowest pole-skipping momentum. \textit{Right:} The first higher pole-skipping momentum. \label{fig:scalar_pole_func_l}}
\end{figure}

Finally, before closing this section, it is interesting to look at what happens when we change the quantum backreaction. In figure~\ref{fig:quBTZ_nudep} we display the lowest pole-skipping points for fixed operator dimension as a function of the quantum backreaction. One can see that as we decrease the quantum backreaction from the largest value displayed as gray dots to the smallest value, displayed as blue dots, the curves in the qCone region begin to deform to a constant line given by $\q_*=0$ before sharply growing near the quantum vacuum geometry towards the values they would take in the uncorrected BTZ geometry. In the qBTZ mass range one can see that the multivalued branches begin to deform, with value of the pole-skipping momentum of branch 1b and branch 2 approaching one another as they both move to meet at the pole-skipping momentum of the uncorrected geometry. This is the same type of behavior as displayed in figure~\ref{fig:qubtz_pole_skipping_ratio}. Another way to display the effect of the quantum correction on the pole-skipping points is to work at fixed temperature as we did with the QNMs. In figure~\ref{fig:scalar_pole_func_l} we display this information on the left for the lowest pole-skipping momentum and on the right for the next highest pole-skipping momentum. Interestingly, one can find through fitting the data that $\text{Im}(\q^{(0)}_*)\propto \ell^{1/3}, \ell^{2} ,\ell$ for branch 1a, 1b and 2 respectively at $\ell\ll 1$ (the dashed vertical line shows the cut in the data taken to find these exponents). This provides another distinction between not only what is cloaked behind the horizon, but the mechanism responsible for the quantum correction. However, something interesting occurs beginning with the first higher pole-skipping point as displayed in the right image. We can see that the blue dots corresponding to branch 1b intersect the black dots representing branch 1a. That is, at this specific value of $\ell$, the pole-skipping momentum is degenerate, and the Green's function skips a pole at the same momentum in the thermal state dual to the qBTZ black hole and the state dual to the qCone. Hence at this momentum the ability to distinguish the state via the Green's functions is partially lost. It is only partially lost, since other pole-skipping momentum are still distinct.

%%%%%%%%%%%%%%%%%%%%%%%%%%%%%%%%%%%%%%%%%%%%%%%%%%%%%%%%%%%%%%%%%%%%%%%%%%%%%%%%%%%%%%%%%%%%%%%%%%%%%%%%%%%%%55
\subsection{Operators with dimension $\Delta$ and spin $s=\pm 1/2$}\label{sec:fermion_pole_skipping}
\begin{figure}[t!]
    \centering
    \includegraphics[width=0.49\textwidth]{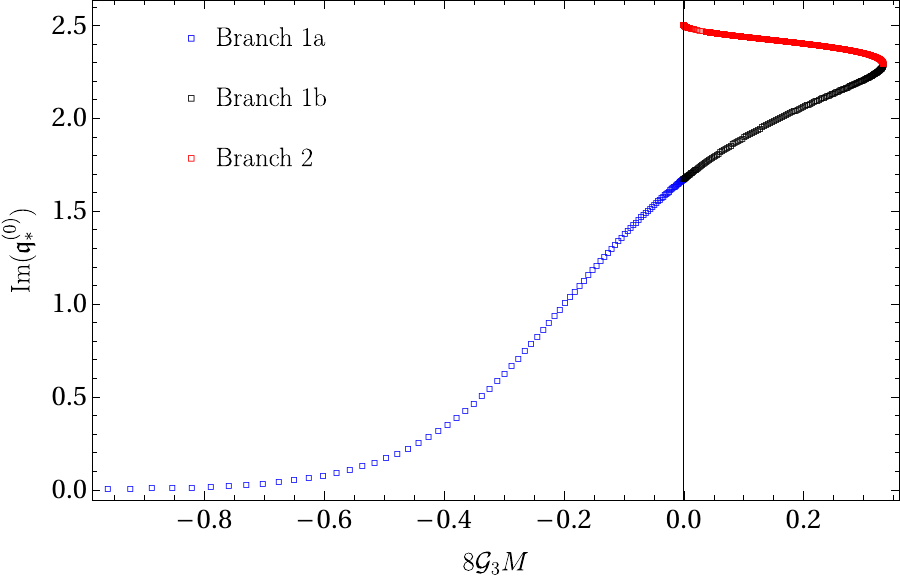} \hfill
    \includegraphics[width=0.49\textwidth]{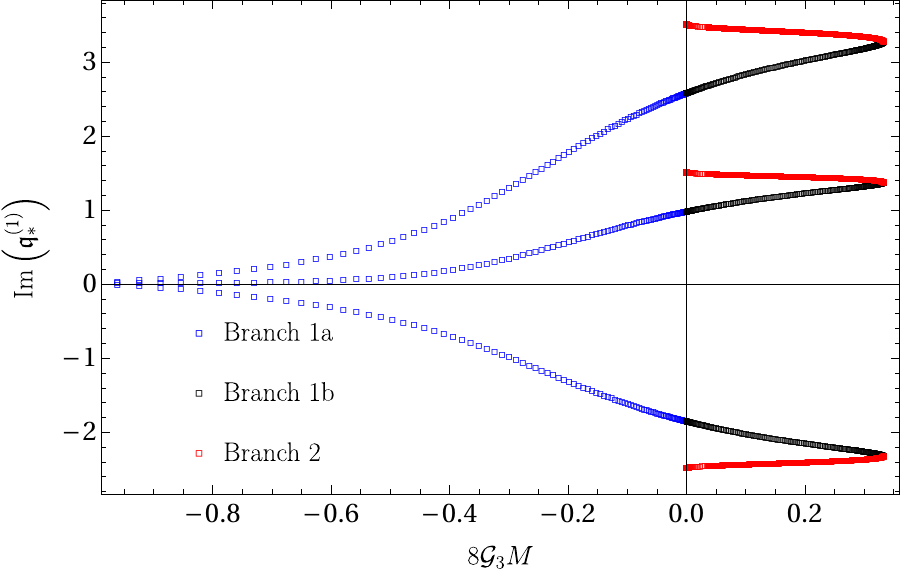} \\
     \includegraphics[width=0.49\textwidth]{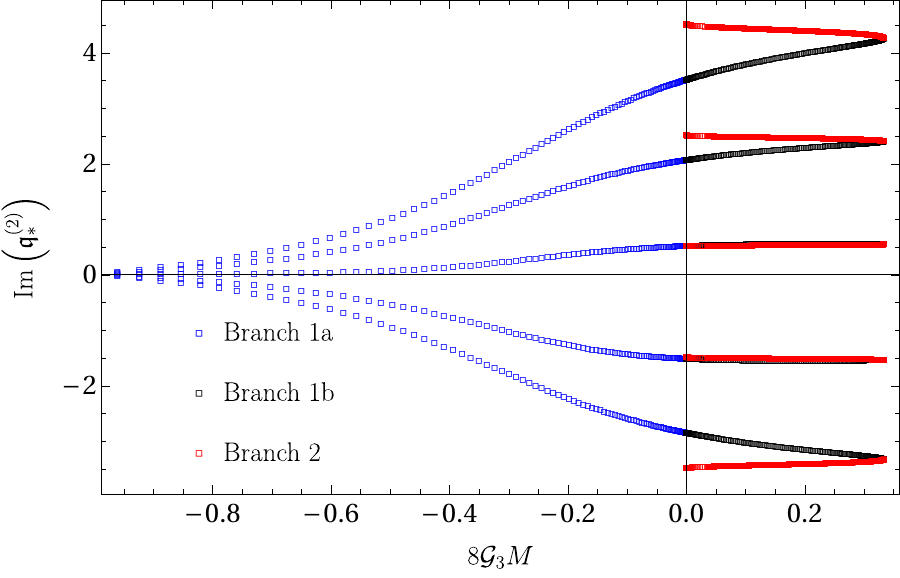}\hfill
      \includegraphics[width=0.49\textwidth]{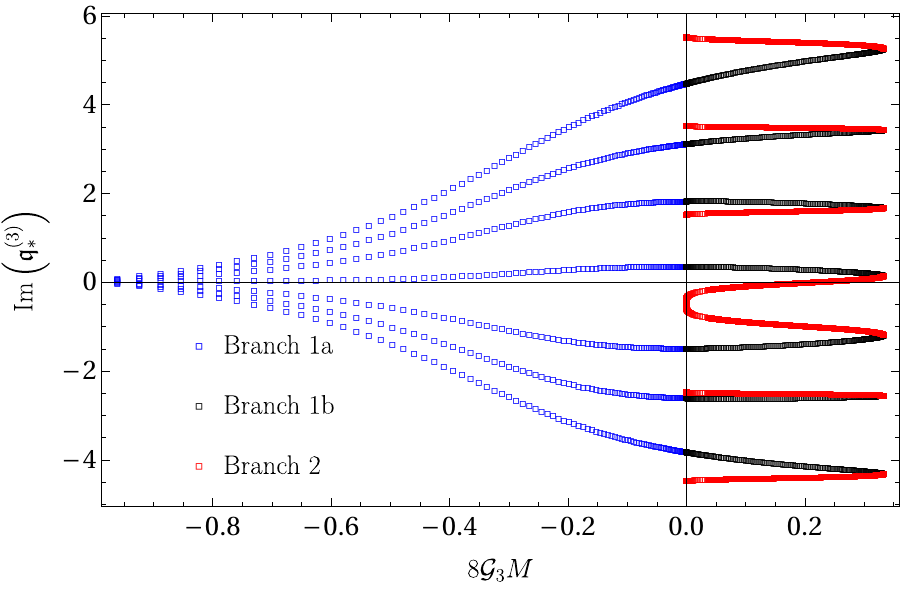}
      \vspace{-4mm}
    \caption{Lowest pole-skipping momentum of Green's functions of single trace fermion operators dual to massive scalar fields in qBTZ geometry. We take $m=\sqrt{1.25}$ and $\ell=1/3$, hence the strength of the quantum backreaction $\nu=1/3$. Each image displays the pole-skipping momentum associated to different Matsubara frequencies for $n=0,1,2,3$, with the top displaying $n=0,1$ from left to right, and the bottom displaying $n=2,3$ from left to right.
    \label{fig:quBTZ_Lowest_pole_skipping_fermion}}
\end{figure}

For spinors, the procedure to obtain the location of pole-skipping points roughly follows the analysis of the scalar field. And like the scalar field the higher curvature corrections of the bulk theory do not leave an imprint on the structure of the fermionic Matsubara frequencies, just as they had not for the scalar Matsubara frequencies.
However, as in the scalar case, the momentum will have corrections. The analytic form of the momentum is not very enlightening. Instead, we simply display the result for the pole-skipping momentum of order $n$ as a function of $8\mathcal{G}_3M$ in figure~\ref{fig:quBTZ_Lowest_pole_skipping_fermion}.
Here we plot the pole-skipping momentum of the lowest four Matsubara frequencies. Just as in the scalar case, we see that in the qCone mass range the number of pole-skipping momentum is the same as what would be expected from a BTZ black hole. i.e. there are $2n+1$ pole-skipping momentum for a given Matsubara frequency $\w_n$. And, that for masses in the qBTZ range the number of pole-skipping momentum doubles to $2n+2$ for a given mass. However unlike the scalar case, we do not see the interchanging of the branches for the first 3 pole-skipping frequencies, i.e. the maximal value, the largest magnitude, of each pole-skipping momentum is obtained in branch 2. However, this begins to occur for the fourth Matsubara frequencies $n=3$, where $\q^{(3)}_j$ for $j=5,6,7,8$ switch their ordering (the magnitude of $\q^{(3)}_j$ is largest for branch 1b) while the remaining pole-skipping momentum have their largest magnitudes in branch 2.

\begin{figure}[t!]
    \centering
    \includegraphics[width=0.85\textwidth,trim={0 1mm 0 0},clip]{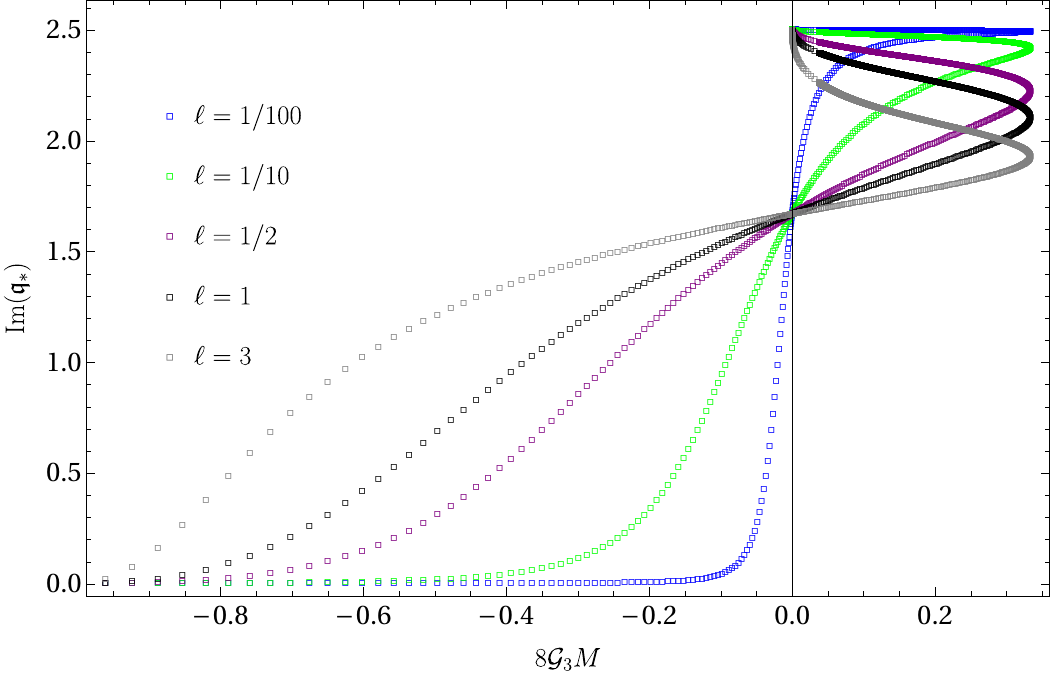}
    \vspace{-3mm}
    \caption{Lowest pole-skipping momentum of Green's functions of single trace fermion operators dual to massive spinor fields in qBTZ geometry. Here we have taken $m=\sqrt{1.25}$, fixing $\ell_3=3$ allowing the strength of the quantum backreaction $\nu=\ell/\ell_3$ to vary.
    \label{fig:quBTZ_nudep_fermion}}
\end{figure}
Finally, before closing this section, we also look briefly at what happens when we vary the quantum backreaction. In figure~\ref{fig:quBTZ_nudep_fermion} we plot the lowest pole-skipping point for fixed operator dimension as a function of the quantum backreaction. Just as in the scalar case, one can see that as we decrease the quantum backreaction from the largest value displayed as gray dots to the smallest value, displayed as blue dots, the curves in the qCone region begin to deform to a constant line given by $\q_*=0$ before sharply growing near the quantum vacuum geometry towards the values they would take in the uncorrected BTZ geometry. In the qBTZ mass range, one can see that the multivalued branch begins to deform, with value of the pole-skipping momentum of branch 1b and branch 2 approaching one another as they both move upward to meet at the pole-skipping momentum of the uncorrected BTZ geometry. This is the same type of behavior as displayed in figure~\ref{fig:qubtz_pole_skipping_ratio} and figure~\ref{fig:quBTZ_nudep}.

%%%%%%%%%%%%%%%%%%%%%%%%%%%%%%%%%%%%%%%%%%%%%%%%%%%%%%%%%%%%%%%%%%%%%%%%%%%%%%%%%%%%%%%%%%%%%%%%%%%%%%%%%%
\section{Critical points}\label{sec:critical_points}
In this section, we will concern ourselves with another aspect of finite-temperature retarded two-point functions of operators $\mathcal{O}$ with operator dimension $\Delta$ and spin $s=0,\pm 1/2$ in the $2d$ CFT dual to the qBTZ geometry. We will concern ourselves primarily with the poles of these Green's functions, each of which defines a mode that obeys a dispersion relation of generic form~\cite{Birmingham:2001pj,Son:2002sd}
\begin{equation}\label{eq:general_dispersion}
    \omega(\mathbf{q})=\sum_{j=0}^\infty a_j (\mathbf{q}-\mathbf{q}_c)^{j/\alpha} \, .
\end{equation}
For $\alpha\neq 1$ this dispersion relation displays non-analytic behavior. For the BTZ geometry $\alpha=1$, while here we will find that the quantum corrections lead to $\alpha=2$ and hence the dispersion relations in the qBTZ geometry display non-analytic behavior. In this expression $\mathbf{q}$ is a wave vector and the coefficients of the series $a_j\in\mathbb{C}$ while the exponent $m\in\mathbb{N}$. For the CFT dual to the BTZ geometry these relations for the poles of $\mathcal{O}$ in the complex momentum plane are known analytically
\begin{equation}\label{eq:BTZ_disp}
    \w(\q)=\pm \q -i (2n+\Delta)\, , \quad n\in\mathbb{Z} \, .
\end{equation}
One can notice that here the series truncates at $j=1$. Naturally, this series has an infinite radius of convergence (the dispersion is only ill-defined when $\q\rightarrow \infty$) but in general, this is not so easy to test. The simplest test of convergence of the series requires a study of the asymptotic behavior of the coefficients $a_j$. Each coefficient can in principle be computed, although the expressions to obtain each $a_j$ can quickly become analytically or numerically intractable. The radius of convergence of such an expression is highly interesting, say, in the case of hydrodynamic dispersion relations since the breakdown of such an expression provides the location where the effective description breaks down. A method for computing the location for such momentum at which the dispersion relation of hydrodynamic modes breaks down was introduced in~\cite{Grozdanov:2019kge} where the authors used it to describe the radius of convergence of the linearized hydrodynamic dispersion relations of $\mathcal{N}=4$ SYM. It has subsequently been used to provide bounds on the hydrodynamic description of a range of different theories~\cite{Grozdanov:2019uhi,Abbasi:2020ykq,Jeong:2021zsv,Jansen:2020hfd,Cartwright:2021qpp,Cartwright:2024rus}.

As already discussed in the literature, and will be seen again in what follows, it is easy to see what these points correspond to, a point of multiplicity $r$ represents the collision of $r+1$ modes~\cite{Grozdanov:2019kge,Grozdanov:2019uhi}. These mode collisions typically occur at complex momentum values and these locations in the complex momentum plane $(\w_c,\q_c)$ signal that the dispersion relations of the two colliding modes are transformed into one another via monodromy and are referred to as QNM level crossing. That is, representing the momentum by $\q_c=|q_c|e^{i\varphi}$ for $\q<\q_c$ the modes travel in closed loops. As the magnitude of the momentum is increased at $\q=\q_c$ the modes collide and the trajectory each individual mode follows degenerates. For momentum larger than the critical momentum $\q>\q_c$ the once separate trajectories followed by the modes have become one large orbit that each mode follows on. And as $\varphi$ varies~\footnote{In a rotationally invariant equilibrium state the implicit function $P(\q,\w)$ can only depend on $\q^2$. Hence the dispersions follow as $\w(\q^2)$, parameterizing $\q_c=|q_c|e^{i\varphi}$ for $\varphi\in [0,\pi]$ we can see that $\q^2$ undergoes a full $2\pi$ rotation in the complex momentum plane. In the following section, we will see that, as one might suspect, for fermions we will need to complete a $4\pi$ rotation to return to the original location in the complex momentum plane. } from $0$ to $\pi$ these modes interchange with one another cyclically (see for instance figure~\ref{fig:level_touching_vs_crossing}). While we will keep the discussion general as much as possible, we will for simplicity display numerical data only for vanishing probe mass $m=0$, and hence $\Delta=2$ for the scalar field and $\Delta=1$ for the spinor. Information about the numerical calculation of these modes can be found in appendix~\ref{app:critical}.

%%%%%%%%%%%%%%%%%%%%%%%%%%%%%%%%%%%%%%%%%%%%%%%%%%%%%%%%%%%%%%%%%%%%%%%%%%%%%%%%%%%%%%%%%%%%%%%%%%%%%%%%5
\subsection{Operators with dimension $\Delta$ and spin $s=0$}\label{sec:scalar_critical}
The critical points of the dispersion relations for scalar operators ($\Delta,s=0$) in the field theory dual to the BTZ geometry have been considered in past works~\cite{Grozdanov:2019uhi,Abbasi:2020xli,Cartwright:2024rus}. In particular~\cite{Cartwright:2024rus} demonstrated how to obtain these directly from the spectral curve and further identified them as singular, rather than, critical points. They are given as
\begin{subequations}
\begin{align}\label{eq:BTZ_crit}
    \q_c&=i(m-n) \, ,\\
    \w_c&=-i(m+n+\Delta)\, , \\
    m,n &\in \mathbb{Z} \,\,\,\, \text{with} \,\,\, m\neq n
\end{align}
\end{subequations}
while they are
\begin{align}\label{eq:BTZ_crit_zero_m}
    \q_c&=\pm in \, ,\\
    \w_c&=-i(n+\Delta)\, , \\
    n &\in \mathbb{Z} \,\,\,\,
\end{align}
if $m=0$.
The identification as singular points is important here. This identification implies that although they are locations where the QNMs collide, they are not locations where QNM ``levels'' cross. That is, they are values of the Fourier parameters where the QNM levels interchange, where, say, the lowest QNM becomes the first overtone and the first overtone now takes on the role of the lowest QNM.  Rather than call them level-crossings they are referred to as level-touching.

\begin{figure}[H]
    \begin{center}
\includegraphics[width=0.85\textwidth]{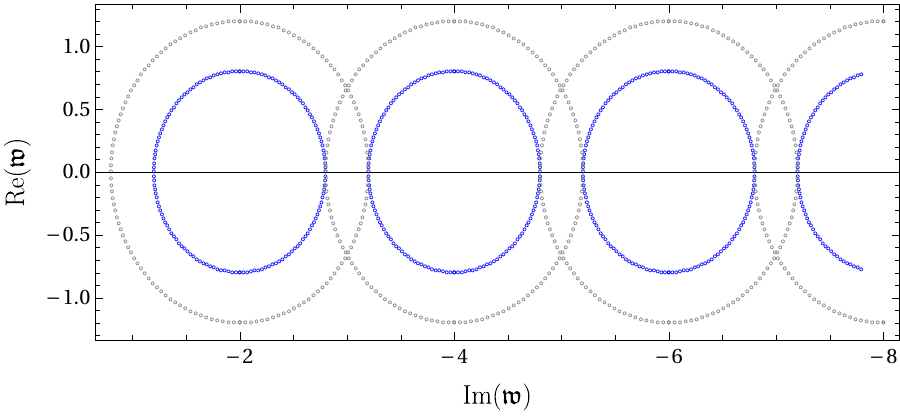} \\
     \includegraphics[width=0.85\textwidth]{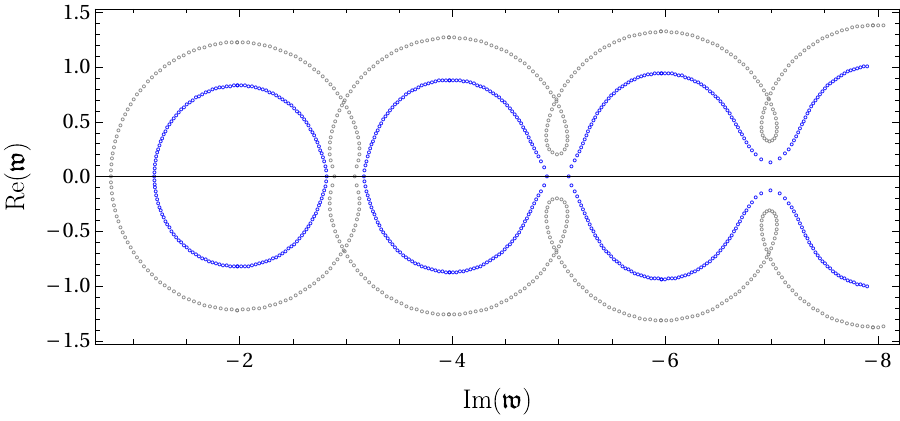}
     \\ \includegraphics[width=0.85\textwidth]{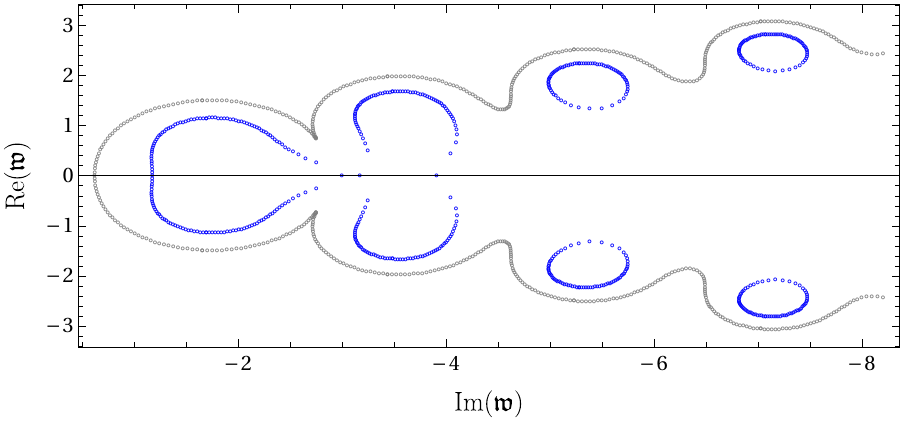}
     \end{center}
    \vspace{-8mm}
    \caption{QNM spectrum of Green's functions of single trace scalar operators dual to massless scalar fields at complex momentum $\q=\q_0 e^{i \varphi}$. As $\varphi$ scans through the range $\varphi\in [0,\pi]$ the curves denote circuits traced out beginning at the QNM located at $\varphi=0$. In all images, the blue curve corresponds to a value of the momentum for which $\q=\q_1<\q_{c}$ while the gray curves denote momentum above the critical momentum $\q_2>\q_c$ and the temperature is held fixed to $4\pi T=1$. \textit{Top:} The modes computed on a BTZ background geometry ($\ell=0$).
    \textit{Middle:} The modes computed on a qBTZ background geometry with $\ell=1/100$ for which the strength of the quantum backreaction.  \textit{Bottom:} The modes computed on a qBTZ background geometry with $\ell=1/10$.
    \label{fig:level_touching_vs_crossing}}
\end{figure}

In terms of the monodromy, one finds that QNM frequencies of the BTZ black hole trace out circles in the complex frequency plane as one traces out circles in the complex momentum plane. From the dispersion relation given in eq.\ (\ref{eq:BTZ_disp}) it is clear that the radius of the circle increases as one increase the magnitude of the momentum. An image of this occurring for the BTZ black hole is displayed in the top image of figure~\ref{fig:level_touching_vs_crossing}. The image displays the behavior of the frequency as we complete a circuit in the phase of the complex momentum $\q$. The blue displays a momentum $\q<\q_c$, while the gray displays a momentum $\q>\q_c$. One can see that for momentum less than the critical value where the ``levels-touch'' the blue curves are separate circles. Likewise, for the BTZ black hole, with momentum larger than the critical momentum the curves are again circles of larger radii. Notice that although the circles intersect, the point of intersection is not at the same value of the phase angle, hence the frequency traces out their own circles. The image displays the lowest value of the critical momentum, hence $n=1$, i.e. $\q_c=\pm 1$ and $\w_c=-3i$ for $m=0$ and standard quantization.
In addition, we will continue to select values of $\ell$ and $x_1$ displayed in table~\ref{tab:parameter_values} which remain on the line of fixed temperature displayed in figure~\ref{fig:temperature}.
In the images, the lower momentum was chosen to be $|\q_1|=1/5$, and the larger momentum was chosen to be $|\q_2|=3/10$.

With some understanding of the mode collisions in the uncorrected BTZ geometry, we now turn to the quantum corrected geometry. We will begin with $\kappa=-1$ so we can smoothly connect to the uncorrected BTZ geometry. We expect that there will be changes to the mode collision locations when we include quantum correction.
Indeed this is the case and furthermore, the location of the mode collision is not the only change. The lower two images of figure~\ref{fig:level_touching_vs_crossing} display the situation in qBTZ black hole.
The middle image shows what happens when we now take $\ell=1/100$ while the bottom image displays the spectrum for $\ell=1/10$. One can see that already at $\ell=1/100$ there has been a dramatic change to the spectrum. What had been disconnected circles traced out by the frequency characterized by level-touching at the mode collision locations has been transformed into a complicated intertwining of the tower of higher QNM overtones. This is made even more apparent in the bottom image at $\ell=1/10$ where the mode spectrum more closely resembles what is seen in higher dimensional cases. The paths traced out by the frequencies under a phase rotation of the momentum is a reflection of the analytic structure of the poles of the retarded Green's function, and this image makes clear that this structure has been significantly more complex by the inclusion of the quantum backreaction in the bulk, or from the CFT perspective, the coupling of the $1+1$ CFT as a defect between two $2+1$ CFTs.
\begin{figure}[t!]
    \centering
    \includegraphics[width=0.85\textwidth,trim={0 2mm 0 0},clip]{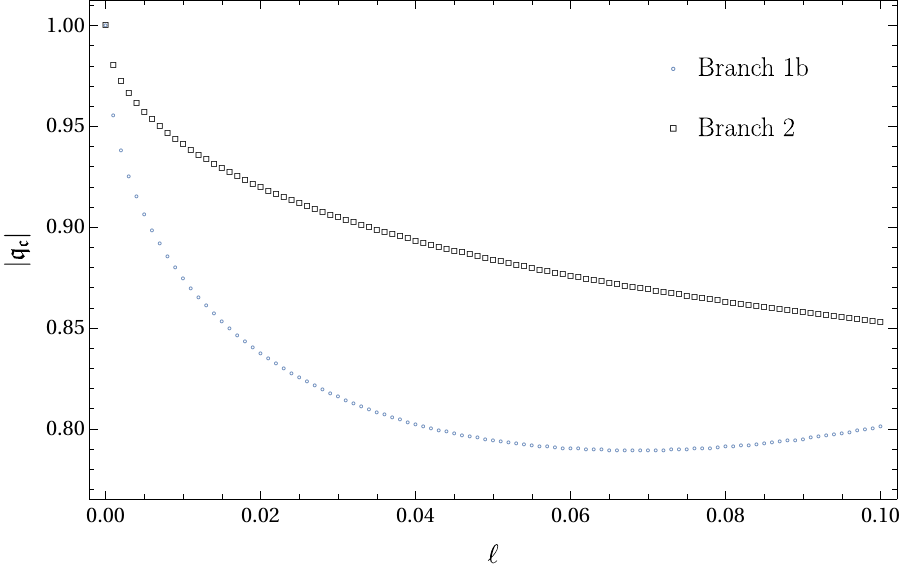}
    \vspace{-3mm}
    \caption{The lowest critical point in the dispersion relation of Green's functions of single trace scalar operators dual to massless scalar fields is tracked as the quantum backreaction $\ell$ is turned on. Here $\kappa=-1$ ensures a smooth connection to the BTZ geometry.
    \label{fig:critical_momentum_quantum_backreaction}}
\end{figure}

Looking carefully at figure~\ref{fig:level_touching_vs_crossing} we can also notice that the location of the critical momentum has shifted, and hence so has the location of the critical frequency. One tracks this directly by making use of the numerical technique described at the beginning of this section. The results of tracking the lowest momentum for mode collisions is displayed in figure~\ref{fig:critical_momentum_quantum_backreaction}. We can see that at zero quantum backreaction, the mode begins as expected at $|\q_c|=1$, however as we increase the quantum backreaction the mode sharply decreases in magnitude before turning to a slow increase as a function of $\ell$. Although not displayed, while the critical momentum for the mode collision closest to the origin in the complex plane is purely imaginary for the uncorrected geometry, with quantum correction the mode collision closest to the origin picks up a non-zero real part. Interestingly, the origin of the quantum correction also has an effect on the mode collisions. Also shown in figure~\ref{fig:critical_momentum_quantum_backreaction} is the lowest critical momentum as computed in Branch 2 of the qBTZ solution. One can see that the decrease in the magnitude of the mode collision is slower, indicating that the mode collision closest to the origin in Branch 2 of the solution occurs at a larger momentum, further from the origin, than in Branch 1b.

\begin{figure}[H]
    \begin{center}
\includegraphics[width=0.85\textwidth]{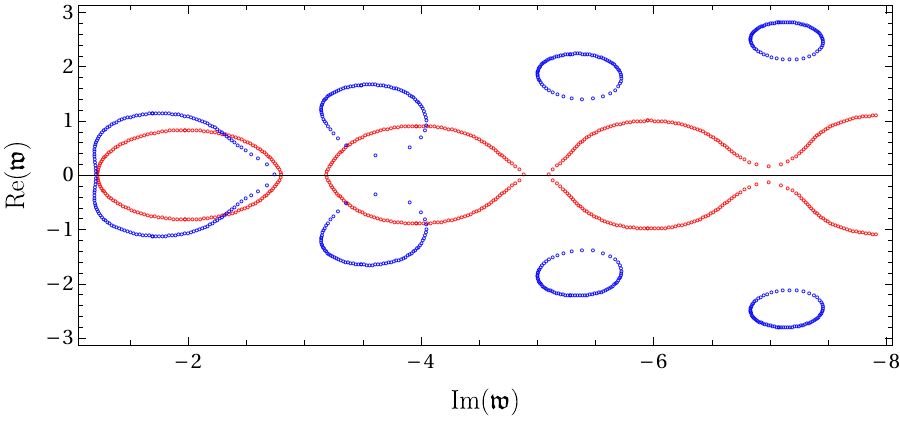} \\
\includegraphics[width=0.85\textwidth]{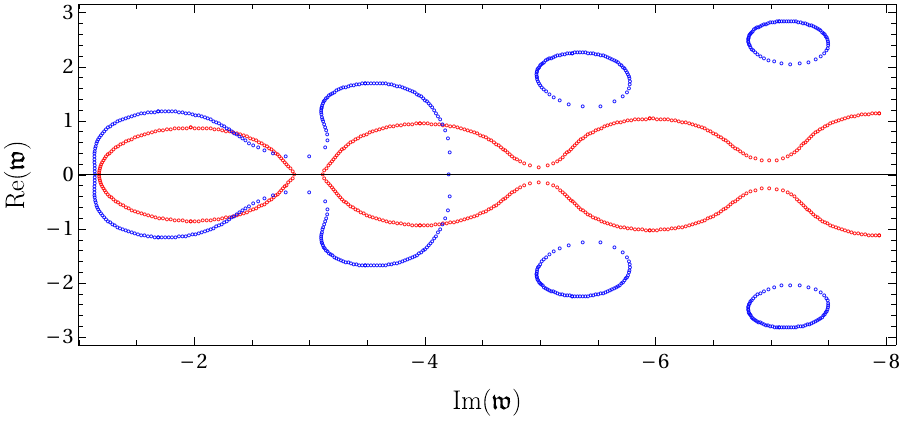} \\
\includegraphics[width=0.85\textwidth]{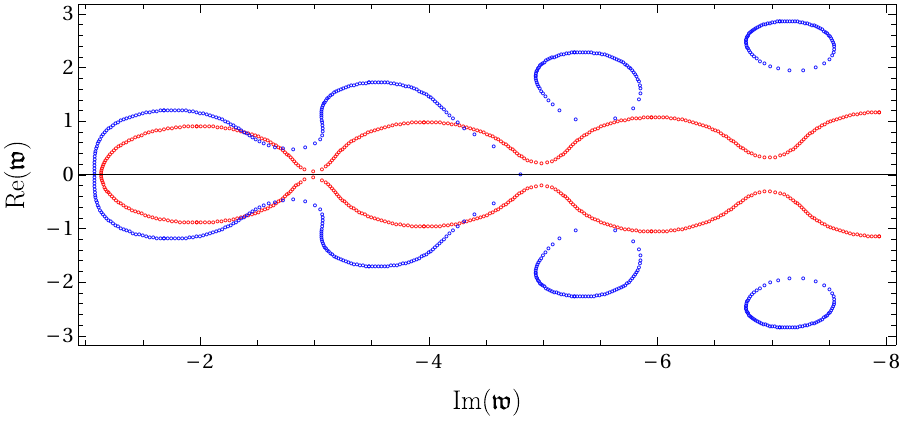}
     \end{center}
     \vspace{-8mm}
    \caption{QNM spectrum of Green's functions of single trace scalar operators dual to massless scalar fields at complex momentum $\q=\q_0 e^{i \varphi}$. As $\varphi$ scans through the range $\varphi\in [0,\pi]$ the curves denote circuits traced out beginning at the QNM located at $\varphi=0$. In all three images the blue curve corresponds Branch 1b while the red curve corresponds to Branch 2. \textit{Top:} $|\q|=|\q_1|<|\q^{(1b)}_{c}|$.
    \textit{Middle:} $|\q^{(1b)}_{c}|<|\q|=|\q_2|<\q^{(2)}_{c}|$. \textit{Bottom:} $|\q^{(2)}_{c}|<|\q|=|\q_3|$. In all three images $\ell=1/10$ and $x_1$ is chosen as in table~\ref{tab:parameter_values} to fix  $4\pi T=1$.\label{fig:level_touching_vs_crossing_compare_qbtz_1b_2}}
\end{figure}

To illustrate this further, we have displayed the monodromy of the QNMs for three different momenta, in Branch 1b (blue) and Branch 2 (red), in figure~\ref{fig:level_touching_vs_crossing_compare_qbtz_1b_2}. From figure~\ref{fig:critical_momentum_quantum_backreaction} we can see that the critical momentum at $\ell=0.1$ for Branch 1b is $|\q^{(1b)}_c|=0.400471$ and for Branch 2 is $|\q^{(2)}_c|=0.426341$. We therefore choose the three momentum $\q_i$ to be such that $|\q_1|<|\q^{(1b)}_c|<|\q_2|<|\q^{(2)}_c|<|\q_3|$. The images are ordered with the top image corresponding to $\q_1$, the middle to $\q_2$ and the bottom to $\q_3$. Besides the obvious visual differences in the spectrum of the qBTZ black hole in these two different branches, we can see that in the top image the mode closest to the origin traces out a closed curve as we rotated $\q^2$ through a $2\pi$ phase angle for both branches. However, in the middle image, with $|\q^{(1b)}|<|\q_2|$, we can now see that the leading mode of branch 1b, depicted in blue, has merged with the first overtone, with each of the modes exchanging locations under monodromy. The leading mode in Branch 2 however, remains closed in its own curve. Increasing the momentum further such that $|\q^{(2)}|<|\q_3|$, in the bottom image we now see that the leading mode computed in Branch 2 has now merged with the first overtone. And in fact has connected itself with the rest of the spectrum which has already merged under the monodromy. This is noticeably distinct from the behavior of the modes in Branch 1b, where the leading mode first joins the first overtone, then the leading mode and the first overtone join the second overtone etc. For branch 2 the, the modes join in the opposite ordering, from deep in the complex plane at infinite overtone number, and progressively linkup until finally connecting with the leading mode.

Following what we did in the previous sections, we can also compare the mode collisions of the scalar QNM in the quantum-dressed conical singularity and the quantum-corrected BTZ geometry. Like before, we will select the value of the parameter $x_1$ such that the temperature remains the same while switching $\kappa=-1$ to $\kappa=+1$.

\begin{figure}[t!]
    \begin{center}
\includegraphics[width=0.85\textwidth]{images/qbtzmom5l110_FT.pdf} \\
\includegraphics[width=0.85\textwidth]{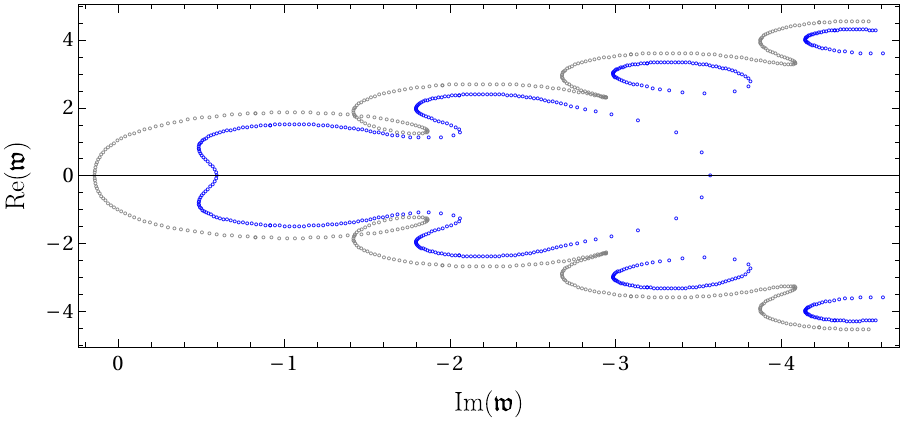}
     \end{center}
\vspace{-8mm}
    \caption{QNM spectrum of Green's functions of single trace scalar operators dual to massless scalar fields at complex momentum $\q=\q_0 e^{i \varphi}$. As $\varphi$ scans through the range $\varphi\in [0,\pi]$ the curves denote circuits traced out beginning at the QNM located at $\varphi=0$. In the image, the blue curves corresponds to a value of the momentum for which $\q=\q_1<\q_{c}$ while the gray curves denote momentum above the critical momentum $\q_2>\q_c$.
    \textit{Top:} The modes computed on a qBTZ background geometry.
    \textit{Bottom:} The modes computed on a qCone background geometry.
  In both images, $\ell=1/10$ and the value of $x_1$ is chosen as in table~\ref{tab:parameter_values} to fix  $4\pi T=1$. \label{fig:level_touching_vs_crossing_compare_qcone_qbtz}}
\end{figure}

The result of doing so is displayed in figure~\ref{fig:level_touching_vs_crossing_compare_qcone_qbtz} with the top image displaying an image of the qBTZ geometry ($\kappa=-1$) and the bottom image displaying qCone geometry ($\kappa=+1$). As before, the blue depicts a momentum with magnitude below the magnitude of a mode collision while the gray depicts a momentum with magnitude above the magnitude of the momentum required for a pole collision. To keep the comparison as close as possible we use the same lower and upper values of the momentum relative to the temperature. One can notice, visually, differences of the monodromy of the curves between the qBTZ and qCone geometry. Between the two, in both cases at the lower momentum we have selected, one mode collision has already occurred for the QNMs, where those modes closest to the origin have already coalesced into a single trajectory under phase rotation, with the higher overtones remaining closed amongst themselves. However, it is clear that the path traced out by the modes in the complex frequency space is clearly different. The curve depicted in figure~\ref{fig:critical_momentum_quantum_backreaction} is clearly not the same between branch 1a and 1b.

\begin{mdframed}
    The location of mode collisions of the poles of the retarded Green's function of single trace scalar operators as a function of the quantum backreaction is not the same in the state of the CFT dual to the quantum dressed conical singularity and the quantum corrected BTZ black hole. One can distinguish between them by tracking modes collision locations, hence by understanding the analytic structure of the poles of the retarded Green's functions in the CFT dual.
\end{mdframed}
Coming back to the image, as we move to the larger of the momentum, we can see in the upper image of figure~\ref{fig:level_touching_vs_crossing_compare_qcone_qbtz} we can see that the trajectory of the lowest QNM has joined with the trajectory of the first overtone, while the higher overtones remain distinct closed loops. However, for the quantum dressed conical singularity the same jump in the momentum has caused both the leading QNM and the first overtone to join, indicating the spacing between the magnitude of pole collision momentum has decreased relative to that in the qBTZ geometry.

%%%%%%%%%%%%%%%%%%%%%%%%%%%%%%%%%%%%%%%%%%%%%%%%%%%%%%%%%%%%%%%%%%%%%%%%%%%%%%%%%%%%%%%%%%%%%%%%%%%%%55
\subsection{Operators with dimension $\Delta$ and spin $s=\pm 1/2$}\label{sec:fermion_critical}
The critical points of the dispersion relations for operators ($\Delta,s=1/2$) in the field theory dual to the BTZ geometry have not been considered in past works. However, following~\cite{Cartwright:2024rus} we can obtain these directly from the spectral curve. The relevant correlation function in the BTZ background
\begin{equation}
   \exd s^2 =- \frac{(r^2-r_+^2)(r^2-r_-^2)}{r^2}\exd t^2 + \frac{r^2 \exd r^2}{(r^2-r_+^2)(r^2-r_-^2)} +r^2 \left(\exd \phi -\frac{r_+r_-}{r^2} \exd t\right)^2
\end{equation}
where the mass, angular momentum and left and right moving temperature are given by
\begin{equation}
    M=\frac{r_+^2+r_-^2}{8G_3}\, , \hspace{.2cm} J=\frac{r_+r_-}{4G_3}\, , \hspace{0.2cm} T_L=\frac{r_+-r_-}{2\pi}\, , \hspace{0.2cm} T_R=\frac{r_++r_-}{2\pi}
\end{equation}
can be obtained in the following closed form~\cite{Birmingham:2001pj,Iqbal:2009fd}
\begin{equation}
    \tilde{G}_R=-i \frac{\Gamma(\frac{1}{2}-m)\Gamma(h_L-i\frac{\omega -q}{4\pi T_L})\Gamma(h_R-i\frac{\omega +q}{4\pi T_R})}{\Gamma(\frac{1}{2}+m)\Gamma(\tilde{h}_L-i\frac{\omega -q}{4\pi T_L})\Gamma(\tilde{h}_R-i\frac{\omega +q}{4\pi T_R})}
\end{equation}
where the conformal weights and associated weights are given by
\begin{equation}
    (h_L,h_R)=\left(\frac{m}{2}+\frac{1}{4},\frac{m}{2}+\frac{3}{4}\right)\, ,\quad (\tilde{h}_L,\tilde{h}_R)=\left(-\frac{m}{2}+\frac{3}{4},-\frac{m}{2}+\frac{1}{4}\right)\, .
\end{equation}
The correlator has a sequence of poles at the locations
\begin{equation}\label{eq:btz_fermion_disp}
    \omega= -q -4\pi T_R(n+h_R)\, , \quad \omega= q -4\pi T_L(n+h_L)\,, n\in\mathbb{Z}_+\, .
\end{equation}
and these poles match CFT expectations~\cite{Gubser:1997cm,Birmingham:2001pj}. Furthermore it should be noted, as discussed in~\cite{Iqbal:2009fd}, that since $h_L-h_R=-1/2$ this implies that $\psi$ corresponds to $\mathcal{O}_{-}$, while if we take $m<0$, then $\psi_-$ is the source and the corresponding operator is $\mathcal{O}_+$.

\begin{figure}[H]
    \begin{center}
    \includegraphics[width=0.85\textwidth]{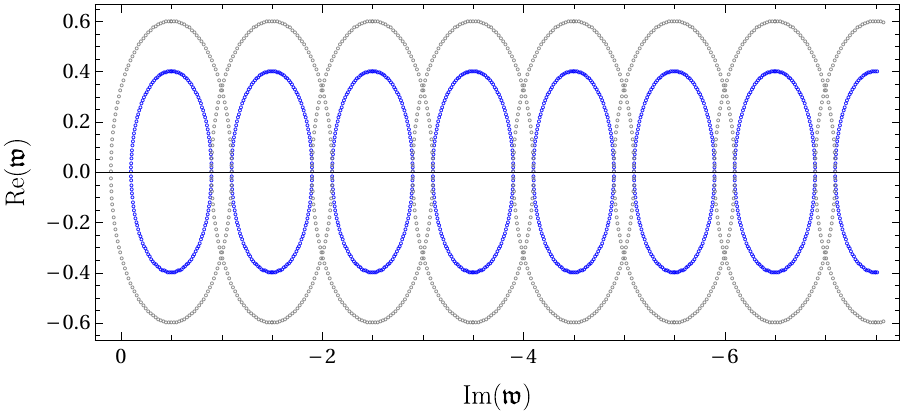} \\
     \includegraphics[width=0.85\textwidth]{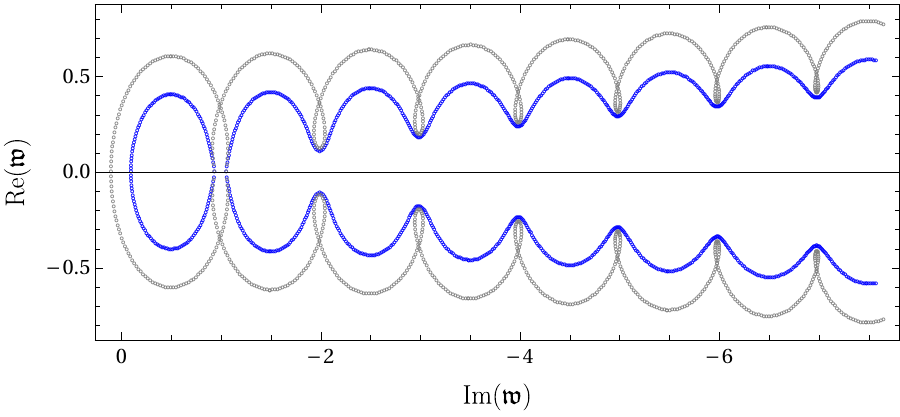}
     \\ \includegraphics[width=0.85\textwidth]{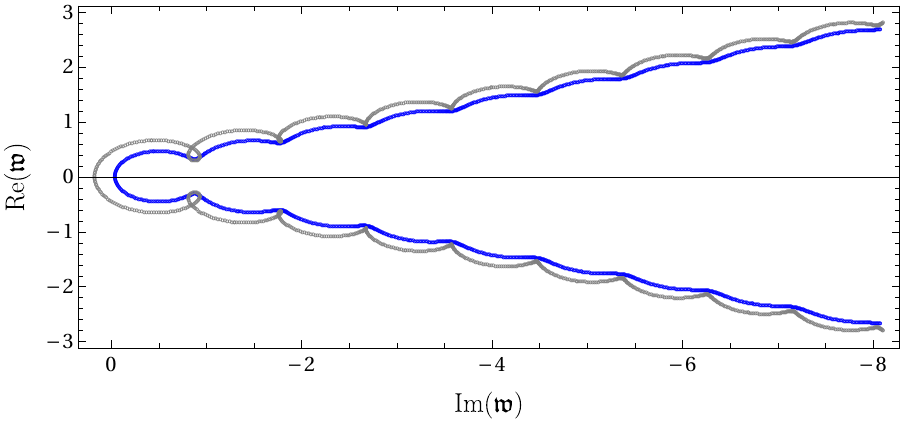}
     \end{center}
     \vspace{-8mm}
    \caption{QNM spectrum of Green's functions of single trace scalar operators dual to massless spinor field at complex momentum $\q=\q_0 e^{i \varphi}$. As $\varphi$ scans through the range $\varphi\in [0,2\pi]$, so that $\q^2$ goes through $4\pi$, the curves display circuits traced out beginning at the QNM located at $\varphi=0$. In all images, the blue curve corresponds to a value of the momentum for which $\q=\q_1<n_{c}$ while the gray curves denote momentum above the critical momentum $\q_2>\q_c$ and the temperature is held fixed to  $4\pi T=1$. \textit{Top:} The modes computed on a BTZ background geometry.
    \textit{Middle:} The modes computed on a qBTZ background geometry with $\ell=1/100$.  \textit{Bottom:} The modes computed on a qBTZ background geometry with $\ell=1/10$.
\label{fig:level_touching_vs_crossing_fermion}}
\end{figure}

We can construct a spectral curve for the interaction of two modes as
\begin{subequations}
\begin{align}
    P_{n,m}(\omega,q)&=P_n(\omega,q)P_m(\omega,q)\, , \\
    P_n(\omega,q)&=(4 i \pi  T_L (h_L+n)-q+\omega ) (4 i \pi  T_R (h_R+n)+q+\omega )\, .
\end{align}
\end{subequations}
Imposing the conditions $P=\partial_\omega P=0$ we obtain the list of singular points contained in table~\ref{tab:fermionic_singular_points}.
\begin{table}[h]
    \centering
    \begin{tabular}{c|c}
    $\omega_c$ & $q_c$ \\ \hline
$ -i \pi \left( T_L (\Delta +2 n+s)+  T_R (\Delta +2 n-s)\right)$&$ i \pi  \left( T_L (\Delta +2 n+s)+  T_R (-\Delta -2 n+s) \right)$ \\
$ -i \pi \left( T_L (\Delta +2 m+s)+  T_R (\Delta +2 n-s)\right) $&$ i \pi  \left( T_L (\Delta +2 m+s)+ T_R (-\Delta -2 n+s) \right)$ \\
$ -i \pi \left( T_L (\Delta +2 n+s)+  T_R (\Delta +2 m-s)\right)$&$ i \pi \left( T_L (\Delta +2 n+s)+   T_R (-\Delta -2 m+s) \right)$ \\
    \end{tabular}
    \caption{Fermionic Singular points, where $s=\pm 1/2$. The first row holds for any $m$.  \label{tab:fermionic_singular_points}}
\end{table}
Imposing that the left and right moving temperatures are the same, as in our system, leads to the results of table~\ref{tab:fermionic_singular_points_left_right_equal}.
One can check directly that
\begin{equation}
\partial_qP(\omega,q)|_{q=q_c,\omega=\omega_c}=0
\end{equation}
and hence these modes should be referred to as singular points~\cite{Cartwright:2024rus}. That is, they correspond to level-touching, not level-crossing.
\begin{table}[h]
    \centering
    \begin{tabular}{c|c}
    $\omega_c$ & $q_c$ \\ \hline
      $ -2 i \pi  T (\Delta +2 n) $& $ 2 i \pi  s T$ \\
$ -2 i \pi  T (\Delta +m+n)$ & $-2 i \pi  T (\mp m\pm n+s)$ \\
    \end{tabular}
    \caption{Fermionic Singular points, where $s=\pm 1/2$ and $T_L=T_R=T$. The first row is true for any $m$. \label{tab:fermionic_singular_points_left_right_equal}}
\end{table}
In terms of the monodromy, just like the scalar QNM, one finds that QNM frequencies of fermions in the BTZ black hole background trace out circles in the complex frequency plane as one traces out circles in the complex momentum plane. From the dispersion relation given in eq.\ (\ref{eq:btz_fermion_disp}) we again see that the radius of the circle increases as one increases the magnitude of the momentum. The top image of figure~\ref{fig:level_touching_vs_crossing_fermion} displays this behavior of the frequency as we complete a circuit in the phase of the complex momentum $\q$. Like the scalar sector, the blue displays a momentum $\q<\q_c$, while the gray displays a momentum $\q>\q_c$. One can see that for momentum less than the critical value where the levels-touch the blue curves are separate circles. Likewise with momentum larger than the critical momentum the curves are again circles of larger radii. The image displays the lowest value of the critical momentum, hence i.e. $\q_c= -i/2$ and $\w_c=-i2(n+1)$ for $m=0$ and standard quantization. In addition, we have also selected $x_1=1$ which implies a horizon radius of $r_h=\ell_3/2$ for $\ell=0$. In the images, the lower momentum was chosen to be $|\q_1|=2/5$ and the larger momentum was chosen to be $|\q_2|=3/5$ and we made the choice of $\ell_3=1$.

Turning now to the quantum corrected geometry we begin again with $\kappa=-1$ so we can smoothly connect to the uncorrected BTZ geometry. We expect that there will be changes to the mode collision locations when we include quantum correction. The lower two images of figure~\ref{fig:level_touching_vs_crossing_fermion} display the QNM spectrum at complex momentum for $\ell\neq 0$, with the middle image displaying $\ell=1/100$ and the lower image displaying $\ell=1/10$. As in the previous sections, in the images, we held the temperature fixed using the parameters displayed in table~\ref{tab:parameter_values}. We can see there is a dramatic change to the spectrum even at $\ell=1/100$ where the would-be level-touching events of the QNMs, both the leading and the overtones, have been intertwined, under a $4\pi$ phase rotation of $\q^2$ all higher overtones are mapped into one another. This is even more pronounced in the bottom image of figure~\ref{fig:level_touching_vs_crossing_fermion} where the mapping and exchange of the leading mode to the higher overtones is clearly displayed.

\begin{figure}[t!]
    \centering
    \includegraphics[width=0.85\textwidth,trim={0 2mm 0 0},clip]{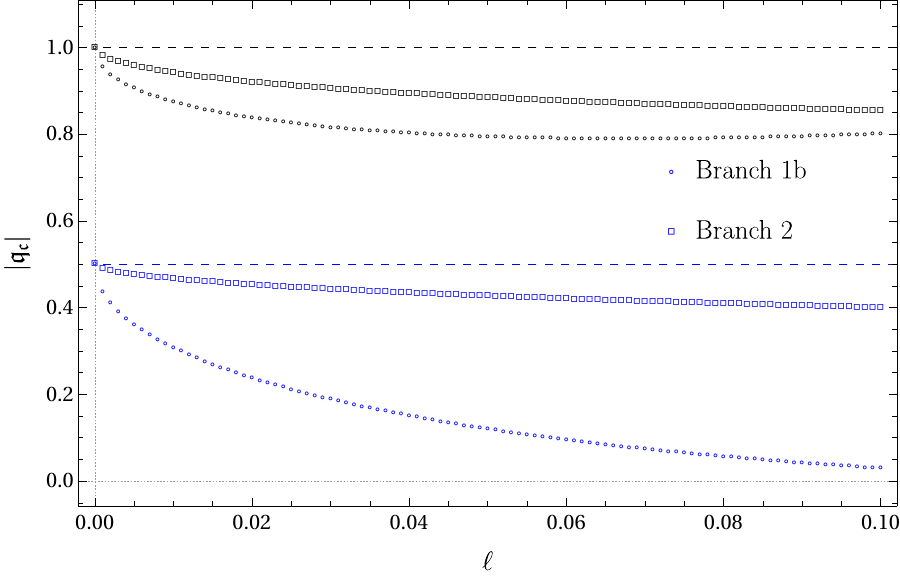}
    \vspace{-2mm}
    \caption{The lowest critical point in the dispersion relation of Green's functions of single trace operators is tracked as the quantum backreaction $\ell$ is turned on. Here $\kappa=-1$ ensures a smooth connection to the BTZ geometry. The black dots correspond to operators with $s=0$ (as displayed in figure~\ref{fig:critical_momentum_quantum_backreaction}) while the blue diamonds correspond to operators with $s=1/2$. The dashed lines show the uncorrected value for $s=0$ (black) and $s=1/2$ (blue) to help guide the eye. The two branches that connect smoothly to the qBTZ solution are displayed as empty circles (branch 1b) and empty squares (branch 2).
\label{fig:critical_momentum_quantum_backreaction_fermion}}
\end{figure}

From the lower two images of figure~\ref{fig:level_touching_vs_crossing_fermion} it is evident that the location of the nearest mode collision to the origin in the complex plane is shifted away from the BTZ value when quantum corrections are taken into account. Using the technique discussed in the introduction to this section we can obtain directly the location of this mode collision in the complex frequency plane. Displayed in figure~\ref{fig:critical_momentum_quantum_backreaction_fermion} is the result of tracking the location of the nearest mode collision to the origin for QNMs of the bulk fermion. In the image we have plotted the magnitude of the collision momentum in blue, and a dashed blue line to display the uncorrected value. There is a clear decrease in the magnitude of the momentum as we increase the quantum correction, hence this mode collision is pushed closer to the origin. Although it is clear from the figure that, as in the scalar case, the mode collision in branch 1b occurs closer to the origin. In the same image, we also display the results for $s=0$ operators for easy comparison. There we see there is actually a minimal value for the magnitude of the momentum of the pole collision (in branch 1b), while for $s=1/2$ operators no minimal value appears.

\begin{figure}[H]
    \begin{center}
\includegraphics[width=0.85\textwidth]{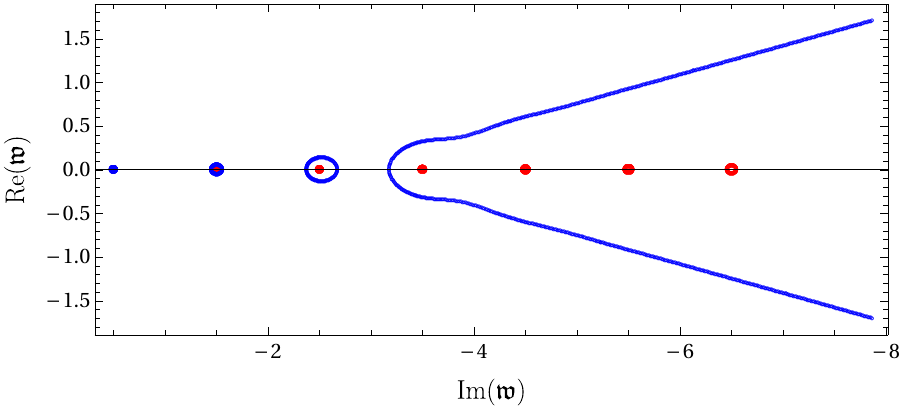} \\
\includegraphics[width=0.85\textwidth]{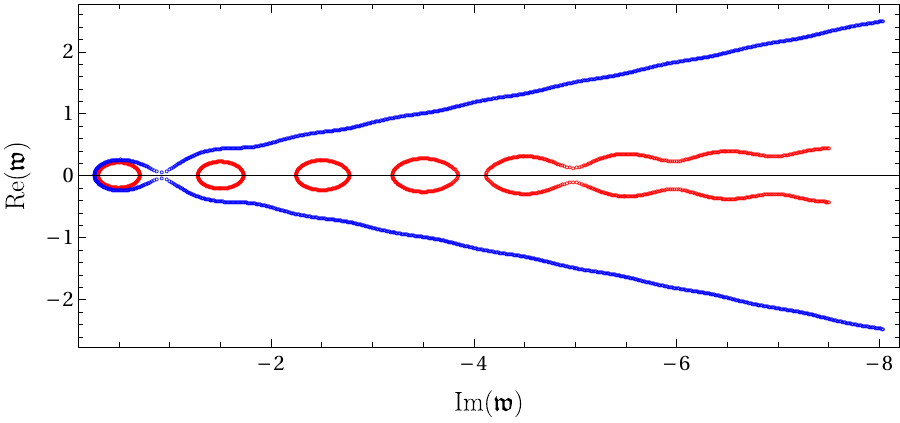} \\
\includegraphics[width=0.85\textwidth]{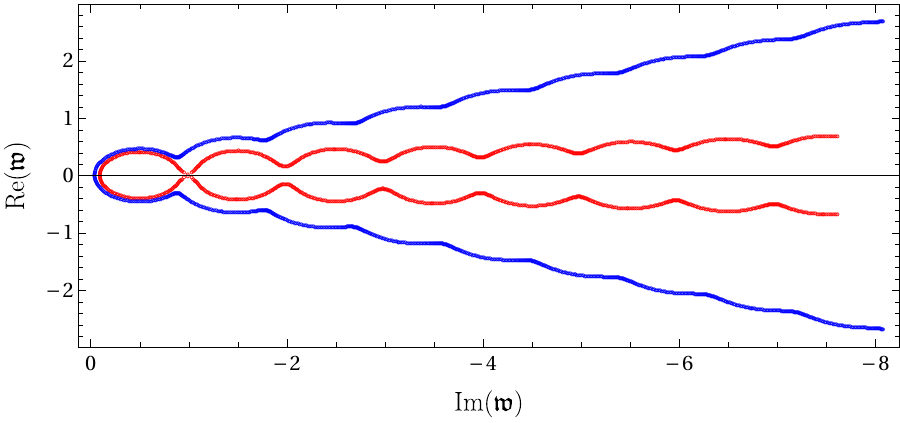}
     \end{center}
     \vspace{-7mm}
    \caption{QNM spectrum of Green's functions of single trace fermion operators dual to massless spinor fields at complex momentum $\q=\q_0 e^{i \varphi}$. As $\varphi$ scans through the range $\varphi\in [0,\pi]$ the curves denote circuits traced out beginning at the QNM located at $\varphi=0$. In all three images the blue curve corresponds Branch 1b while the red curve corresponds to Branch 2. \textit{Top:} $|\q|=|\q_1|<|\q^{(1b)}_{c}|$
    \textit{Middle:} $|\q^{(1b)}_{c}|<|\q|=|\q_2|<\q^{(2)}_{c}|$. \textit{Bottom:} $|\q^{(2)}_{c}|<|\q|=|\q_3|$. In all three images $\ell_3=1$, $\ell=1/10$ and $x_1$ is chosen as in table~\ref{tab:parameter_values} to fix $4\pi T=1$.\label{fig:level_touching_vs_crossing_compare_qbtz_1b_2_fermion}}
\end{figure}

To further illustrate the differences between the two branches of the qBTZ solution, we have displayed the monodromy of the QNMs for three different momenta, in Branch 1b (blue) and Branch 2 (red), in figure~\ref{fig:level_touching_vs_crossing_compare_qbtz_1b_2_fermion}. From figure~\ref{fig:critical_momentum_quantum_backreaction_fermion} we can see that the critical momentum at $\ell=0.1$ for Branch 1b is $|\q^{(1b)}_c|=0.0293734$ and for Branch 2 is $|\q^{(2)}_c|=0.399202$. We therefore choose the three momentum $\q_i$ to be such that $|\q_1|<|\q^{(1b)}_c|<|\q_2|<|\q^{(2)}_c|<|\q_3|$. The images are ordered with the top image corresponding to $\q_1$, the middle to $\q_2$ and the bottom to $\q_3$. Beginning with $|\q_1|=0.0199$, in the top image, we recall that for zero momentum the fermionic QNMs did not change i.e.\ they were identical to the BTZ QNMs and they appeared all at purely imaginary momentum. Since $|\q_1|\ll |\q^{(2)}_c|$, it is not surprising that the leading QNM and all the overtones in branch 2 are confined close to the imaginary axis while the modes in branch 1b show a more complicated set of features. In particular the third and all higher overtones have already coalesced under the monodromy, while the leading QNM and the first two overtones remain closed loops. Increasing the momentum to $|\q_c^{(1b)}|<|\q_2|=0.199$ we can see that the mode collision closest to the origin has occurred for branch 1b, connecting the leading QNM mode to the rest of the infinite tower of overtones. Meanwhile, we can see the modes of branch 2 have begun to merge, with the leading QNM and the first three overtones remaining closed under monodromy, and the rest of the tower of QNMs already exchanging their levels. Finally, increasing the momentum further such that $|\q^{(2)}|<|\q_3|=0.4$, in the bottom image we now see that the leading mode computed in Branch 2 has merged with the rest of the tower of overtones.

\begin{figure}[t!]
    \begin{center}
\includegraphics[width=0.85\textwidth]{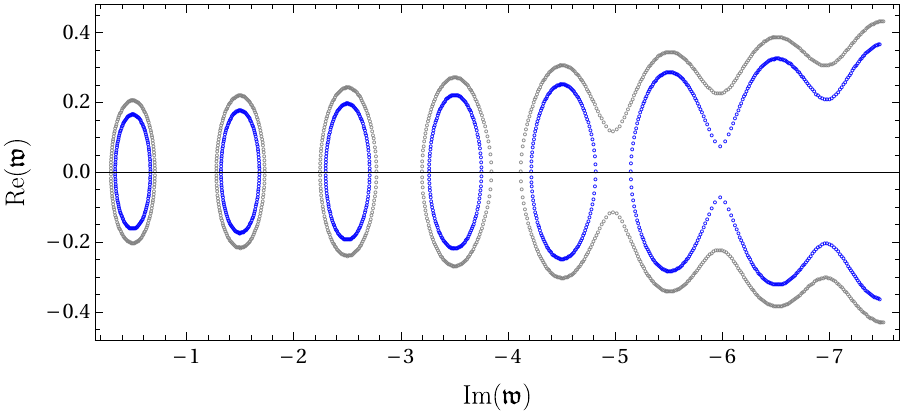} \\
\includegraphics[width=0.85\textwidth]{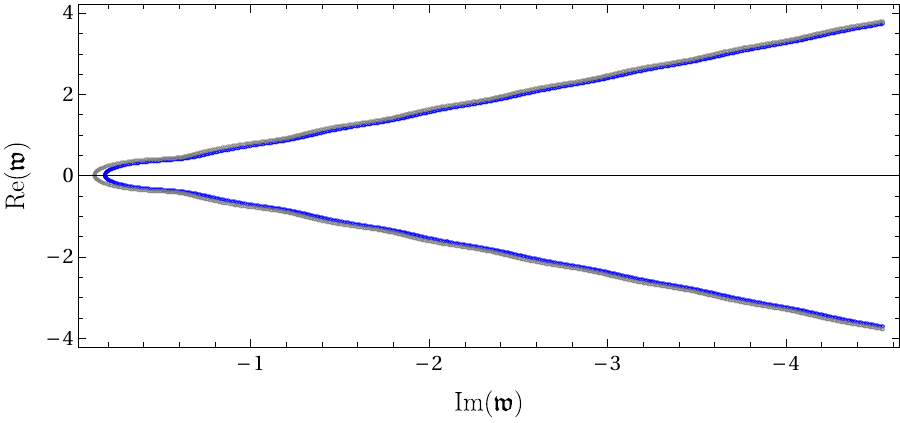}
     \end{center}
    \vspace{-8mm}
    \caption{QNM spectrum of Green's functions of single trace fermion operators dual to massless spinor fields at complex momentum $\q=\q_0 e^{i \varphi}$. As $\varphi$ scans through the range $\varphi\in [0,\pi]$ the curves denote circuits traced out beginning at the QNM located at $\varphi=0$. In the image the blue curves corresponds to a value of the momentum for which $\q=0.16=\q_1<\q^{(1b)}_{c}$ while the gray curves denote momentum above the critical momentum $\q_2=0.199>\q^{(1b)}_{c}$.
    \textit{Top:} The modes computed in a branch 1b qBTZ background geometry.
    \textit{Bottom:} The modes computed in branch 1a, qCone, background geometry.
  In both images $\ell=1/10$ and the value of $x_1$ is chosen as in table~\ref{tab:parameter_values} to fix $4\pi T=1$\label{fig:level_touching_vs_crossing_compare_qcone_qbtz_fermion}}
\end{figure}

Finally, we can also compare the level-crossing that occurs between the qCone and the qBTZ geometry. In figure~\ref{fig:level_touching_vs_crossing_compare_qcone_qbtz_fermion} the top image displays the qBTZ geometry ($\kappa=-1$) and the bottom image displaying qCone geometry ($\kappa=+1$). As before, the blue depicts a momentum with magnitude below the magnitude of a mode collision while the gray depicts a momentum with magnitude above the magnitude of the momentum required for a pole collision. In both cases this refers to a potential pole collision in the branch 1b qBTZ geometry. To keep the comparison as close as possible, the chosen momentum is the same for both geometries, and the temperature of the backgrounds is held fixed, while $\ell=0.1$ and $\ell_3=1$. One can see, even visually, major differences in the behavior of the spectrum. Already at this value of the momentum, all QNM of the qCone have joined into one trajectory under monodromy while the change in the magnitude of the momentum causes the fourth overtone to join with the rest of the tower of overtones when computed in the qBTZ geometry. This dramatic difference further displays the sensitivity of simple field theory observables to differences in the CFT state dual to these two distinct quantum corrected geometries.

%%%%%%%%%%%%%%%%%%%%%%%%%%%%%%%%%%%%%%%%%%%%%%%%%%%%%%%%%%%%%%%%%%%%%%%%%%%%%%%%%%%%%%%%%%%%%%%%%
\section{Discussion}\label{sec:discussion}
In this article, we have initiated an exploration into the out-of-equilibrium properties of quantum field theory duals to semi-classical black hole geometries. Focusing on the qBTZ black hole, an exact solution derived via braneworld holography, we computed the poles of the retarded Green's functions for operators with dimension $\Delta$ and spin $s=0,\pm 1/2$ in the brane's dual field theory, characterizing their approach to thermal equilibrium. These poles correspond to the QNM spectrum of probe scalar fields and spinors. We traced the evolution of this spectrum from the uncorrected BTZ geometry to a solution incorporating exact quantum backreaction and an infinite tower of higher curvature corrections. Additionally, we investigated pole-skipping points, locations where the Einstein equations near the horizon do not yield a unique solution, and analyzed the poles' analytic structure.

A remarkable feature of the qBTZ geometry is its realization of quantum censorship, the idea that naked singularities should not exist and that quantum effects may cloak singularities behind a horizon. While the classical theory predicts conical singularities, separated by a gap from the otherwise continuous spectrum of black holes, the semi-classical theory dresses these singularities, thereby continuously connecting them to the black hole branch of solutions. From the boundary theory perspective, the appearance of these horizons is natural. That is, upon turning on backreaction, the DCFT$_2$ (dual to brane degrees of freedom) couples with the BCFT$_3$ (dual to bulk degrees of freedom), resulting in a state described by density matrix. Thus thermality emerges from entanglement.

The qBTZ metric, therefore, encompasses both quantum-dressed conical singularities and quantum-corrected BTZ geometries, which are smoothly connected within the parameter space. Interestingly, even simple observables, such as two-point correlators of single-trace operators, can distinguish between the two dual CFT states (see figure~\ref{fig:quBTZ_TO_QDConical}). Moreover, quantum backreaction does not merely shift the spectrum and dispersion relations but also directly modifies their analytic structure (see for instance figure~\ref{fig:level_touching_vs_crossing}).

One of the simplest ways to benchmark the quantum corrected geometry is by measuring the thermalization times of the dual field theory in the different states. Figure~\ref{fig:BTZ_TO_QUBTZ_Thermalization_time} displays the thermalization time given by the leading QNM. For branch 1b and 2 of the qBTZ black hole, we find consistency with the standard holographic lore~\cite{Danielsson:1999fa,Giddings:2001ii,Abajo-Arrastia:2010ajo,Balasubramanian:2010ce,Balasubramanian:2011ur,Aparicio:2011zy,Sekino:2008he,Lashkari:2011yi}: field theories with classical gravity duals are `fast' thermalizers, or scramblers of information, and quantum corrections increase the time required for thermalization. However, somewhat unexpectedly, the situation is markedly different for the qCone (branch 1a). In this case, increasing quantum corrections actually reduces the thermalization time of the dual field theory. Currently, we lack a clear intuition for what should be expected in thermal states dual to quantum dressed conical singularities, where thermality emerges purely from entanglement. Understanding this reduction in thermalization time is intriguing, particularly in the context of the double-holographic interpretation of the qBTZ geometry, where a large number of light operators should be coupled to the boundary DCFT$_2$ (upon integrating out the BCFT$_3$). It seems that this increase in light operators provides more channels for small perturbations to dissipate. Interestingly, while it takes longer to excite these channels in the thermal state dual to the quantum-corrected black hole, it appears to take less time in the thermal state dual to the qCone.

A remarkably similar phenomenon of scaling behaviors is observed in the pole-skipping momentum. As illustrated in figure~\ref{fig:scalar_pole_func_l} the momentum at the pole-skipping point exhibits distinct behaviors as a function of $\ell$ (the strength of the backreaction in units of the AdS radius) depending on the black hole branch. Specifically, for $\ell \ll 1$, the momentum scales approximately as $\ell^{1/2}, \ell^2$, and $\ell^1$ for branches 1a, 1b, and 2, respectively (see figure~\ref{fig:scalar_pole_func_l}). These differing dependencies on the quantum backreaction not only provide further insight into the nature of the CFT duals to semi-classical geometries but also shed light on the mechanisms driving the quantum backreaction. Intriguingly, we also observe that at higher Matsubara frequencies, the ordering of the magnitude of pole-skipping momentum changes. While branch 2 consistently exhibits the largest absolute momentum value, the next smallest pole-skipping momentum occurs in branch 1b (see the discussion near figure~\ref{fig:quBTZ_Lowest_pole_skipping}).

In addition to analyzing the Matsubara frequencies and their dependence on the branch of the solution and the strength of quantum backreaction, we also investigated the ratio of the pole-skipping frequency to the momentum. Although we expect energy-energy correlations to be described by a scalar mode, the near-horizon behavior of scalar fields yields pole-skipping points closest to the origin in the negative half of the complex frequency plane. This contrasts with the pole-skipping points for gravitational perturbations, which appear in the upper half of the frequency plane \cite{Natsuume:2019sfp}. Interestingly, or perhaps coincidentally, the absolute value of the frequency-to-momentum ratio is the same in both cases.
For gravitational perturbations, a positive frequency of the mode leads to a waveform characteristic of quantum chaotic behavior and, in holographic theories with a classical dual, this is robustly related to the analysis of graviton scattering in the bulk (dual to OTOCs)~\cite{Schalm:2018lep}. While one might expect that energy-energy fluctuations would exhibit a qualitatively similar behavior form to those of a massless scalar field ---given that the graviton on the brane is massive but behaves approximately as massless\footnote{The massless limit of massive gravity is subtle and requires careful consideration~\cite{vanDam:1970vg,Zakharov:1970cc,Kogan:2000uy,Bergshoeff:2009hq,Karch:2000ct,Emparan:2020znc}.} for small $\ell$--- this must be confirmed to ensure that the ratio accurately reflects the information derived from gravitational fluctuations.
Previous studies have examined both near-horizon perturbations and gravitational scattering to extract butterfly velocities and quantum Lyapunov exponents for higher derivative gravity theories, including the massive gravity theory derived from this braneworld construction at leading order in $\ell$ \cite{Alishahiha:2016cjk, Qaemmaqami:2017jxz, Huang:2018snb, Bergshoeff:2009hq}. However, these studies did not account for quantum backreaction effects. Therefore, a comprehensive analysis of the pole-skipping locations of metric perturbations and a direct calculation of shockwave solutions are still needed to provide a reliable diagnostic of quantum chaos for the field theory dual to the qCone and qBTZ solutions. We expect to report on this matter in the near future~\cite{Cartwright:2025tes}.

One final marker of the imprint of quantum corrections on boundary correlators is found in the analytic structure of the poles in the retarded Green's functions. The poles of the Green's functions in the field theory dual to the uncorrected BTZ solution are analytic functions. However, the quantum corrections lead to new non-analytic behavior of the poles. This is evident from the behavior of QNM collisions in the complex frequency plane, where they shift from closed trajectories (level-touching events) to complex level-crossings due to quantum corrections. These mode collisions offer a distinct signature of the CFT states dual to semi-classical geometries and should have significant implications for the reconstruction program \cite{Abbasi:2020xli, Grozdanov:2023tag, Grozdanov:2022npo}. In particular, the doubly holographic interpretation of these braneworld constructions as a $2+1$ BCFT coupled to a $1+1$ DCFT should allow for a direct CFT analysis of the dispersion relations (see for instance~\cite{Barrat:2022psm,Bianchi:2022ppi}). We intend to investigate these ideas further in future work.

%%%%%%%%%%%%%%%%%%%%%%%%%%%%%%%%%%%%%%%%%%%%%%%%%%%%%%%%%%%%%%%%%%%%%
\acknowledgments
We are very grateful to Rafael Carrasco, Roberto Emparan, Antonia Frassino, Robie Hennigar, Hyun-Sik Jeong, Emanuele Panella, Ayan Patra and Andrew Svesko for useful discussions and correspondence.
CC, UG and GPP are supported by the Netherlands Organisation for Scientific Research (NWO) under the VICI grant VI.C.202.104. JFP is supported by the `Atracci\'on de Talento' program grant 2020-T1/TIC-20495 and by the Spanish Research Agency through the grants CEX2020-001007-S and PID2021-123017NB-I00, funded by MCIN/AEI/10.13039/501100011033 and by ERDF A way of making Europe.

%%%%%%%%%%%%%%%%%%%%%%%%%%%%%%%%%%%%%%%%%%%%%%%%%%%%%%%%%%%%%%%%%%
\appendix

%%%%%%%%%%%%%%%%%%%%%%%%%%%%%%%%%%%%%%%%%%%%%%%%%%%%%%%%%%%%%%%%%%%%%%%%%
\section{Details on the BTZ and qBTZ black holes}\label{app:black_holes_on_brane}
\begin{figure}[t!]
    \centering
\includegraphics[scale=0.8,trim={0 0 0 0},clip]{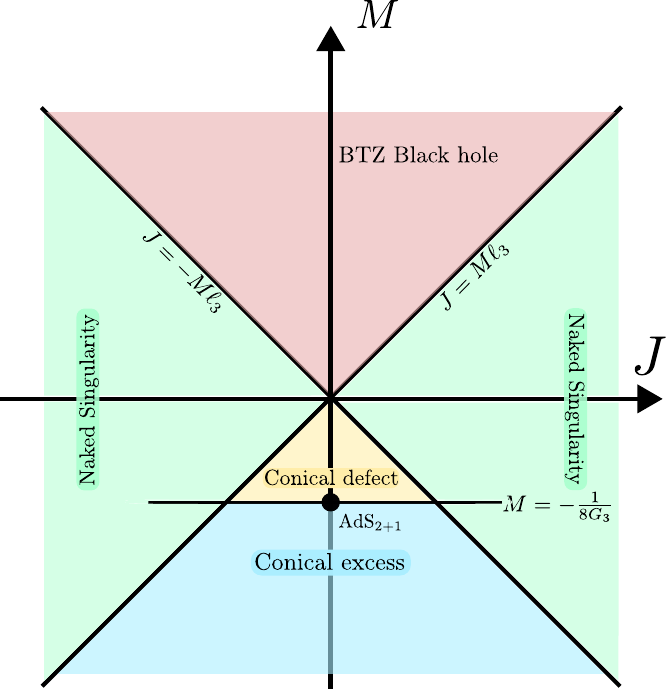}
\vspace{-2mm}
\caption{\textbf{Classical BTZ spectrum:} A depiction of the classical spectrum of 2+1 dimensional solution given in eq.\ (\ref{eq:BTZ})
    \label{fig:btz_spectrum}}
\end{figure}

\textit{A BTZ Primer:} The BTZ black hole~\cite{Banados:1992wn,Banados:1992gq} is an asymptotically AdS$_3$ black hole solution to the equations of motion derived from
\begin{equation}\label{action1}
    S=\frac{1}{16\pi G_3}\int \exd^3 x \sqrt{-g}\left(R-2\Lambda\right)\, ,
\end{equation}
given by the following line element
\begin{subequations}
\begin{align}
    ds^2 &= -N^{\perp}(r)\exd t^2 + N^{\perp}(r)^{-1}\exd r^2 + r^2 \left(\exd\phi+N^\phi\exd t\right)^2 \, \label{eq:BTZ},\\
    N^{\perp} &= -8G_3 M +\frac{r^2}{\ell_3^2}+\frac{(4G_3J)^2}{r^2}\, , \quad N^\phi = -\frac{4G_3J}{r^2} \, ,
\end{align}
\end{subequations}
where $N^{\perp}$ and $N^\phi$ are referred to as the lapse and the shift, and $M$ is the mass and $J$ is the angular momentum. This solution has been described in detail in many locations, so we will not dwell on its many interesting facets.
Of particular importance to us is the spectrum of the classical solution as displayed in figure~\ref{fig:btz_spectrum}.
For $(0<|J|< M \ell_3)$
the solution is a rotating black hole, with inner and outer horizon radii given by
\begin{equation}
  r^2_{\pm}= \frac{\ell_3^2}{2}\left(8G_3 M\pm\sqrt{(8G_3 M)^2-\left(\frac{8G J}{\ell_3}\right)^2}\right)\, ,
\end{equation}
while for $|J|=M\ell_3$ the solution is an extremal black hole with $r_+=r_-$ and vanishing surface gravity. If we go to the vacuum solution defined by vanishing mass and $J=0$ we arrive at
 \begin{equation}
      ds^2_{vac}= -\left(\frac{r}{\ell_3}\right)^2\exd t^2 + \left(\frac{r}{\ell_3}\right)^{-2}\exd r^2 + r^2 \exd\phi^2\, .
 \end{equation}
One can notice that this geometry is regular everywhere within the coordinate range, has no singularities and no horizons. Continuing now at $J=0$ for $8 G_3 M<0$ we have
\begin{equation}
    ds^2 = -\left(\frac{r^2}{\ell_3^2}-8G_3M\right)\exd t^2 + \left(\frac{r^2}{\ell_3^2}-8G_3M\right)^{-1}\exd r^2 + r^2 \exd\phi^2 \, ,
\end{equation}
which under a quick coordinate transformation gives
\begin{equation}\label{eq:conical_defect}
 ds^2 =- \left(\frac{r^2}{\ell_3^2}+1\right)\exd t^2 + \left(\frac{r^2}{\ell_3^2}+1\right)^{-1}\exd r^2 + (1-\alpha)^2r^2 \exd\phi^2 \, ,
\end{equation}
where the angular deficit is given by $\Delta\phi=2\pi \alpha$, with $8 G M=-(1-\alpha)^2$. Hence the geometry has a conical defect, with no horizon, and is therefore a naked singularity. However, when $\alpha=0$ i.e.\ $M=-1/8G_3$, there is no angular deficit, and the geometry is global AdS$_3$. Continuing to small values of the parameter $\alpha<0$ leads to angular excess, or conical excess. Hence we can see that the ground state $M=-1/(8G_3)$ is separated from the vacuum solution by a sequence of naked singularities. Additionally, for negative mass and non-zero angular momentum, or mass not satisfying the relation $(0<|J|\leq M \ell_3)$ one also finds naked singularities. However, in this work we will be primarily concerned with the line $J=0$, with $-1/8G_3<M$, as this region, up to some small caveats, will also be included in the quantum black hole solution. We refer the interested reader to~\cite{Miskovic:2009dd,Miskovic:2009uz} for further information about the behavior/interpretation of the line element in the other regions.

\textit{C-Metric and parameters:} Let's begin with recalling the asymptotically anti-de-Sitter form of the C-metric. This metric will be used to construct quantum-corrected versions of the static BTZ geometry discussed above. Here we emphasize that although we will concern ourselves with the static geometry, quantum corrections of the stationary BTZ geometry can obtained from the rotating C-metric~\cite{Emparan:2020znc}. While this metric has been discussed in different forms in the literature, we find the following form~\cite{Emparan:2020znc} most useful for our current purpose
\begin{equation}\label{eq:cmetric}
    \exd s^2= \frac{\ell^2}{\ell+x r}\left(   -H(r)\exd t^2 +\frac{\exd r^2}{H(r)}+r^2\left(\frac{\exd x^2}{G(x)}+G(x)\exd \phi^2 \right)\right)
\end{equation}
where the functions $H(r)$ and $G(x)$ are defined as,
\begin{align}
    H(r) &= \frac{r^2}{\ell_3^2}+\kappa-\frac{\mu \ell}{r}\, , \label{eq:CmetricCompH}\\
    G(x)&=1-\kappa x^2 -\mu x^3 \, .\ \label{eq:CmetricCompG}
\end{align}
Although computationally we will find this form to be the most useful, we will gain a deeper understanding of the various parameters introduced in eq.\ (\ref{eq:CmetricCompH}) and eq.\ (\ref{eq:CmetricCompG}) through an additional parameterization of the metric. It is to the various meanings of these parameters that we turn to now. The metric (\ref{eq:cmetric}) is a solution to Einstein's equations in $3+1$ dimensions with cosmological constant $\Lambda=-3/\ell_4^2$. Inserting eq.\ (\ref{eq:cmetric}) into Einstein's equations yields a relation between the four-dimensional AdS radius $\ell_4$ and the parameters $\ell$ and $\ell_3$ given by
\begin{equation}
    \ell_4=\left(\frac{1}{\ell^2}+\frac{1}{\ell_3^2}\right)^{-1/2}
\end{equation}
It is also inversely proportional to the acceleration of the black holes $A=1/\ell$. Furthermore, we will take this parameter to be real, to rule out de-Sitter branes, with a range of $0\le \ell < \infty$. The parameter $\ell$ as we will soon see directly controls the tension of the brane (see eq.\ (\ref{eq:brane_tension}) ).
The parameter $\ell_3$ is the radius of the AdS geometry which will be induced on the brane. As we will see the parameter $\mu$ can be removed in favor of the horizon radius or seen as a derived parameter given in terms of the roots of metric component $G$ which ensures the regularity of the symmetry axes of $\partial_\phi$. It is also simply given as a combination of the acceleration parameter $A$ and the mass of the black hole $\mu=2 m A$. The final parameter, $\kappa$, is a discrete parameter taking the values $\kappa=\pm 1,0$. As we will see, the choice $\kappa=-1$ leads to a quantum-corrected BTZ geometry, while the choice $\kappa=+1$
will lead to another interesting quantum corrected geometry, the quantum dressed conical singularity.

Before moving further it is worth looking at the regularity of the symmetry axis $\partial_\phi$ in more detail. Fixed points of translation generators for compact parameter ranges correspond to rotations. That is $\partial_\phi^2=G(x)=0$,
hence the roots $x_i$ of $G(x)$ define the location of fixed points of the symmetry generator. The change of variables $x=z-3\kappa/\mu$ further simplifies the equation as~\cite{Panella:2024sor}
\begin{equation}
    z^3+p z+ q =0\, ,\quad p=-\frac{\kappa ^2}{3 \mu ^2}\, ,\quad q=\frac{2 \kappa ^3}{27 \mu ^3}-\frac{1}{\mu } \, .
\end{equation}
The types of roots are determined by the discriminant $\varsigma=-(4 p^3+27q^2)=(4 \kappa ^3-27 \mu ^2)/\mu ^4$ and can be categorized as
\begin{equation}
  \varsigma  \hspace{0.1cm}\begin{cases}
        > 0 & \text{3 real roots} \\
        = 0 & \text{2 repeated roots} \\
        < 0 & \text{1 real root, 2 complex roots}
    \end{cases}
\end{equation}
For our purposes we are interested in having at least one real root, if we make the assumption that $\mu\ge 0$ then for $\kappa=-1,0$, $\varsigma$ is strictly less than zero and there is one real root. For $\kappa=1$ for $\mu<2/(3\sqrt{3})$ there are 3 real roots and for $\mu>2/(3\sqrt{3})$ there is 1 real, positive, root and 2 complex roots. In all cases, we see that there exist at least one real root, we will label the smallest positive real root of $G(x)$ as $x_1$. In what follows we find it significantly simpler to use $x_1$ as the primary parameter, leaving $\mu$ as a derived parameter fixed in terms of $x_1$ as
\begin{equation}
    \mu=\frac{1-\kappa x_1^2}{x_1^3}
\end{equation}
The range of the parameter $x_1$ is then apparent from this equation considering that we assumed that $\mu\ge 0$ with
\begin{equation}
    x_1\in \begin{cases}
        (0,1] & \text{for} \hspace{0.1cm} \kappa=1 \\
         (0,\infty) & \text{for} \hspace{0.1cm} \kappa=-1,0
    \end{cases}
\end{equation}
Notice in each parameter range the derived parameter $\mu\rightarrow 0$ as $x_1$ approaches the upper range of the domain and the $\mu\rightarrow \infty$ as $x_1\rightarrow 0$. Furthermore, there is a transition of roots at $\mu=2/(3\sqrt{3})$ as $\mu$ moves from $\infty$ to $0$ where the root structure begins as a system of 1 real root and 2 complex roots for $\mu>2/(3\sqrt{3})$, then becomes a system of 3 real roots with one set of repeated roots which are arranged as $x_2=x_3<0<x_1$ for $\mu=2/(3\sqrt{3})$ and finally becomes 3 distinct real roots when $\mu<2/(3\sqrt{3})$ which can be arranged as $x_3<x_2<0<x_1$. Importantly, the real root denoted as $x_1$ changes continuously across this range of parameters.

Again following~\cite{Panella:2024sor} a quick calculation shows there is a conical singularity at any of the real roots, however for our purposes only $x=x_1$ will play a role since we will excise the negative $x$-region of the geometry. We can express the metric near this point as
\begin{equation}
    \frac{\exd x^2}{G(x)}+G(x)\exd\phi^2\approx \frac{\exd x^2}{G'(x)(x-x_1)}+G'(x)(x-x_1)\exd\phi^2\, ,
\end{equation}
after a coordinate transformation $x'^2=4(x-x_1)/G'(x_1)$ one finds
\begin{equation}
  \frac{\exd x^2}{G'(x)(x-x_1)}+G'(x)(x-x_1)\exd\phi^2=\exd x'^  2+ \frac{G'(x_1)^2}{4}x'^2 \exd \phi^2
\end{equation}
Computing the length of a curve around the point $x'=0$ at constant radius $x'=R$ at fixed $r,t$ of period $T$ one obtains
\begin{equation}
    \Delta S=\int_0^T\sqrt{g_{\phi\phi}}\exd \lambda = T R |G(x_1)|/2
\end{equation}
Hence to maintain regularity we impose the periodicity of $\phi$ as
\begin{equation}\label{PhiDelta}
    \phi\sim \phi+2\pi \Delta \, ,\quad \Delta= \frac{2 }{|G'(x_1)|}=\frac{2 x_1}{3-\kappa  x_1^2}
\end{equation}
\textit{Brane metric and junction conditions:} To construct the qBTZ geometry, we need to place a brane within the ambient C-metric geometry. Hence we must ask where we should place this brane. It turns out that there is a particularly simple location in the bulk geometry to place it, the surface $x=0$. This is because this surface is referred to as totally umbilic, i.e. the extrinsic curvature $K_{ab}$ and the induced metric on the brane
\begin{equation}\label{eq:induced_metric}
    h_{ab}\exd x^a\exd x^b= \frac{\exd r^2}{H(r)}-\exd t^2 H(r)+r^2\exd\phi^2
\end{equation}
satisfy the relation
\begin{equation}
    K_{ab}=-\frac{1}{\ell}h_{ab} \, .
\end{equation}
For this reason, the location $x=0$ will be where we place the brane. In doing so we will have split our space in two, above and below the brane. Recall that splitting our space ``above'' and ``below'' a hypersurface $\Sigma$ results in us studying a distributional version of the Einstein equations where we have decomposed our metric as,
\begin{equation}\label{eq:metric_glue}
g_{\mu\nu}=g^+_{\mu\nu}\Theta(x)+g^-_{\mu\nu}\Theta(-x), \qquad \Theta(x)=\begin{cases}
1 & \text{for}\hspace{0.2cm} x>0 \\
0 & \text{for} \hspace{0.2cm}x<0 \\
DNE & \text{for} \hspace{0.2cm}x=0
\end{cases}\, .
\end{equation}
Since derivatives of this distribution are defined as the Dirac delta one searches for conditions that lead to a well-posed set of distribution-valued Einstein equations. Requiring that $\Theta(l)\delta(l)$, which is not well defined, does not appear in the Einstein equations leads us to the Israel junction conditions~\cite{Darmois:1927,Israel:1966rt,lanczos1922bemerkung,lanczos1924},
\begin{subequations}
\begin{align}
    [g_{\mu\nu}]&=0 \label{eq:junction_1} \\
    -\frac{\epsilon }{8\pi G_4} \left([K_{ab}]-h_{ab}[K]\right)&= S_{ab} \label{eq:junction_2}
\end{align}
\end{subequations}
where we have defined, $[A]\equiv \left(A^+ -A^-\right) |_\Sigma$ and $\epsilon=\pm 1$, with $+1$ for $\Sigma$ timelike and $-1$ when $\Sigma$ is spacelike~\cite{Poisson_2004}.  The term $S_{ab}$ is the stress-energy tensor tangential to the surface and arises from the inclusion of the brane action
\begin{equation}
    I_{brane}=\tau \int \exd^4x \delta(x) \sqrt{-h}\, .~\label{eq:brane_action}
\end{equation}
The brane will be two sided, with $\mathbb{Z}_2$ orbifold conditions, hence by direct calculation one finds that
\begin{equation}
  S_{ab}  = -\frac{\epsilon }{8\pi G_4} \left([K_{ab}]-h_{ab}[K]\right) = -\frac{1}{2\pi G_4 \ell}h_{ab}
\end{equation}
Comparing this with the stress-energy tensor computed from eq.\ (\ref{eq:brane_action}) one finds that the brane tension is given by
\begin{equation}\label{eq:brane_tension}
    \tau=\frac{1}{2\pi G_4\ell }
\end{equation}

\textit{Effective action on the brane:} Taken together with the original action, the total effective action on the brane can be constructed simply by computing
\begin{equation}
    I_{eff}=2I_{ct}+I_{brane}+I_{CFT}
\end{equation}
where $I_{ct}$ and $I_{CFT}$ are the counterterm and the CFT actions which come from integrating out the non-normalizable and normalizable modes respectively. This results in the following structure
\begin{equation}
    I_{eff}=\frac{\ell_4}{8\pi G_4}\int \exd^3x \sqrt{-h}\left[\frac{4}{\ell_4^2}\left(1-\frac{\ell_4}{\ell}\right)+R+\ell_4^2\left(\frac{3}{8}R^2-R_{ab}R^{ab} \right)+\cdots  \right] + I_{CFT}
\end{equation}
The curvatures are those induced on the brane.
The dots indicate higher powers of $\ell_4$. Now, it is useful to define an effective three-dimensional Newton's constant and ``quantum corrected'' three-dimensional cosmological constant as
\begin{equation}
    G_3=\frac{1}{2\ell_4} G_4 \, , \quad \frac{1}{L_3^2}=\frac{2}{\ell_4^2}\left(1-\frac{\ell_4}{\ell}\right)\, .
\end{equation}
Recalling that the four-dimensional AdS radius is given by $\ell_4^{-2}=\ell^{-2}+\ell_3^{-2}$ we can expand the quantum corrected AdS radius in a series in $\ell$ as
\begin{equation}\label{eq:corrected_ads_radius}
    \frac{1}{L_3^2}=\frac{1}{\ell_3^2}\left(1+\frac{\ell^2}{4\ell_3^2}+\cdots\right)\, .
\end{equation}
Inserting these definitions into the effective action we arrive at
\begin{equation}\label{eq:effective_action}
     I_{eff}=\frac{1}{16\pi G_3}\int \exd^3x \sqrt{-h}\left[\frac{2}{L_3^2}+R+\ell^2\left(\frac{3}{8}R^2-R_{ab}R^{ab} \right)+\cdots  \right] + I_{CFT}
\end{equation}
Varying the action with respect to the brane metric $h$ we arrive at the following equations of motion on the brane
\begin{align}
    8\pi G_3 \vev{T_{ab}}&= R_{ab}-\frac{1}{2}h_{ab}\left(R+\frac{2}{L_3^2}\right) \nonumber\\
    &+ \ell^2 \left[4\tensor{R}{_a^c}R_{bc}-\frac{9}{4}RR_{ab}-\nabla^2R_{ab}+\frac{1}{4}\nabla_a\nabla_b R \right. \nonumber\\
     &+ \left. \frac{1}{2}h_{ab}\left(\frac{13}{8}R^2-3R_{cd}R^{cd}+\frac{1}{2}\nabla^2R \right)\right] + O(\ell^4)
\end{align}
Notice that although the gravitational theory is a perturbative expansion in $\ell$, the metric on the brane is known exactly.

\textit{Branches of quantum black holes on the brane:} The metric $h$, induced on the brane, given in eq.\ (\ref{eq:induced_metric}) is clearly asymptotically AdS, but it is not canonically normalized~\cite{Emparan:2020znc}. This is due to the periodicity of the coordinate $\phi$, a simple coordinate transformation can fix this
\begin{equation}\label{eq:cannonical_transformation}
    t=\Delta \bar{t}\, , \quad \phi=\Delta\phib \, ,\quad \Delta r =\rb\, ,
\end{equation}
restoring $\phi$ to be $2\pi$ periodic, and leaving the metric in the following form
\begin{equation}\label{eq:quantum_blackhole}
   \exd s^2= \frac{\exd \rb^2}{ H\left(\rb\right)}-\exd \bar{t}^2 H\left(\rb\right)+\rb^2\exd\phib^2\, , \quad H(\rb)= \frac{\rb^2}{\ell_3^2}+\Delta ^2 \kappa -\frac{\Delta^3  \ell \mu }{\rb}\, .
\end{equation}
Our line element is noticeably different than that in~\cite{Emparan:2020znc}. First, what is given in eq.\ (\ref{eq:quantum_blackhole}) is precisely what one obtains from the procedure described in the previous discussion, finding the induced metric on the brane, eliminating the conical singularity, and putting the asymptotic form of the brane metric in canonical form. The function referred to as $F(M)$ in~\cite{Emparan:2020znc} is precisely given as $F(M)=\Delta^3\mu$, matching eq.\ (\ref{eq:quantum_blackhole}). The subleading coefficient, $\Delta^2\kappa$ requires more care. The authors of~\cite{Emparan:2020znc} identify this as
\begin{equation}
    8 \mathcal{G}_3M= -\Delta^2 \kappa
\end{equation}
appealing to the identification, in a theory given by the Einstein-Hilbert action, of the subleading term in $g_{tt}$ as the mass of the solution $8G_3 M$. However, this relation changes when one considers higher curvature terms~\cite{Cremonini:2009ih}. Making use of their results\footnote{In their notation, $d=3$, $g^2=\frac{1}{L_3^2} $, and $(\tilde{\alpha}_1,\tilde{\alpha}_2,\alpha_3)=(-3\ell^2/8,\ell^2,0)$. Notice the difference in the overall sign of the actions when making the comparison.} gives
\begin{equation}
    M=-\frac{1}{8 G_3} \Delta^2 \kappa \left(1+\frac{\ell^2}{2 L_3^2}\right)\approx-\frac{1}{8 G_3 \left(1-\frac{\ell^2}{2 L_3^2}\right)} \Delta^2 \kappa
\end{equation}
providing motivation for introducing a ``renormalized'' Newton's constant given by
\begin{equation}
   \mathcal{G}_3= G_3 \left(1-\frac{\ell^2}{2 L_3^2}+O\left(\frac{\ell}{L_3}\right)^4\right)\, ,
\end{equation}
which is equivalent to $\mathcal{G}_3=G_4/(2\ell)$ up to, but not including, $O(\ell/L_3)^4$ as found in~\cite{Emparan:1999fd}. The interpretation is that $\mathcal{G}_3$ is an all orders resummation of higher derivative corrections while $G_3$ is the bare Newton constant.

Finally, as can be seen from the analysis of the roots of the function $G(x)$, the solutions can be characterized into 2 distinct branches, with further subdivision for the three branches, as discussed in~\cite{Emparan:2020znc}, given by,
\begin{subequations}
\begin{align}
    \text{Branch 1a: } \quad & \kappa=+1,\quad 0< x_1 \leq 1\, , \\
    \text{Branch 1b: } \quad & \kappa=-1,\quad 0< x_1 < \sqrt{3}\, , \\
    \text{Branch 2: } \quad & \kappa=-1,\quad \sqrt{3}< x_1 < \infty \, .
\end{align}
\end{subequations}
As can be seen by inspecting the mass relationship these branches cover the range $-\frac{1}{8\mathcal{G}_3}\leq M \leq \frac{1}{24\mathcal{G}_3}$. Beginning with the negative mass range, for which $\kappa=+1$ and $x_1\in (0,1]$, it can be seen that the range contains the AdS spacetime, ($x_1=1$ where $F(M) =0$ and $M=-1/(8\mathcal{G}_3)$), however this is true for all values of $\ell$ and hence can be labeled as a \textit{quantum} anti-de Sitter spacetime~\footnote{The geometry is AdS, but the quantum corrections have renormalized the Newton constant.}. For values of the mass $-1/(8\mathcal{G}_3)<M<0$, at vanishing $\ell$, the solution corresponds to AdS with a conical defect. Hence for non-vanishing $\ell$ the solution in this range can be thought of as quantum-corrected conical singularities. Of note here is that unlike the classical solutions, which are horizonless, here the backreaction of the quantum matter on the geometry dresses the singularity with a horizon. Classically, when $M=0$, the solution is referred to as the vacuum, or vacuum AdS, which is separated from AdS at $M=-\frac{1}{8G_3}$ by a mass gap. The quantum dressing acts to smoothly connect the ground state to the qBTZ by a smooth family of quantum corrected conical singularities which again join at the location $M=0$ for which $F(M)=-8\ell/3$. Hence this vacuum AdS geometry also sees corrections, and will be referred to as the quantum AdS vacuum geometry. In other works the branch 1a is still referred to under the umbrella of quantum black holes and as part of the qBTZ black hole geometry. Here we will distinguish this branch by referring to it as the qCone solution. While the positive branch of masses will be referred to as the qBTZ solution in an effort to distinguish whats hiding behind the horizon. To avoid confusion, we will not rename the different branches, with the qCone covering branch 1a, the qBTZ covering branch 1b and 2.

The remaining branches of the solution, branch 1b and 2, exist within the range $0<M\leq1/(24\mathcal{G}_3)$. They are separated into two distinct branches due to the primary source of corrections to the BTZ black hole. The corrections to the BTZ black hole in branch 1b is dominated by the backreaction of Casimir stress-energy. While along branch 2 the correction is due to Hawking radiation in thermal equilibrium with the black hole. Further details this separation into 3 branches and on the physical interpretation of the origin of the corrections can be found in~\cite{Emparan:1999wa,Emparan:1999fd,Emparan:2002px,Emparan:2020znc}.

%%%%%%%%%%%%%%%%%%%%%%%%%%%%%%%%%%%%%%%%%%%%%%%%%%%%%%%%%%%%%%%%%%%%%%%%
\section{Field Equations}\label{app:fieldEQ}

%%%%%%%%%%%%%%%%%%%%%%%%%%%%%%%%%%%%%%%%%%%%%%%%%%%%%%%%%%%%%%%%%%%%%%%%%
\subsection{Scalar field equations}\label{app:fieldEQ_Scalar}
Consider a probe scalar field in the quantum backreacted AdS$_{2+1}$ geometry described in section~\ref{sec:background_geometry}. We can take the action of the probe scalar field $\Phi(\rb,v,\bar\phi)$ to be,
\begin{equation}
    S_{scalar}=\int \exd \rb \exd^2x \sqrt{-g} \left(\partial_{\mu}\Phi\partial^{\mu}\Phi+m^2\Phi^2\right) \, .
\end{equation}
The equation of motion for $\Phi$ is given by
\begin{equation}\label{eq:scalar_field_equation}
    \frac{1}{\sqrt{-g}}\partial_{\mu}\left(\sqrt{-g}g^{\mu\nu}\partial_\nu \Phi\right)=m^2\Phi
\end{equation}
One can solve eq.\ (\ref{eq:scalar_field_equation}) order by order near the asymptotic boundary of the AdS$_{2+1}$ spacetime to find that the scalar field behaves as
\begin{equation}~\label{eq:scalar_field_asymp}
    \Phi(\rb,x^i)\sim \Phi_{(0)}(x^i)\rb^{-\Delta_-}+\Phi_{(+)}(x^i)\rb^{-\Delta_+},\qquad \Delta_{\pm}=1\pm \sqrt{1+m^2\ell_3^2}\, ,
\end{equation}
where $x^i=(v,\phib)$ and $\Delta_{\pm}$ are the operator dimensions associated with a scalar operator $\mathcal{O}$ with conformal weights $(h_L,h_R)$ such that $h_L+h_R=\Delta_{\pm}$ and $h_R-h_L=0$. We will consider the source of the CFT to be $\Phi_{(0)}$ and the vev to be $\Phi_{(+)}$. Our goal will then to be compute QNM associated with the poles of the retarded Green's functions $G(\omega,n)$. Although in the previous section we described that AdS radius has received corrections, as seen in eq.\ (\ref{eq:corrected_ads_radius}), the near boundary analysis reveals that conformal weights of the scalar operators dual to $\Phi$ are not sensitive to this. Solving order by order near the horizon reveals the near horizon behavior of $\Phi=(\bar{r}-\bar{r}_h)^{\alpha}\bar{\Phi}$ is determined by the exponents
\begin{equation}
    \alpha=\left\{0,\frac{i \omega }{2 \pi  T}\right\}\, .
\end{equation}
The choice of $\alpha=0$ represents the infalling mode, while $\alpha=i\omega/(2\pi T)$ represents the outgoing mode. Since our goal is to study the QNMs, we will focus only on the choice of $\alpha=0$. Furthermore, since we have a static solution with Killing vectors $\partial_v$, $\partial_{\phib}$ we will use a Fourier ansatz for $\Phi$ given by
\begin{equation}\label{eq:mode_expansion}
    \Phi(\rb,v,\phib)=\sum_{n= -\infty}^{\infty}\int d\omega e^{-i (\omega v - n \phib)}\bar{\Phi}_n(\omega,\rb) \, .
\end{equation}
Inserting the Fourier expansion into eq.\ (\ref{eq:scalar_field_equation}) one finds following equation of motion for $\bar{\Phi}$
\begin{equation}
\bar{r} \left(\bar{r} H(\bar{r}) \bar{\Phi}_n ''(\bar{r})+\bar{\Phi}_n '(\bar{r}) \left(\bar{r} H'(\bar{r})+H(\bar{r})-2 i \bar{r} \omega \right)\right)-\bar{\Phi}_n (\bar{r}) \left(n^2+ \bar{r}\left(m^2 \bar{r}+i \omega \right)\right)=0 \, .
\end{equation}
When one takes the quantum correction parameter $\ell=0$ the equation of motion reduces to the equation one finds for the BTZ geometry~\cite{Cardoso:2001hn}. In this case one can find analytic solutions to the equations of motion, and the QNM are given by~\cite{Cardoso:2001hn}
\begin{equation}\label{eq:Exact_BTZ_omega}
       \frac{\omega}{\ell_3}=\pm n -2i M^{1/2}(n_z+1) \, , \quad n_z\in \mathbb{Z} \, ,
    \end{equation}
where $n_z$ is the overtone, and $M$ is the black hole mass. Indeed, organizing the field equation as an expansion in $\ell$ and taking $\bar{\Phi}=\bar{\Phi}_0+\ell^2 \bar{\Phi}_2+O(\ell)^4$ one can again find analytic solutions. However, these solutions, even at leading $\ell^2$ order are very complicated combinations of hypergeometric functions. As a result, we will need to resort to numerical methods to obtain the QNM frequencies. To this end, we will find it useful to work with a dimensionless radial coordinate given by $z=\bar{r}_h/\bar{r}$ such that the horizon is located at $z=1$. The procedure to obtain QNMs numerically is well documented in many works, see for instance~\cite{Jansen:2017oag}
as such we will only briefly describe the numerical construction~\footnote{The reference~\cite{Jansen:2020hfd} also provides a Mathematica notebook for the calculation of QNM. We do not use his notebook, rather we use our own numerical construction}. We begin by addressing the Dirichlet boundary condition, this can be easily incorporated by working with a scaled field $\bar{\Phi}_n=z^{1+\sqrt{1 + \ell_3^2 m^2} }\tilde{\Phi}_n$ resulting in a field equation for $\tilde{\bar{\Phi}}$ which, implicitly, includes all of the boundary conditions on $\Phi$. We then arrange the equation of motion for the function $\tilde{\bar{\Phi}}$ as a generalized eigenvalue problem
\begin{equation}\label{eq:eigenvalue_problem_for_phi}
    B_0\tilde{\Phi}_n=\omega B_1\tilde{\Phi}_n
\end{equation}
and discretize by making a truncated Chebyshev approximation of the field $\tilde{\bar{\Phi}}$. The equation of motion in this form is then represented by a matrix equation which is solved numerically, with Mathematica's Eigenvalues command. In what follows we will display the QNM frequencies as dimensionless quantities normalized to the temperature i.e. $\mathfrak{w}=\frac{\omega}{2\pi T}$.

%%%%%%%%%%%%%%%%%%%%%%%%%%%%%%%%%%%%%%%%%%%%%%%%%%%%%%%%%%%%%%%%%%%%%%%%%%%%%%%55
\subsection{Spinor field equations}\label{app:fieldEQ_Fermion}
In this appendix, we will consider a probe fermion field in the quantum backreacted AdS$_3$ geometry described in section~\ref{sec:background_geometry}. The analysis of fermions in the AdS/CFT correspondence requires some additional machinery~\cite{Henneaux:1998ch,Iqbal:2009fd,Ceplak:2019ymw}. We will follow the analysis of~\cite{Ceplak:2019ymw}. We can take the action of the probe fermion field $\psi(\rb,v,\phib)$ to be,
\begin{equation}
    S_{scalar}=\int \exd \rb \exd^2x \sqrt{-g} \left(i\bar{\psi}\left(\Gamma^MD_M-m\right)\psi \right)
\end{equation}
Here the conjugate spinor is given by $\bar{\psi}=\psi^\dagger \Gamma^0$. To work with fermions we will need to introduce a local Lorentz frame $\theta^a$ where the lowercase Latin index denotes the local Lorentz, or flat space, quantity and goes $a=0,1,2$ corresponding to the flat spacetime associated with $(v,\rb,\phib)$. The frame can be expanded in the coordinate cotangent space as $\theta^a=\theta^a_M \exd x^M$ where indexes given by capital Latin will denote the curved space quantities. The covariant derivative is given by
\begin{equation}
    D_M=\partial_M+\frac{1}{4} \omega_{abM}\Gamma^{ab}, \quad \Gamma^{ab}=\frac{1}{2}[\gamma^a,\gamma^b] \, .
\end{equation}
The Dirac equation which then follows from the action is
\begin{equation}
    (\Gamma^MD_M-m)\psi=0 \, .
\end{equation}
To solve the equation we begin by specifying the frame. Although a simple choice would be to take the ``square root of the metric" it will be easiest to work with a frame like that in~\cite{Ceplak:2019ymw}. To accomplish this factor out two powers of $\rb$ from the function $H$ and write the metric in Eddington-Finkelstein coordinates as
\begin{equation}
   \exd s^2 =\exd v (2 \exd \rb-\exd v \frac{\bar{r}^2}{\ell_3^2} \bar{H}(\bar{r}))+\rb^2\exd \phi^2
\end{equation}
where $\rb^2/\ell_3^2 \bar{H}=H$. The frame, or non-holonomic basis can be used to reconstruct the line element
\begin{equation}
    \exd s^2 = \eta_{ab}\theta^a\theta^b\, , \quad \eta=\text{diag}(-1,1,1)
\end{equation}
We will denote the inverse of the frame field as $e$ (the dreibein) in an effort to avoid confusion with the indexes and hence
\begin{equation}
    \theta^a_M e^M_b =\delta^a_b\, ,\quad \theta^a_M e^N_a =\delta^N_M \, .
\end{equation}
We take the following frame choice
\begin{equation}
    \theta^{0}=\frac{1+H(\rb)}{2\ell_3}\rb \exd v-\ell_3\frac{\exd\rb}{\rb},\quad  \theta^{1}=\frac{1-H(\rb)}{2\ell_3}\rb\exd v+\ell_3\frac{\exd\rb}{\rb},\quad  \theta^{2}=\rb\exd\phi \, .
\end{equation}
The spin connection can be generated from the frame field, dreibein and Chirstoffel connection or from the first Cartan equation (vanishing Torsion) as
\begin{equation}
    \tensor{\omega}{^a_b_P} =\theta^a_M e^N_b \Gamma^{M}_{NP} -e^M_b \partial_P \theta^a_M
\end{equation}
The components of this object are given by
\begin{subequations}
\begin{align}
    \omega_{10}&=\frac{\exd \rb}{\rb}+\frac{\exd v}{\ell_3^2} \left(\frac{1}{2} \rb^2 \bar{H}'(\rb)+\rb \bar{H}(\rb)\right)\, ,\quad   \omega_{12}=-\frac{1}{2\ell_3} \exd \phib  \rb (\bar{H}(\rb)+1)\, , \\
    \quad \omega_{02}&=-\frac{1}{2\ell_3} \exd \phib  \rb (\bar{H}(\rb)-1)
\end{align}
\end{subequations}
and all other components not given by symmetry vanish. We take the gamma matrices as
\begin{equation}
    \Gamma^a=(i \sigma^2,\sigma^3,\sigma^1) \, ,
\end{equation}
where $\sigma^i$ are the standard Pauli matrices. In asymptotically AdS$_{2+1}$ spacetime the number of components of a spinor is 2 and can be decomposed in terms of eigenvectors of the operator $\Gamma^{\bar r}$
\begin{equation}
    \Gamma^{\bar r} \psi_\pm =\pm \psi_\pm\, , \quad P_\pm=\frac{1}{2}(1\pm \Gamma^{\bar r})\, ,
\end{equation}
where $\psi_{\pm}$ are complex scalar functions and $\psi$ is given by
\begin{equation}
    \psi=\begin{pmatrix}
        \psi_+ \\
        \psi_-
    \end{pmatrix}\, .
\end{equation}
As with the scalar case we will make use of symmetries of the geometry to take a Fourier ansatz for $\psi$ given by
\begin{equation}
    \psi(\bar{r},v,\phib)=\sum_{k= -\infty}^{\infty}\int d\omega e^{-i \omega v +i n \phib}\psi_n(\omega,\rb)
\end{equation}
Inserting the Fourier ansatz into the equations of motion one finds, after some manipulation, the following coupled set of first-order differential equations
\begin{align}
   0&= \frac{1}{4} \psi_{-}(\rb) \left(\rb^2 H'(\rb)+2\ell_3 i H(\rb) (n+i m \rb)+2  \ell_3 (i n+ m \rb-2 i \ell_3\omega) \right) \nonumber \\
    &+\frac{1}{4} \psi_{+}(\rb) \left(\rb^2 H'(\rb)+H(\rb) (2 i n\ell_3-2 m\ell_3 \rb+4 \rb)-2 i\ell_3( n-i m \rb+2  \ell_3\omega \right)+\rb^2 H(\rb) \psi_{+}'(\rb)  \label{eq:psi_plus_eq}\\
    0&=\frac{1}{4} \psi_{-}(\rb) \left(\rb^2 H'(\rb)+2 H(\rb) ((\ell_3 m+2) \rb-i n\ell_3)+2 \ell_3(i n+ m \rb-2 i \ell_3\omega \right)\nonumber \\
    &+\frac{1}{4} \psi_{+}(\rb) \left(\rb^2 H'(\rb)+2\ell_3(- i n + m \rb) H(\rb)-2 i \ell_3(n- i m \rb+2 \ell_3\omega \right)+\rb^2 H(\rb) \psi_{-}'(\rb) \label{eq:psi_minus_eq}
\end{align}
which reduces to the equations of motion for a probe fermion in the BTZ geometry when $\ell=0$. Near the AdS boundary, the equation of motion can be solved term by term, leading to the near boundary expansion given by
\begin{equation}
  \psi_+=A(n) \rb^{-1+m\ell_3}  + B(n) \rb^{-2-m\ell_3}\, , \quad \psi_-=C(n) \rb^{-2+m\ell_3}  + D(n) \rb^{-1+m\ell_3}
\end{equation}
The leading contribution near the boundary is identified as the source $A(n)$ while the response is given by $D(n)$. The operator dimensions associated with the fermion field $\psi$ are given by $\Delta=m \ell_3+ 1$ corresponding to an operator $\mathcal{O}_{\psi}$ with conformal weights $(h_L,h_R)$ such that $h_L+h_R=\Delta$ and $h_R-h_L=\pm 1/2$. Our goal then, will be to compute QNM associated with the poles of the retarded Green's functions $G(\omega,n)=\vev{\mathcal{O}_{\psi}\mathcal{O}_{\psi}}$. Notice again, just as in the scalar case, the near boundary analysis reveals that conformal weights of the operators dual to $\psi$ are not sensitive to the ``quantum corrected'' AdS radius. Solving order by order near the horizon one finds the near horizon behavior $\psi=(\bar{r}-\bar{r}_h)^{\alpha}\tilde{\psi}$ determined by the exponents
\begin{equation}
    \alpha=\left\{0,-\frac{1}{2}+\frac{i \omega }{2 \pi  T}\right\}\, .
\end{equation}
The choice of $\alpha=0$ represents the infalling mode, while $\alpha=-1/2+i\omega/(2\pi T)$ represents the outgoing mode. Since our goal is to study the QNMs, we will focus only on the choice of $\alpha=0$.

When we take the quantum backreaction to zero one can find analytic solutions to the equations of motion, and the QNM are given by~\cite{Cardoso:2001hn,Birmingham:2001pj,Iqbal:2009fd}
\begin{equation}\label{eq:btz_fermion_poles}
    \omega= -n -4\pi i T_R(n_z+h_R)\, , \quad \omega= n -4\pi i T_L(n_z+h_L)\,, \quad n_z\in\mathbb{Z}_+\, .
\end{equation}
where the conformal weights and associated weights are given by
\begin{equation}
    (h_L,h_R)=\left(\frac{m\ell_3}{2}+\frac{1}{4},\frac{m\ell_3}{2}+\frac{3}{4}\right)\,
\end{equation}
and $T_L,T_R$ are left and right moving temperatures which are equal when there is no rotation ($J=0$) i.e. $T_L=T_R=T$. The more general case will be discussed briefly in section~\ref{sec:fermion_critical}.
As in the scalar case, one can organize an expansion in the quantum backreaction $\psi=\psi_0+\ell^2\psi_2+O(\ell)^4$ and perturbatively find analytic solutions to the equations of motion. However, they are a very complicated combination of hypergeometric functions, with no clear structure. As a result analytic analysis here is also restricted, hence we again turn to numerical solutions for the QNMs. Given that we take $\alpha=0$ the remaining issue is imposing a Dirichlet boundary condition at the conformal boundary. The Green's function is given roughly by $G_R\sim D(n)/A(n)$ and hence to locate poles of the correlator we need to impose that the source vanishes. The simplest method to impose such a condition and obtain QNM is to work with scaled fields such that
\begin{equation}
   \psi_+(\rb)=\rb^{-2-m\ell_3}\psi_{+s}(\rb) \, , \quad \psi_-(\rb)=  \rb^{-1+m\ell_3}\psi_{-s}(\rb)\, .
\end{equation}
With these choices of scaling we obtain a coupled set of first-order differential equations with all boundary conditions implicitly imposed in the equations of motion. The rest of the procedure follows much like the scalar case, we make use of a truncated Chebyshev representation of the field equations to represent the continuous problem as a matrix system. And we will find it convenient again to work with a dimensionless radial coordinate given by $z=\bar{r}_h/\bar{r}$ such that the horizon is located at $z=1$. We can then rearrange the field equation for $\psi_{\pm s}$ into the form
\begin{equation}\label{eq:eigenvalue_problem_for_psi}
  (B_0(n)+\omega B_1(n))\Psi=0,
  \quad  B_i=B_i^{(0)} I+B_i^{(1)}\frac{\exd }{\exd z}
\end{equation}
where $\Psi=(\psi_{+s},\psi_{-s})$ which can be solved as a generalized eigenvalue problem using Mathematica's Eigenvalues command.

%%%%%%%%%%%%%%%%%%%%%%%%%%%%%%%%%%%%%%%%%%%%%%%%%%%%%%%%%%%%%%%%%%%%
\section{Details on pole-skipping}\label{app:pole}
\subsection{Scalar field}\label{app:pole_scalar}
Recall the equation of motion for the probe scalar matter given by,
\begin{equation}
\bar{r} \left(\bar{r} H(\bar{r}) \Phi ''(\bar{r})+\Phi '(\bar{r}) \left(\bar{r} H'(\bar{r})+H(\bar{r})-2 i \bar{r} \omega \right)\right)-\Phi (\bar{r}) \left(q^2+ \bar{r}\left(m^2 \bar{r}+i \omega \right)\right)=0
\end{equation}
Near $\bar{r}=\bar{r}_h$ we have
\begin{align}
    H(\bar{r})&=(\bar{r}-\bar{r}_h)H'(\bar{r}_h)+O\left((\bar{r}-\bar{r}_h)^2\right)\, , \\ \Phi&=\Phi^{(0)}+(\bar{r}-\bar{r}_h)\Phi^{(1)}+O\left((\bar{r}-\bar{r}_h)^2\right)\, .
\end{align}
Inserting this behavior of the field $\Phi$ and the blackening factor into the differential equation for $\Phi$ and expanding one finds
\begin{equation}
   \Phi^{(1)} \left(H'(\bar{r}_h)-2 i \omega \right)+\Phi^{(0)} \left(-m^2-\frac{q^2}{\bar{r}_h^2}-\frac{i \omega }{\bar{r}_h}\right) +O\left((\bar{r}-\bar{r}_h)\right) =0
\end{equation}
In principle this equation relates the coefficients $ \Phi^{(0)}$ and  $\Phi^{(1)}$ together and order by order would relate all higher $\Phi^{(m)}$ to  $\Phi^{(0)}$. However, one can notice immediately that this is only true provided the bracketed quantities remain non-zero. There is a separate solution which leaves $\Phi^{(1)}$ undetermined, this is given by,
\begin{equation}\label{eq:pole_skipping_scalar_on_the_brane}
    \omega_* = -\frac{1}{2} i H'(\bar{r}_h)\, , \qquad q_*=\pm \frac{\sqrt{-\bar{r}_h \left(H'(\bar{r}_h)+2 m^2 \bar{r}_h\right)}}{\sqrt{2}}
\end{equation}
Using the relation between the blackening factor $H$ and the temperature as well as the relation between the mass of the bulk scalar field and the operator dimension we can rewrite this more succinctly as,
\begin{equation}
  \omega_*= -2\pi i T\, , \qquad q_*^2= \rb_h \left(-\frac{(\Delta -2) \Delta  \rb_h}{\ell_3^2}-2 \pi  T\right)
\end{equation}

This story is directly related to the singularity structure of the ODE. One can notice that the ODE can be written in a canonical form as,
\begin{equation}
  \Phi ''(\bar{r})+  \Phi '(\bar{r}) \left(\frac{H'(\bar{r})-2 i \omega }{H(\bar{r})}+\frac{1}{\bar{r}}\right)-\Phi (\bar{r}) \left(\frac{m^2 \bar{r}^2+q^2+i \omega \bar{r} }{ H(\bar{r})}\right)=0
\end{equation}
The event horizon, at the point $\bar{r}=\bar{r}_h$ is a regular singular point, near the horizon the equation takes the form
\begin{equation}
\Phi''+\frac{H_1-2 i \omega }{H_1 (\bar{r}-\bar{r}_h)}\Phi'  -\frac{m^2 \bar{r}_h^2+q^2+i \bar{r}_h \omega }{H_1 \bar{r}^2 (r-\bar{r}_h)}\Phi =0
\end{equation}
Similar to the BTZ black hole, the singularity at $\bar{r} = \bar{r}_h$ is mild, with the coefficient multiplying $\Phi(\bar{r})$ behaving as $(r - \bar{r}_h)^{-1}$ rather than $(r - \bar{r}_h)^{-2}$. Additionally, the horizon becomes a regular point, rather than a regular singular point, when the momentum $q$ and frequency $\omega$ take the values in eq.\ (\ref{eq:pole_skipping_scalar_on_the_brane}). Therefore, from a brane observer's perspective, the backreaction of quantum fields, leading to a semi-classical metric, preserves the singularity structure of massive scalar probes, consistent with previous studies on higher curvature corrections to pole-skipping points~\cite{Natsuume:2019vcv}. Pole-skipping thus occurs for correlators of the scalar operators dual to $\Phi$.

\textit{Higher Matsubara frequencies:} Working with the infalling solution where we take
\begin{equation}
    \Phi=\sum_{a=0}^\infty \Phi_a (\bar{r}-\bar{r}_h)^a
\end{equation}
and writing the equations order by order near the horizon one finds,
\begin{align}
    0&= L_{11} \Phi_0+ (\bar{r}_h^2 H'(\bar{r}_h)-2 i \bar{r}_h^2 \omega)\Phi_1 \\
     0&= L_{21} \Phi_0+L_{22} \Phi_1+ (4\bar{r}_h^2 H'(\bar{r}_h)-4 i \bar{r}_h^2 \omega)\Phi_2 \\
   0&= L_{31} \Phi_0+L_{32} \Phi_1+L_{33} \Phi_2+ (18\bar{r}_h^2 H'(\bar{r}_h)-12 i \bar{r}_h^2 \omega)\Phi_3  \\
   \vdots&=\hspace{1cm} \vdots \nonumber\\
   0&= \left(\sum_{l=0}^{b}{L_{b \, l}\Phi_{l-1}}\right)+ M_b \Phi_{b}
\end{align}
Where $M_b$ is a relation like the previous coefficients of the highest terms. What one sees is at a given level if $M_b=0$ is satisfied it is not possible to uniquely determine a solution. The condition $M_b=0$ determines $\omega_b=-2\pi b i T$ for $b\in\mathbb{Z}$. Here it is interesting to note that, as discussed in~\cite{Natsuume:2019vcv}, pole-skipping occurs at the higher Matsubara frequencies, and is not corrected by the higher-curvature theory obeyed on the brane. It is suggested in~\cite{Natsuume:2019vcv} that the appearance of this structure is the result of the strong coupling limit of the field theory. While it is still unclear the precise source of this occurrence of pole-skipping at Matsubara frequencies in holographic theories, it appears that a crucial criterion is that the near horizon behavior of the field (or master field in the context of other perturbations) takes the form
\begin{equation}\label{eq:near_horizon_requirement}
    \Phi \sim (\rb- \rb_h)^{\pm \frac{i \omega}{4\pi T}}
\end{equation}
in the ``Poincare'' coordinates used in eq.\ (\ref{eq:quantum_blackhole}). This case provided that the equation of motion takes a form
\begin{equation}
    0=\Phi''+P(\rb)\Phi'+Q(\rb)\Phi \, ,\quad
    P=\sum_{i=-1}^{\infty}P_{i}(\rb-\rb_h)^i\, , \quad  Q=\sum_{i=-1}^{\infty}Q_{i}(\rb-\rb_h)^i\, ,
\end{equation}
writing schematically in Eddington-Finkelstein coordinates. Hence it is necessary that the corrections to the geometry, or the solution to the equations of motion itself does not destroy this structure.

\subsection{Spinor field}\label{app:pole_spinor}
For spinors, the procedure to obtain the location of pole-skipping points roughly follows the analysis above. There are two ways we could approach the problem, we could either, work with the first-order system, or we can decouple the coupled first-order system for two second-order differential equations. We will use the first method and work directly with the first-order equations of motion. This analysis was originally done in~\cite{Ceplak:2019ymw} and is nicely explained there. We will follow their work closely, and before providing the results, we will briefly review their method.

We take a general expansion near the horizon as
\begin{equation}
      \psi_+=\sum_{a=0}^\infty \psi_{+}^{(a)} (\bar{r}-\bar{r}_h)^a\, ,\quad   \psi_-=\sum_{a=0}^\infty \psi_{-}^{(a)} (\bar{r}-\bar{r}_h)^a
\end{equation}
inserting this into the equations of motion, and expanding near the horizon the equations of motion schematically take the following form
\begin{equation}
    S_+= \sum_{a=0}^\infty S_{+}^{(a)} (\bar{r}-\bar{r}_h)^a=0\, ,\quad   S_-=\sum_{a=0}^\infty S_{-}^{(a)} (\bar{r}-\bar{r}_h)^a=0
\end{equation}
where $S_{\pm}$ is the equation of motion for the two independent spinor components given in eq.\ (\ref{eq:psi_plus_eq}) and eq.\ (\ref{eq:psi_minus_eq}). Combining these equations we can express the near-horizon expansion in a generic form
\begin{equation}\label{eq:nearHorizon_fermion}
    \begin{pmatrix}
        S_{+}^{(n)} \\ S_{-}^{(n)}
    \end{pmatrix}=M^{(nn)}
    \begin{pmatrix}
          \psi_{+}^{(n)} \\ \psi_{-}^{(n)}
    \end{pmatrix}
    +\cdots +
    M^{(n0)} \begin{pmatrix}
          \psi_{+}^{(0)} \\ \psi_{-}^{(0)}
    \end{pmatrix}
\end{equation}
The pole-skipping constraint, which ensures a single independent constraint on $\psi^{(n)}_{\pm}$, is given by
\begin{equation}
    \text{det} M^{(nn)}(\omega,q)=0\, .
\end{equation}
That is, if this is satisfied then we are unable to obtain a relation for both $\psi^{(n)}_{\pm}$ in terms of $\psi^{(0)}_{\pm}$. The matrix takes the same general form given by
\begin{equation}
     M^{(nn)}(\omega,q)=\begin{pmatrix}
 -\frac{i q \ell_3}{2}+\pi  \ell_3^2 (4n+1) T-i \ell_3^2 \omega -\frac{\ell_3 m\rb_h}{2} & \frac{i q \ell_3}{2}+\pi  \ell_3^2 T-i \ell_3^2 \omega +\frac{\ell_3 m\rb_h}{2} \\
 -\frac{i q \ell_3}{2}+\pi  \ell_3^2 T-i \ell_3^2 \omega -\frac{\ell_3 m\rb_h}{2} & \frac{i q \ell_3}{2}+\pi  \ell_3^2 (4n+1) T-i \ell_3^2 \omega +\frac{\ell_3 m\rb_h}{2} \\
     \end{pmatrix} \, ,
\end{equation}
as seen in~\cite{Ceplak:2019ymw} (notice we leave the AdS radius intact, and it is not the quantum corrected radius that appears, but the bare AdS radius $\ell_3$).  As such, the determinant of this matrix, which vanishes at the fermionic Matsubara frequencies, is precisely the same
\begin{equation}
    \text{det}(M^{(nn)})= 8 \pi n \ell_3^4 T \left((2n+1) \pi  T-i \omega \right)=0 \, ,
\end{equation}
and we therefore find
\begin{equation}
    \omega_n=-2\pi iT \left(n+\frac{1}{2}\right)\, .
\end{equation}
Before obtaining the momentum we pause and observe that the higher curvature corrections of the bulk theory have not left an imprint on the structure of the fermionic Matsubara frequencies, just as they had not for the scalar Matsubara frequencies.
However, as in the scalar case, the momentum will have corrections.

As is familiar in the pole-skipping story, and as stated above, at level $n$, a frequency given by $\omega_n$ above reduces the total number of equations by one, and hence one can only obtain a constraint on a linear combination of $\psi_\pm^{(n)}$. Without loss of generality~\footnote{In principle one should consider $\psi_c^{(n)}=\psi_+^{(n)}-\Gamma^0\psi_-^{(n)}$, although since only the combination is constrained, we are free to set one of them to zero.} one can take $\psi^{(n)}_-=0$ and assemble the equations up to order $n$, evaluated at $\omega_n$ as
\begin{equation}
    \begin{pmatrix}
        S^{(0)}_{+} \\
       S^{(1)}_{+} \\
       \vdots \\
       S^{(n)}_{+}
\end{pmatrix}=\mathcal{M}\begin{pmatrix}
         \psi^{(0)}_{+} \\
       \psi^{(0)}_{-} \\
       \vdots \\
       \psi^{(n)}_{+}
    \end{pmatrix}=\left.\begin{pmatrix}
        M^{(00)}_{++} & M^{(00)}_{++} & 0 & \multicolumn{3}{c}{\cdots} & 0 \\
         M^{(10)}_{++} & M^{(10)}_{+-} &  M^{(11)}_{++} & M^{(11)}_{+-}  & 0 & \cdots & 0 \\
         \vdots & \vdots & \vdots & \vdots & \vdots & \vdots & \vdots \\
         M^{(n0)}_{++} & M^{(n0)}_{+-} &  0 & 0 & 0 & \cdots & M^{(nn)}_{++}  \\
    \end{pmatrix} \right|_{\omega=\omega_n}\begin{pmatrix}
         \psi^{(0)}_{+} \\
       \psi^{(0)}_{-} \\
       \vdots \\
       \psi^{(n)}_{+}
    \end{pmatrix}=0
\end{equation}
where we have indexed the matrices that appear in eq.\ (\ref{eq:nearHorizon_fermion}) as
\begin{equation}
    M^{(nn)}=\begin{pmatrix}
         M^{(nn)}_{++} &  M^{(nn)}_{+-}\\
         M^{(nn)}_{-+} &  M^{(nn)}_{--}
    \end{pmatrix}
\end{equation}
If the matrix~\footnote{Notice that the second line is constructed with the first order equation for the coefficient $\psi_-^{(0)}$. This is no mistake, the first order system for $
\psi$ requires two boundary conditions. } $\mathcal{M}$ is invertible, then we will find all of the coefficients vanish, but if this matrix is not invertible we will be in the position of having two independent regular solutions at the horizon, hence we then must solve
\begin{equation}
    \text{det}(\mathcal{M})=0
\end{equation}
As noted in~\cite{Ceplak:2019ymw}, this matrix is linear in $q$ in each entry, hence the determinant is a polynomial in $q$ of order $2n+1$ with $2n+1$ solutions which represent the pole-skipping points of order $n$.

%%%%%%%%%%%%%%%%%%%%%%%%%%%%%%%%%%%%%%%%%%%%%%%%%%%%%%%%%%%%%%%%%%%%%%%%
\section{Details on critical points}\label{app:critical}
Roughly speaking, the dispersion relations, upon complexifying the momentum and frequency, can be assumed to arise from an implicit equation $P(\q,\w)=0$, a curve in $\mathbb{C}^2$ whose would-be solutions determine $\w(\q)$. Hence the question to ask is when, and therefore where in the complex plane, a local solution around the point $z_0=(\q_0,\w_0)$ exists. As discussed in~\cite{walker:1950alg,wall_2004,Amano:2023bhg} the analysis of the local behavior of the curve around $z_0$ begins by considering the intersections of the curve and a line parameterized by $z_0+(\alpha_0,\alpha_1) t$. Of particular interest are single intersections determined by $dP/dt|_{t=0}=0$. If both of the partial derivatives of $P$ vanish then this is referred to as a singular point, while if only $\partial_\w P=0$ this is referred to as a critical point. Local solutions, expansions of $\w$ as a function of $\q$, are guaranteed by the analytic implicit function theorem in the form of a Taylor expansion, for non-singular points, and an extension of the analytic function theorem in the form of Puiseux series, in the case of singular or critical points. To be more precise, let $r$ denote the order of differentiation i.e. $d^rP/dt^r$, and by extension all possible partial derivatives up to order $r$. A point is a singular point if, for all $r\ge 2$, all the $(r-1)$ partial derivatives vanish and at least one partial derivative at order $r$ is non-zero. This is referred to as a point of multiplicity $r$. A critical point is of multiplicity $r=1$, and only one of its partial derivatives of order $r=1$ vanishes. That is we find critical points when the conditions that $P(\q_c,\w_c)$ be analytic and
\begin{equation}\label{eq:mode_crossing_conditions}
    P(\q_c,\w_c)=0\, , \quad \partial_\w P(\q_c,\w_c)=0 \, ,
\end{equation}
are satisfied for a point $(\q_c,\w_c)$ in the complex momentum plane. Obtaining these solutions to these equations analytically is typically not possible and hence we turn to numerical solutions. Our method is well documented in~\cite{Cartwright:2021qpp,Cartwright:2024rus} so we will only briefly mention how we numerically construct solutions. Recall that we obtained the QNM for $\Phi$ and $\psi$ by representing the equations of motion in the form of a generalized eigenvalue problem in eq.\ (\ref{eq:eigenvalue_problem_for_phi}) and eq.\ (\ref{eq:eigenvalue_problem_for_psi}). While we can solve this problem directly for the eigenvalues and eigenvectors, by discretizing this problem by means of a Chebyshev representation of the fields, the mode spectrum itself is contained in the operator $M=M_0(q)+\omega M_1(q)$ where we have already extended the momentum from $n$ to $q$. Hence the determinant of this matrix furnishes a representation of the spectral curve $P(\q,\w)=\text{det}(M)$. Then, the mode crossing conditions can be solved by means of a Newton-Raphson root-finding scheme, where the derivative of the curve can be constructed with finite differences.

%%%%%%%%%%%%%%%%%%%%%%%%%%%%%%%%%%%%%%%%%%%%%%%%%%%%%%%%%%%%%
\section{Quasi-normal modes for other fixed parameters}
In this appendix, we collect some additional QNMs computed in the qCone and qBTZ geometry.
%%%%%%%%%%%%%%%%%%%%%%%%%%%%%%%%%%%%%%%%%%%%%%%%%%%%%%%%%%%%%%%%%%%%%%%%%%%%
\subsection{Scalar fields}\label{app:extra_modes}
\begin{figure}[h]
    \centering
\includegraphics[width=0.75\textwidth]{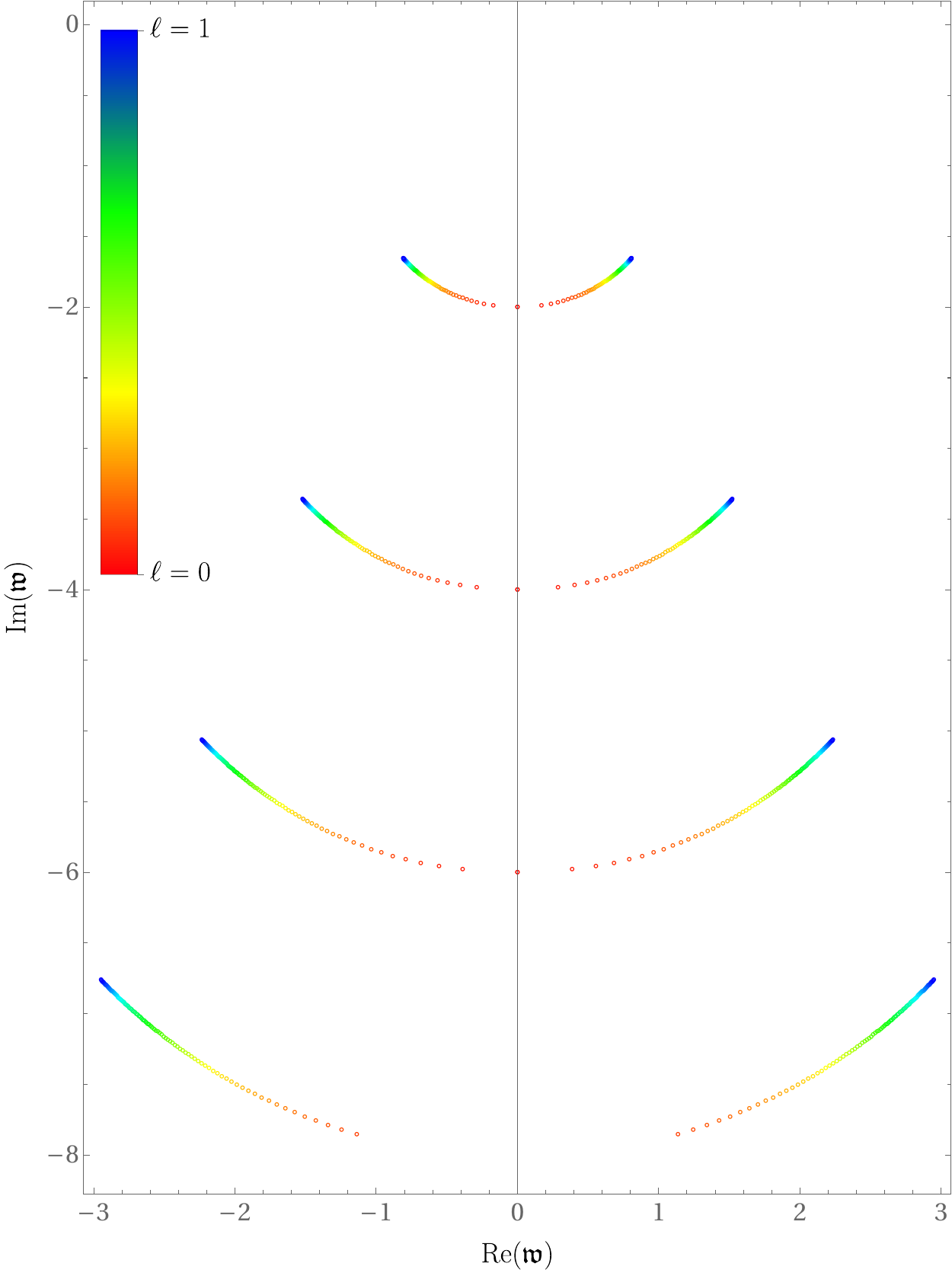}
    \caption{The QNM frequencies of a BTZ black hole are displayed as one slowly turns on quantum backreaction with each different curve representing a different overtone. The modes are displayed for fixed $\ell_3=1$, $n=0$ and $x_1=1/2$ as $\ell$ varies from $\ell=0$ to $\ell=1$. The figure appears to display that there are no modes for $\ell=0$ of the third overtone, however, this is just a reflection of the numerics breaking down.
    \label{fig:BTZ_TO_QUBTZ_appendix}}
\end{figure}

\begin{figure}[t]
    \centering
\includegraphics[width=0.9\textwidth]{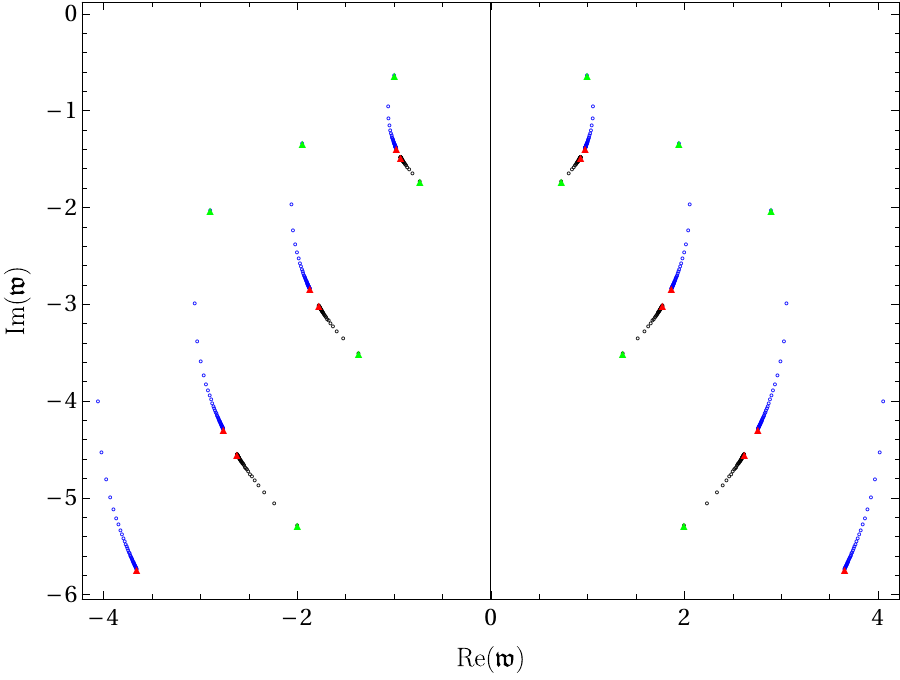}
    \caption{The lowest QNM frequencies are displayed as one slowly turns on quantum backreaction. The modes are displayed for fixed $\ell_3=1$, $n=0$ and $x_1=1/2$ as $\ell$ varies from $\ell=1/10$ (green triangles) to $\ell=4$ (red triangles)
    Blue dots represent the quantum dressed conical singularity ($\kappa=1$) while the black dots represent the quantum corrected BTZ geometry ($\kappa=-1$).
\label{fig:quBTZ_TO_QDConical_appendix}}
\end{figure}

For $\kappa=-1$, we display in figure~\ref{fig:BTZ_TO_QUBTZ_appendix} the behavior the QNMs as one switches on the quantum backreaction at zero momentum ($n=0$) not holding the temperature fixed.
The color coding is used to encode the value of the quantum correction parameter $\ell$, interpolating between red, denoting $\ell=0$, and blue, denoting $\ell=1$. We give the values of the QNM for the leading and first three overtones at zero momentum in table~\ref{tab:qubtzModes_appendix}. For $\kappa=+1$ we display the values of the QNMs for the leading and first three overtones at zero momentum of the qCone in table~\ref{tab:conSingModes_appendix}, again while not holding the temperature fixed. In figure~\ref{fig:quBTZ_TO_QDConical_appendix} the motion of the lowest QNMs, at zero momentum, in quantum BTZ black hole and the quantum dressed conical singularity as we increase the quantum backreaction.
The blue dots correspond to the quantum-dressed conical singularity and black dots to the quantum-corrected BTZ black hole. Beginning with here $\ell=1/10$, the green triangles, the QNMs are well separated. However, we see that as we increase the quantum back reaction, the QNMs move towards one another ending at $\ell=4$ (red triangles).
\newpage
\begin{landscape}
\begin{table}[t!]
    \centering
    \begin{tabular}{c |c c c c}
    $\ell$ & $n_z=0$ &  $n_z=1$&  $n_z=2$&  $n_z=3$\\
    \hline
       $0$  & $-2 i$ & $-4 i$ & $-6 i$  & $-8 i$ \\
      $0.1$   &   $\pm 0.471482 - 1.90323 i$ & $\pm 0.845842 - 3.83921 i$ & $\pm 1.21146 - 5.77052 i$  & $\pm 1.5772 - 7.70095 i$ \\
       $0.2$  &  $\pm 0.591629 - 1.83902 i$  & $\pm 1.08333 - 3.71752 i$ &  $\pm 1.57103 - 5.59054 i$ & $\pm 2.05911 - 7.46312 i $\\
      $0.3$   &  $\pm 0.657848 - 1.79381 i$ & $ \pm 1.2162 - 3.63019 i $ & $\pm 1.77239 -  5.46124 i $ & $ \pm 2.32886 - 7.29199 i $ \\
       $0.4$  & $\pm 0.700739 - 1.76003 i $ & $ \pm 1.30287 - 3.56451 i $ & $ \pm 1.90374 -  5.36392 i $ & $ \pm  2.5048 - 7.16306  $ \\
      $0.5$   &  $\pm 0.731106 - 1.73367 i$ & $\pm 1.36448 - 3.51307 i$ & $\pm 1.99715 -  5.28764 i$ & $\pm 2.62992 - 7.06196 i$ \\
      $0.6$  & $\pm  0.753888 - 1.71241 i$ & $\pm 1.41085 - 3.47149 i$ & $\pm 2.06743 -  5.22594 i$ & $\pm 2.72409 - 6.98015 i$\\
      $0.7$   &$\pm  0.771694 - 1.69481 i$ & $\pm 1.44716 - 3.43705 i$ & $\pm 2.1225 -  5.17481 i$ & $\pm 2.79788 - 6.91232 i$  \\
       $0.8$  & $\pm  0.786045 - 1.67997 i$ & $\pm 1.47649 - 3.40795 i$ & $\pm 2.16697 - 5.13159 i$ & $\pm 2.85747 - 6.85498 i$\\
      $0.9$   & $\pm 0.797892 - 1.66723 i$ & $\pm 1.50073 - 3.38296 i$ & $\pm 2.20373 - 5.09447 i$ & $\pm 2.90675 - 6.80572 i$  \\
       $1$   & $\pm  0.807859 - 1.65616 i$ & $\pm 1.52115 - 3.36123 i$ & $\pm 2.23471 - 5.06218 i$ & $\pm 2.94827 - 6.76286 i$ \\
    \end{tabular}
    \caption{\textbf{QNM of the qBTZ black hole ($s=0$):} The QNM frequencies, $\mathfrak{w}=\omega/(2\pi T)$, displayed here were computed with $\ell_3=1$, $x_1=1$ and for $\kappa=-1$ at zero momentum.
    \label{tab:qubtzModes_appendix}}
\end{table}
\begin{table}[b!]
    \centering
    \begin{tabular}{c |c c c c}
    $\ell$ & $n_z=0$ &  $n_z=1$&  $n_z=2$&  $n_z=3$\\
    \hline
      $0.1$   &   $\pm 0.995676 - 0.641248  i$ & $\pm 1.94569 - 1.33932  i$ & $\pm 2.89771 - 2.03577 i$  & $\pm 3.85013 - 2.73203 i$ \\
       $0.2$  &  $\pm 1.05855 - 0.95753 i$  & $\pm 2.05672 - 1.97526i$ &  $\pm 3.05621 - 2.99122 i$ & $\pm 4.05594 - 4.0069 i $\\
      $0.3$   &  $\pm 1.05442 - 1.08955 i$ & $ \pm 2.04208 - 2.23903i $ & $\pm 3.03095 - 3.38666 i $ & $ \pm 4.01997 - 4.53396 i $ \\
       $0.4$  & $\pm 1.04485 - 1.16113 i $ & $ \pm 2.01942 - 2.38169 i $ & $ \pm 2.99517 - 3.60031 i $ & $ \pm  3.97102 - 4.8186  $ \\
      $0.5$   &  $\pm 1.03607 - 1.2063 i$ & $\pm 1.99965 - 2.47159i$ & $\pm 2.96437 - 3.73487 i$ & $\pm 3.92918 - 4.9978 i$ \\
      $0.6$  & $\pm  1.02872 - 1.2376 i$ & $\pm 1.98339 - 2.53382 i$ & $\pm 2.9392 - 3.82798 i$ & $\pm 3.89508 - 5.12179  i$\\
      $0.7$   &$\pm  1.02261 - 1.26069 i$ & $\pm 1.97003 - 2.57971 i$ & $\pm 2.91857 - 3.89661  i$ & $\pm 3.86717 - 5.21316  i$  \\
       $0.8$  & $\pm 1.01749 - 1.27851 i$ & $\pm 1.9589 - 2.61509 i$ & $\pm 2.90144 - 3.94952 i$ & $\pm 3.84402 - 5.2836 i$\\
      $0.9$   & $\pm 1.01315 - 1.29272  i$ & $\pm 1.94951 - 2.64331 i$ & $\pm 2.88699 - 3.99171 i$ & $\pm 3.82451 - 5.33975 i$  \\
       $1$   & $\pm  1.00942 - 1.30436 i$ & $\pm 1.94148 - 2.6664 i$ & $\pm 2.87465 - 4.02623i$ & $\pm 3.80785 - 5.38571 i$ \\
    \end{tabular}
    \caption{\textbf{QNM of the qCone ($s=0$):} The QNM frequencies, $\mathfrak{w}=\omega/(2\pi T)$, displayed here were computed with $\ell_3=1$, $x_1=1/2$ and for $\kappa=1$ at zero momentum. Notice, there is no horizon for $\ell=0$, hence this row is omitted.
    \label{tab:conSingModes_appendix}}
\end{table}
\end{landscape}

\subsection{Spinor Fields}
For $\kappa=-1$, we display in table~\ref{tab:qubtzModes_s_1/2_non-zero-momentum_appendix} the behavior the QNMs as one switches on the quantum backreaction at zero momentum ($n=1$) not holding the temperature fixed. For $\kappa=+1$ in table~\ref{tab:quconeModes_s_1/2_non-zero-momentum_appendix} we display the values of the QNMs for the leading and first three overtones at non-zero momentum of the qCone. In an effort to avoid confusion, notice the normalization, for $x_1=1$ we have $M=1/4$ then $r_h^2/\ell_3^2-M=0$ gives $r_h=\ell_3/2$. Therefore $T=\frac{1}{4\pi}(\frac{2r_h}{\ell_3^2})=\frac{1}{2\pi} (\frac{ \ell_3}{2\ell_3^2})=\frac{1}{4\pi \ell_3}$ hence $2\pi T=\frac{1}{2\ell_3}$. Finally, from eq.\ (\ref{eq:btz_fermion_disp}) we have $\w=\pm\frac{n}{2\pi T}-2i(n_z+h_{R,L})$ or for $\ell_3=1$ we have $\w=\pm 2n -2i(n_z+h_{R,L})$. Therefore the lowest mode occurs at $n_z=0$ and is given by $\w=2-i/2$

\begin{figure}[t]
    \centering
\includegraphics[width=0.9\textwidth]{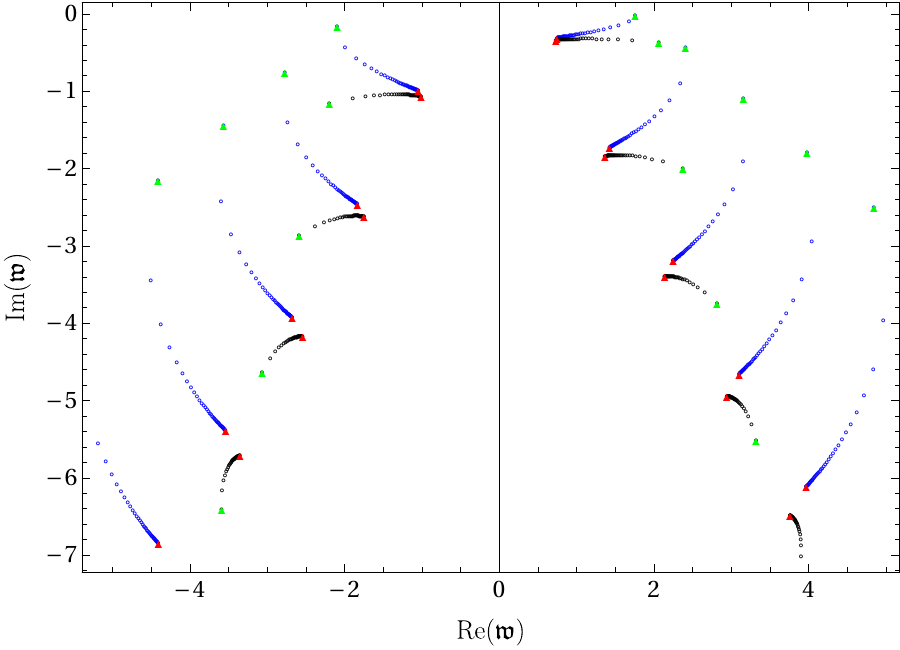}
    \caption{\textbf{Mode transition from quantum dressed conical singularity to quantum corrected BTZ: }The lowest QNM frequencies are displayed as one slowly turns on quantum backreaction for operators obeying $h_R-h_L=\pm 1/2$. The modes are displayed for fixed $\ell_3=1$, $n=1$ and $x_1=1/2$ as $\ell$ varies from $\ell=1/10$ (green triangles) to $\ell=4$ (red triangles). Blue dots represent the quantum dressed conical singularity ($\kappa=1$) while the black dots represent the quantum corrected BTZ geometry ($\kappa=-1$).
\label{fig:quBTZ_TO_QDConical_s_1/2_appendix}}
\end{figure}

In figure~\ref{fig:quBTZ_TO_QDConical_s_1/2_appendix} the motion of the lowest QNMs, at non-zero momentum in qBTZ black hole and the qCone geometry is displayed.
The blue dots correspond to the quantum-dressed conical singularity and black dots to the quantum-corrected BTZ black hole. The general trend of the modes is similar, beginning with here $\ell=1/10$, the green triangles, the QNMs are well separated. And as we increase the quantum back reaction, the QNMs move towards one another ending at $\ell=1$ (red triangles).
\newpage
\begin{landscape}
\begin{table}[t!]
    \centering
    \begin{tabular}{c |c c c c}
    $\ell$ & $n_z=0$ &  $n_z=1$&  $n_z=2$&  $n_z=3$\\
    \hline
 $0$ & $2\, -0.5 i $ & $ -2-1.5 i $ & $ 2\, -2.5 i $ & $ -2-3.5 i $ \\
 $0.1 $ & $ 1.72568\, -0.451197 i $ & $ -1.78723-1.37044 i $ & $ 1.87926\, -2.31323 i $ & $ -1.98829-3.26629 i $\\
 $0.2 $ & $ 1.56092\, -0.425003 i $ & $ -1.65847-1.30197 i $ & $ 1.80067\, -2.21483 i $ & $ -1.9655-3.14087 i $ \\
 $0.3 $ & $ 1.44714\, -0.40866 i $ & $ -1.56892-1.25945 i $ & $ 1.74357\, -2.15327 i $ & $ -1.94331-3.0609 i $\\
 $0.4 $ & $ 1.3621\, -0.397538 i $ & $ -1.50159-1.23047 i $ & $ 1.69935\, -2.11087 i $ & $ -1.92342-3.00485 i $\\
 $0.5 $ & $ 1.29518\, -0.389529 i $ & $ -1.44836-1.2095 i $ & $ 1.66364\, -2.0798 i $ & $ -1.90584-2.96315 i $\\
 $0.6 $ & $ 1.24059\, -0.383529 i $ & $ -1.40477-1.19367 i $ & $ 1.63391\, -2.05602 i $ & $ -1.89027-2.93078 i $\\
 $0.7 $ & $ 1.19487\, -0.378902 i $ & $ -1.36813-1.18135 i $ & $ 1.60859\, -2.03724 i $ & $ -1.87637-2.90486 i $\\
 $0.8 $ & $ 1.15578\, -0.375255 i $ & $ -1.33672-1.17152 i $ & $ 1.58665\, -2.02203 i $ & $ -1.86389-2.88361 i $\\
 $0.9 $ & $ 1.12181\, -0.372331 i $ & $ -1.30934-1.16353 i $ & $ 1.56736\, -2.00947 i $ & $ -1.85258-2.86586 i $\\
 $1 $ & $ 1.0919\, -0.369957 i $ & $ -1.28519-1.15694 i $ & $ 1.55019\, -1.99893 i $ & $ -1.84228-2.85079 i $\\
    \end{tabular}
    \caption{\textbf{QNM of the qBTZ black hole ($s=1/2$):} The QNM frequencies, $\mathfrak{w}=\omega/(2\pi T)$, displayed here were computed with $\ell_3=1$, $x_1=1$, $\kappa=-1$ and momentum $n=1$ for $s=1/2$.
    \label{tab:qubtzModes_s_1/2_non-zero-momentum_appendix}}
\end{table}
\begin{table}[b!]
    \centering
    \begin{tabular}{c |c c c c}
    $\ell$ & $n_z=0$ &  $n_z=1$&  $n_z=2$&  $n_z=3$\\
    \hline
$ 0.1 $&$ 1.75881\, -0.0244129 i $&$ -2.09516-0.165148 i $&$ 2.41061\, -0.435789 i $&$ -2.77191-0.757536 i $ \\
$ 0.2 $&$ 1.68071\, -0.108348 i $&$ -1.9925-0.443793 i $&$ 2.34249\, -0.910954 i $&$ -2.73661-1.40413 i $\\
$ 0.3 $&$ 1.54569\, -0.156861 i $&$ -1.847-0.582091 i $&$ 2.20433\, -1.12787 i $&$ -2.60237-1.69354 i $\\
$ 0.4 $&$ 1.43851\, -0.186052 i $&$ -1.7346-0.662749 i $&$ 2.09485\, -1.25148 i $&$ -2.49369-1.85665 i $\\
$ 0.5 $&$ 1.35486\, -0.205733 i $&$ -1.64816-0.716305 i $&$ 2.01025\, -1.33244 i $&$ -2.40936-1.96258 i $\\
$ 0.6 $&$ 1.28791\, -0.220109 i $&$ -1.57967-0.754987 i $&$ 1.94323\, -1.3903 i $&$ -2.34251-2.03776 i $\\
$ 0.7 $&$ 1.23289\, -0.231212 i $&$ -1.52382-0.784574 i $&$ 1.88865\, -1.43416 i $&$ -2.2881-2.09439 i $\\
$ 0.8 $&$ 1.18664\, -0.240141 i $&$ -1.47716-0.808156 i $&$ 1.84315\, -1.46882 i $&$ -2.24277-2.13891 i $\\
$ 0.9 $&$ 1.14704\, -0.247545 i $&$ -1.43741-0.827539 i $&$ 1.80447\, -1.49709 i $&$ -2.20428-2.17504 i $\\
$ 1 $&$ 1.11261\, -0.25383 i $&$ -1.40301-0.843853 i $&$ 1.77105\, -1.52072 i $&$ -2.17106-2.2051 i $\\
  \end{tabular}
    \caption{\textbf{QNM of the qCone ($s=1/2$):} The QNM frequencies, $\mathfrak{w}=\omega/(2\pi T)$, displayed here were computed with $\ell_3=1$, $x_1=1/2$, $\kappa=1$ and momentum $n=1$ for $s=1/2$.
    \label{tab:quconeModes_s_1/2_non-zero-momentum_appendix}}
\end{table}
\end{landscape}

%%%%%%%%%%%%%%%%%%%%%%%%%%%%%%%%%%%%%%%%%%%%%%%%%%%%%%%%%%%%%%%%%%%%%%%%%%%%%%%%%%%%%%%%%
\section{Numerical checks}
\begin{figure}[t!]
    \centering
    \includegraphics[width=0.48\textwidth]{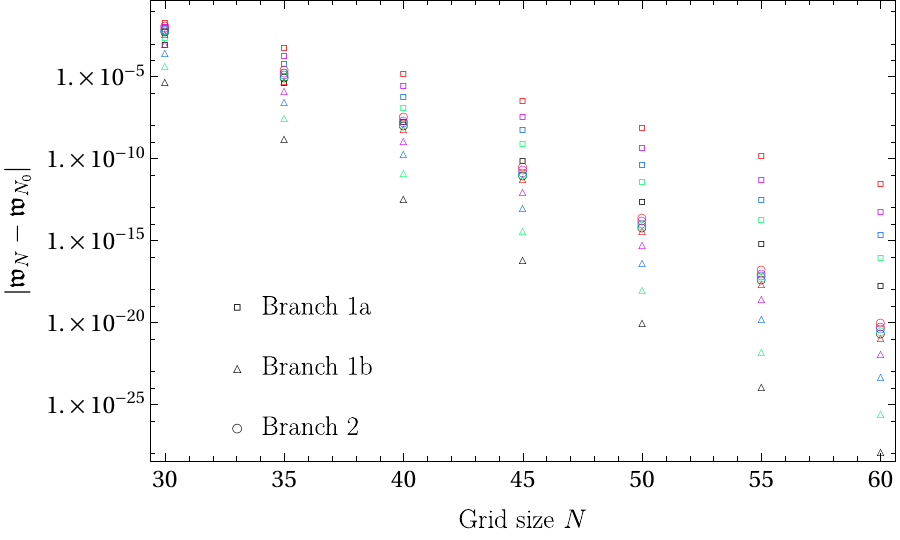} \hfill
    \includegraphics[width=0.48\textwidth]{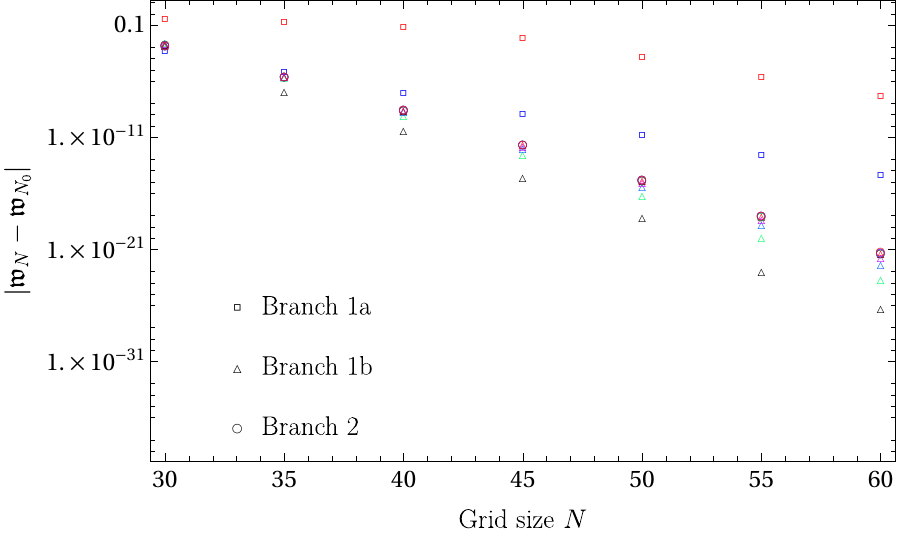}
    \caption{\textbf{Convergence:} The QNMs were calculated with $n=0$ and $4\pi T=1$. The various shaped markers indicate the different branches and the colors indicate different values of $\ell$ ranging from $\ell=0.1$ in red to $\ell=0$ in black. In the image, the rainbow-colored markers correspond to the code used throughout the draft, while the dots correspond to modes obtained from QNMspectral.
    \label{fig:error}}
\end{figure}

Here we give some additional information about the quality of the numerical approach. The QNMs computed in section~\ref{sec:probe_matter_and_QNMs} and section~\ref{sec:critical_points} were computed with 80-digit precision in Mathematica. The modes were computed for three grid sizes, $N=45,50,55$, and spurious modes where filtered removing modes which varied by more than $10^{-3}$ across the different grid sizes. A more accurate check is to solve for the eigenvalue $\w$ as well as the eigenfunction, which we will denote here as $X$ to collectively denote $\Phi$ and $\psi$. The eigenvalue and eigenfunction can be substituted back into the discretized differential equation obeyed by $X$ and the residual value of the equation is tabulated. Let $Y^i$ denote the residual value of the differential equation at each grid point $i\in 1 \cdots N$. We compute
\begin{equation}
    \frac{1}{N m}\sum_{i=0}^N |Y^i|
\end{equation}
where $m$ is the number of differential equations ($m=1$ for the scalar modes and $m=2$ for the spinor modes). This quantity denotes the residual value of the differential equation on average over the grid points. We have checked that this value is on the order $10^{-80}$ for the lowest few modes, indicating we are indeed finding a good solution to the differential equation. Furthermore, each mode can then be checked over various grid sizes, and the difference between the mode $\w_i$ at each of these grid sizes with the mode computed at the largest grid size $\w_{N_{0}}$, i.e. $|\w_N-\w_{N_{0}}|$, should decrease. This measure then indicates the scheme is converging to a fixed value. We display this measure in figure~\ref{fig:error} for a variety of $\ell$ values with $4\pi T=1$ and $N_0=65$. One can see that as we increase the grid size all modes converge towards the modes found at $N_0$. All three checks together indicate the mode is not spurious, represents a good solution to the differential equation and is converging to a fixed value.

\begin{figure}[t!]
    \centering
    \includegraphics[scale=0.6]{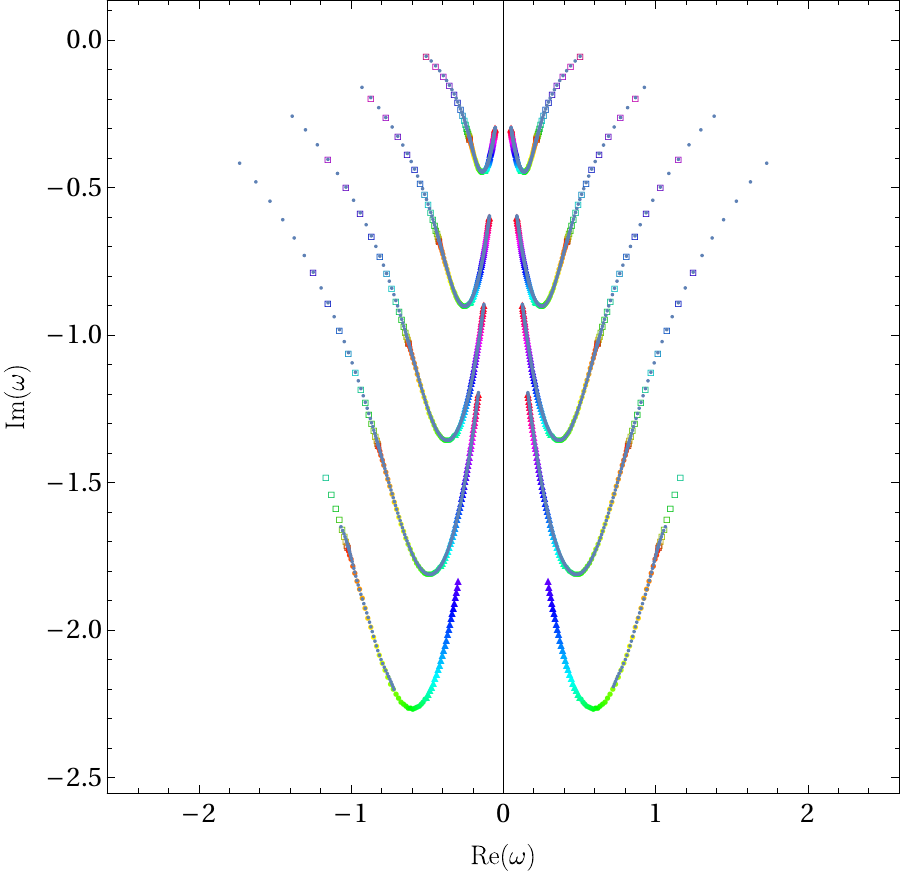}
    \caption{\textbf{Comparison of results from QNMspectral:} The QNMs were calculated with $n=0$, $\ell_3=3$ and $\ell=1$. In the image, the rainbow-colored markers correspond to the code used throughout the draft, while the dots correspond to modes obtained from QNMspectral.
    \label{fig:comparison}}
\end{figure}

As a further test of our numerics, we have also used the code QNMspectral~\cite{Jansen:2017oag} to test the code written for this work. The image in figure~\ref{fig:comparison} visually demonstrates agreement between our code and QNMspectral for the scalar QNMs.

\bibliographystyle{JHEP}
\bibliography{bib}

\end{document}